\documentclass[pra,longbibliography,twocolumn,showpacs,nofootinbib,superscriptaddress,notitlepage]{revtex4-2}

\pdfoutput=1 

\usepackage[ruled,vlined]{algorithm2e}

\usepackage{dsfont}
\usepackage{amsmath}
\usepackage{amssymb,bm}
\usepackage{amsthm}
\usepackage{mathtools}
\usepackage{url}

\usepackage[protrusion=true,expansion=true]{microtype}
\usepackage{booktabs}

\usepackage{array, makecell}
\usepackage{boldline}
\usepackage{enumitem}

\usepackage{color,dsfont}
\usepackage{graphicx}
\usepackage{ragged2e}
\usepackage[colorlinks=true, hyperindex, breaklinks, linkcolor=blue, urlcolor=blue, citecolor=blue]{hyperref} 
\usepackage[normalem]{ulem}
\usepackage[capitalise]{cleveref}
\usepackage{mathrsfs}
\usepackage[caption=false]{subfig}
\usepackage{mathtools}
\usepackage{float}
\usepackage{verbatim}
\usepackage{latexsym}
\usepackage{amsmath}
\usepackage{amssymb}
\usepackage{setspace}
\usepackage{amsfonts}
\usepackage{stmaryrd}
\usepackage{xcolor}
\usepackage{enumitem}
\usepackage{lipsum}

\theoremstyle{definition}
\newtheorem{definition}{Definition}[section]

\newcommand{\CC}[1]{\textcolor{red} {#1}}
\newcommand{\LG}[1]{\textcolor{blue} {#1}}

\newcommand{\SG}[1]{\textcolor{teal} {#1}}

\newcommand{\ket}[1]{| {#1} \rangle }

\usepackage{multirow}
\usepackage{booktabs}

\begin{document}
\title{Techniques for combining fast local decoders with global decoders under circuit-level noise}
\author{Christopher Chamberland}
\thanks{C.C was the main contributor to this work}
\affiliation{AWS Center for Quantum Computing, Pasadena, CA 91125, USA}
\affiliation{IQIM, California Institute of Technology, Pasadena, CA 91125, USA}
\author{Luis Goncalves}
\affiliation{AWS Center for Quantum Computing, Pasadena, CA 91125, USA}
\author{Prasahnt Sivarajah}
\affiliation{AWS Center for Quantum Computing, Pasadena, CA 91125, USA}
\author{Eric Peterson}
\affiliation{AWS Center for Quantum Computing, Pasadena, CA 91125, USA}
\author{Sebastian Grimberg}
\affiliation{AWS Center for Quantum Computing, Pasadena, CA 91125, USA}
\begin{abstract}
Implementing algorithms on a fault-tolerant quantum computer will require fast decoding throughput and latency times to prevent an exponential increase in buffer times between the applications of gates. In this work we begin by quantifying these requirements. We then introduce the construction of local neural network (NN) decoders using three-dimensional convolutions. These local decoders are adapted to circuit-level noise and can be applied to surface code volumes of arbitrary size. Their application removes errors arising from a certain number of faults, which serves to substantially reduce the syndrome density. Remaining errors can then be corrected by a global decoder, such as Blossom or Union Find, with their implementation significantly accelerated due to the reduced syndrome density. However, in the circuit-level setting, the corrections applied by the local decoder introduce many vertical pairs of highlighted vertices. To obtain a low syndrome density in the presence of vertical pairs, we consider a strategy of performing a syndrome collapse which removes many vertical pairs and reduces the size of the decoding graph used by the global decoder. We also consider a strategy of performing a vertical cleanup, which consists of removing all local vertical pairs prior to implementing the global decoder. Lastly, we estimate the cost of implementing our local decoders on Field Programmable Gate Arrays (FPGAs). 
\end{abstract}
\maketitle

\section{Introduction}
\label{sec:Intro}

Quantum computers have the potential to implement certain families of algorithms with significant speedups relative to classical computers \cite{S01p,Gr01p,Gr01a}. However, one of the main challenges in building a quantum computer is in mitigating the effects of noise, which can introduce errors during a computation corrupting the results. Since the successful implementation of quantum algorithms require qubits, gates and measurements to fail with very low probabilities, additional methods are required for detecting and correcting errors when they occur. Universal fault-tolerant quantum computers are one such strategy, where the low desired failure rates come at the cost of substantial extra qubit and gate overhead requirements \cite{shor1996fault,ChamberlandPRL,chamberland2017overhead,fowler2012surface,paetznick2013universal,Anderson14,Yoder2016,fowler2018low,litinski2019game,litinski2019magic,chamberland2020very,chamberland2020building,CC21}.

The idea behind stabilizer based error correction is to encode logical qubits using a set of physical data qubits. The qubits are encoded in a state which is a $+1$ eigenstate of all operators in a stabilizer group, which is an Abelian group of Pauli operators \cite{G01a}. Measuring operators in the stabilizer group, known as a syndrom measurement, provides information on the possible errors afflicting the data qubits. The results of the syndrome measurements are then fed to a classical decoding algorithm whose goal is to determine the most likely errors afflicting the data qubits. In recent decades, a lot of effort has been made towards improving the performance of error correcting codes and fault-tolerant quantum computing architectures in order to reduce the large overhead requirements arising from error correction. An equally important problem is in devising classical decoding algorithms which operate on the very fast time scales required to avoid exponential backlogs during the implementation of a quantum algorithm \cite{terhal2015quantum}. 

Several decoders have been proposed with the potential of meeting the speed requirements imposed by quantum algorithms. Cellular automata and renormalization group decoders are based on simple local update rules and have the potential of achieving fast runtimes when using distributed hardware resources \cite{HarringtonAutomata,BreuckmanHarrington,Herold_2017,KubicaAutomatav1,KubicaAutomatav2,DuclosRen1,DuclosRen2}. However, such decoders have yet to demonstrate the low logical failure rates imposed by algorithms in the circuit-level noise setting. Linear-time decoders such as Union Find (UF) \cite{DelfosseUnionFind} and a hierarchical implementation of UF with local update rules \cite{DelfosseHierarchical} have been proposed. Even with the favorable decoding complexity, further work is needed to show how fast such decoders can be implemented using distributed classical resources in the circuit-level noise regime at the high physical error rates observed for quantum hardware. Lastly, many NN decoders have been introduced, with varying goals \cite{TorlaiNN,JiangNN,VarsamopoulosNN1,Baireuther2018machinelearning,Breuckmann2018scalableneural,CR18NN,Sweke_2020,VarsamopoulosNN2,Andreasson2019quantumerror,Wagner_NN_Symmetries,VarsamopoulosV3,FitzekNN1,ShethNN1,XiaotongNiNN2,LaiaNN,GermanyNN,UsmanNN}. For NN decoders to be a viable candidate in universal fault-tolerant quantum computing, they must be fast, scalable, and exhibit competitive performance in the presence of circuit-level noise. 

In this work, we introduce a scalable NN decoding algorithm adapted to work well with circuit-level noise. Our construction is based on fully three-dimensional convolutions and is adapted to work with the rotated surface code \cite{TomitaSvore}. Our NN decoder works as a local decoder which is applied to all regions of the spacetime volume. By local decoder, we mean that the decoder corrects errors arising from a constant number of faults, with longer error chains left to be corrected by a global decoder. The goal is to reduce the overall decoding time by having a fast implementation of our local decoder, which will remove a large number of errors afflicting the data qubits. If done correctly, removing such errors will reduce the syndrome density, resulting in a faster implementation of the global decoder\footnote{Although sparser syndromes results in faster implementations of global decoders such as MWPM and UF, we leave the problem of optimizing such implementations using distributed resources for future work.}. We note that in the presence of circuit-level noise, the corrections applied by our local NN decoders can result in the creation of vertical pairs of highlighted syndrome vertices (also referred to as defects in the literature), which if not dealt with could result in an \textit{increase} in the error syndrome density rather than a reduction. To deal with this problem, we consider two approaches. In the first approach, we introduce the notion of a syndrome collapse, which removes a large subset of vertical pairs while also reducing the number of error syndromes used as input to the global decoder. Our numerical results show that competitive logical error rates can be achieved when performing a syndrome collapse after the application of the local NN decoders, followed by minimum-weight-perfect-matching (MWPM) \cite{Edmonds65} used as a global decoder. We achieve a threshold of approximately $p_{\text{th}} \approx 5 \times 10^{-3}$, which is less than the threshold of $p_{\text{th}} \approx 7 \times 10^{-3}$ obtained by a pure MWPM decoder due to information loss when performing the syndrome collapse. However, we observe a significant reduction in the average number of highlighted vertices used by the global decoder. On the other hand, a syndrome collapse reduces the surface codes timelike distance and would thus not be performed during a lattice surgery protocol.

The second approach consists of directly removing all vertical pairs after the application of the local decoder, but prior to the implementation of the global decoder. When removing vertical pairs, we observe a threshold which is greater than $5 \times 10^{-3}$ when MWPM is used as a global decoder. We also observe a reduction in the error syndrome density by almost two orders of magnitude in some physical noise rate regimes. This outperforms the reduction achieved by the syndrome collapse strategy, although the size of the decoding graph remains unchanged. We conclude our work with a resource cost estimate of the implementation of our NN decoders on FPGA's, and discuss room for future improvements. 

Our manuscript is structured as follows. In \cref{sec:SurfRev} we give a brief review of the rotated surface code and it's properties, and introduce some notation used throughout the manuscript. In \cref{section:AlgoRun}, we discuss how buffer times during the implementation of algorithms depend on decoding throughput and latency, and use such results to further motivate the need for fast decoders. \cref{sec:NNdecsLocal} is devoted to the description of our local NN decoder and numerical results. In \cref{subsec:NNdescription} we show how NN's can be used as decoders for quantum error correcting codes in the presence of circuit-level noise, and provide the details of our NN architectures and training methodologies. We discuss how representing the data can significantly impact the performance of our NN's, with more details provided in \cref{app:DataRep,appendix:HomEquivConv}. In \cref{subsec:SyndromeCollapse} we show how local decoders can introduce vertical pairs of highlighted vertices in the presence of circuit-level noise models, even when correcting all the data qubit errors resulting from such fault mechanisms. We then describe how we perform a syndrome collapse to remove vertical pairs and reduce the number of syndromes needed by the global decoder. In \cref{subsec:RemoveVerticalPairs} we provide an example correction from a local NN decoder, which illustrates the creation of vertical pairs of highlighted vertices. We then describe the vertical cleanup scheme for removing vertical pairs. We conclude \cref{sec:NNdecsLocal} by providing numerical results of our decoding protocols applied to the surface code in \cref{subsec:Numerics}. Lastly, in \cref{section:LatencyReduction}, we discuss the resource costs of implementing our local decoders on classical hardware.

\section{Brief review of the surface code}
\label{sec:SurfRev}

\begin{figure}
    \centering
    \includegraphics[width=0.8\columnwidth]{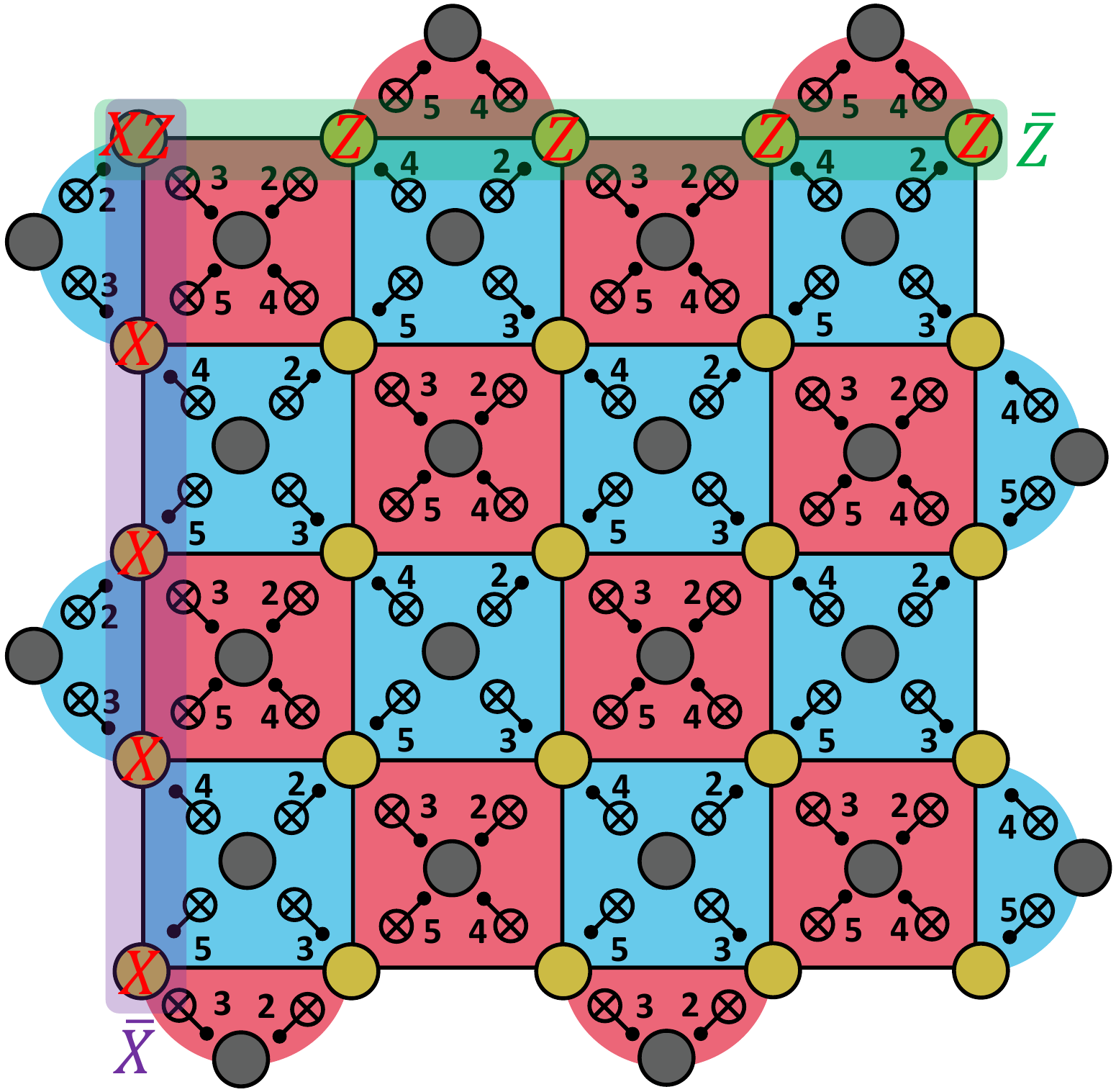}
    \caption{Example of a $d_x = d_z = 5$ surface code. The distances $d_x$ and $d_z$ correspond the minimum-weights of logical $X$ and $Z$ operators of the surface code. Minimum-weight representatives for the logical $\bar{X}$ and $\bar{Z}$ operators are shown in the figure, and form vertical and horizontal string-like excitations.  Data qubits correspond to the yellow vertices in the figure, and ancilla qubits (which are used to store the stabilizer measurement outcomes) are represented by grey vertices. Red plaquettes correspond to $X$-type stabilizers of the surface code, and blue plaquettes correspond to $Z$-type stabilizers. Numbers incident to CNOT gates used to measure the stabilizers indicate the time steps in which such gates are applied.}
    \label{fig:SurfaceCoded5}
\end{figure}

In this work we consider the surface code as the code used to correct errors during a quantum computation. An excellent introduction to the surface code is provided in Ref.~\cite{fowler2012surface}. In this section, we briefly review the properties of the rotated surface code \cite{TomitaSvore} and focus on the main features pertaining to the implementation of our scalable NN decoder. 

The surface code is a two-dimensional planar version of the toric code \cite{kitaev2003fault,dennis02}. The code parameters of the surface code are  $[\![d_xd_z, 1, \min{(d_x,d_z)} ]\!]$, where $d_x$ and $d_z$ are the distances of minimum-weight representatives of the logical $X$ and $Z$ operators of the code (which we refer to as the $X$ and $Z$ distance of the code). The logical $\bar{X}$ and $\bar{Z}$ operators of the code form vertical and horizontal string-like excitations. The surface code belongs to the family of Calderbank-Shor-Steane (CSS) codes \cite{CaldShorCSSv1,Steane95}, with the $X$ and $Z$-type stabilizers in the bulk of the lattice corresponding to weight-four operators. There are additional weight-two operators along the boundary of the lattice. An example of a $d_x=d_z=5$ surface code is shown in \cref{fig:SurfaceCoded5}. The weight-four $X$ and $Z$-type stabilizers correspond to the red and blue plaquettes in the figure, with the weight-two stabilizers being represented by semi-circles. We also define the error syndromes for CSS codes as follows:

\begin{definition}[Error syndrome]
Let $\mathcal{S_X} = \langle g^{(X)}_1, g^{(X)}_2, \cdots, g^{(X)}_{r_1} \rangle$ and $\mathcal{S_Z} = \langle g^{(Z)}_1, g^{(Z)}_2, \cdots, g^{(Z)}_{r_2} \rangle$ be the generating set of $X$ and $Z$-type stabilizers of a CSS code $\mathcal{C}$, and suppose the stabilizer measurements are repeated $d_m$ times. We define $s_X(d_m)$ to be a bit string $(e^{(1)}_{X} e^{(2)}_{X}  \cdots e^{(d_m)}_{X})$ where $e^{(k)}_{X}$ is a bit string of length $r_2$ with $e^{(k)}_{X}(j) = 1$ iff $g^{(Z)}_j$ is measured non-trivially in the $k$'th syndrome measurement round, and is zero otherwise. Similarly, we define $s_Z(d_m)$ to be a bit string $(e^{(1)}_{Z} e^{(2)}_{Z}  \cdots e^{(d_m)}_{Z})$ where $e^{(k)}_{Z}$ is a bit string of length $r_1$ with $e^{(k)}_{Z}(j) = 1$ iff $g^{(X)}_j$ is measured non-trivially in the $k$'th syndrome measurement round, and is zero otherwise.
\label{defXZsyns}
\end{definition}
Note that the $s_X(d_m)$ and $s_Z(d_m)$ syndromes in \cref{defXZsyns} can have non-zero bits due to both the presence of data qubit errors as well as measurement errors. We will also be particularly interested in syndrome differences between consecutive rounds, which are defined as follows:
\begin{definition}[Syndrome differences]
Given the syndromes $s_X(d_m) = (e^{(1)}_{X} e^{(2)}_{X}  \cdots e^{(d_m)}_{X})$ and  $s_Z(d_m) = (e^{(1)}_{Z} e^{(2)}_{Z}  \cdots e^{(d_m)}_{Z})$ for the code $\mathcal{C}$ defined in \cref{defXZsyns}, we set $s^{\text{diff}}_X(d_m) = (e^{(1)}_{X} \tilde{e}^{(2)}_{X}  \cdots \tilde{e}^{(d_m)}_{X})$, where $\tilde{e}^{(k)}_{X}$ is a bit string of length $r_2$ and $\tilde{e}^{(k)}_{X}(j) = 1$ iff the measurement outcome of $g^{(Z)}_j$ in round $k$ is \textbf{different} than the measurement outcome of $g^{(Z)}_j$ in round $k-1$ (for $k > 1$). Similarly, we define $s^{\text{diff}}_Z(d_m) = (e^{(1)}_{Z} \tilde{e}^{(2)}_{Z}  \cdots \tilde{e}^{(d_m)}_{Z})$, where $\tilde{e}^{(k)}_{Z}$ is a bit string of length $r_1$ and $\tilde{e}^{(k)}_{Z}(j) = 1$ iff the measurement outcome of $g^{(X)}_j$ in round $k$ is \textbf{different} than the measurement outcome of $g^{(X)}_j$ in round $k-1$ (for $k > 1$).
\label{defXZsyndiff}
\end{definition}

The standard decoding protocol used to correct errors with the surface code is by performing MWPM using Edmonds Blossom algorithm \cite{Edmonds65}. In particular, a graph $G$ is formed, with edges corresponding to the data qubits (yellow vertices in \cref{fig:SurfaceCoded5}) and vertices associated with the stabilizer measurement outcomes (encoded in the grey vertices of \cref{fig:SurfaceCoded5}). In order to distinguish measurement errors from data qubit errors, the error syndrome (measurement of all stabilizers) is repeated $r$ times (with $r$ being large enough to ensure fault-tolerance, see for instance the timelike error analysis in Ref.~\cite{CC21}). Let $m^{(k)}(g_i) = 1$ if the stabilizer $g_i$ in round $k$ is measured non-trivially and zero otherwise. Prior to implementing MWPM, a vertex $v^{(k)}(g_i)$ in $G$ associated with a stabilizer $g_i$ in the $k$'th syndrome measurement round is highlighted iff $m^{(k)}(g_i) \neq m^{(k-1)}(g_i)$, i.e. the syndrome measurement outcome of $g_i$ changes between rounds $k-1$ and $k$. More generally, for any fault location $l_k$ in the circuits used to measure the stabilizers of the surface code (for instance CNOT gates, idling locations, state-preparation and measurements), we consider all possible Pauli errors $P^{l_k}(j)$ at location $l_k$ (with $k$ indexing through all possible Pauli's) and propagate such Pauli's. If propagating the Pauli $P^{l_k}(j)$ results in two highlighted vertices $v^{(k_1)}(g_{j_1})$ and $v^{(k_2)}(g_{j_2})$, an edge $e$ incident to $v^{(k_1)}(g_{j_1})$ and $v^{(k_2)}(g_{j_2})$ is added to the matching graph $G$\footnote{For the surface code, a Pauli $Y$ error can result in more than two highlighted vertices, thus requiring hyperedges. Such hyperedges can then be mapped to edges associated with $X$ and $Z$ Pauli errors.}. For a distance $d_x=d_z=d$ surface code with $d$ rounds of syndrome measurements, the decoding complexity of MWPM is $\mathcal{O}(n^3)$ where $n \propto d^2$ and corresponds to the number of highlighted vertices in $G$ (see Ref.~\cite{Edmonds65} and \cref{subsec:DownstreamPerfReqs} for more details). The UF decoder, another graph based decoder, has decoding complexity of $\mathcal{O}(\alpha n)$ where $\alpha$ is the inverse of Ackermann’s
function. Remarkably, UF is able to achieve near linear time decoding while maintaining good performance relative to MWPM \cite{HuangWeightedUnionFind}.

Although MWPM and UF have polynomial decoding time complexities, decoders will need to operator on $\mu s$ time scales for many practical quantum hardware architectures (see \cref{section:AlgoRun}). Achieving such fast decoding times using MWPM and UF appears to be quite challenging~\cite{FowlerOhOne,FowlerTimingAnalysis,DasMicroarchitecture,DelfosseHierarchical}. To this end, in \cref{sec:NNdecsLocal} we use \textbf{scalable} NN's as local decoders that have an effective distance $d'$ and which can thus correct errors $E$ of weight $\text{wt}(E) \le (d'-1)/2$. MWPM and UF can then be used as a \textit{global} decoder to correct any remaining errors which were not corrected by the local decoder. The effect of the local decoder is to reduce the value of $n$ by removing many of the errors afflicting the data qubits. NN's have already been used as local decoders in the setting of code capacity noise (where only data qubits can fail, and error syndromes only have to be measured once) and phenomenological noise (where measurements can fail in adition to data qubits) \cite{UsmanNN,GermanyNN}. However, the presence of circuit-level noise introduces significant new challenges which require new methods to cope with the more complex fault patterns. 

Throughout the remainder of this manuscript, we consider the following \textit{circuit-level} depolarizing noise for our numerical analyses:
\begin{enumerate}
    \item Each single-qubit gate location is followed by a Pauli $X,Y$ or $Z$ error, each with probability $\frac{p}{3}$.
	\item With probability $p$, each two-qubit gate is followed
    by a two-qubit Pauli error drawn uniformly and
    independently from $\{I, X, Y, Z\}^{\otimes 2} \backslash \{I \otimes I\}$.
	\item With probability $\frac{2p}{3}$, the preparation of the $\ket{0}$ state is replaced by $\ket{1}=X\ket{0}$. Similarly, with probability $\frac{2p}{3}$, the preparation of the $\ket{+}$ state is replaced by $\ket{-}=Z\ket{+}$.
	\item With probability $\frac{2p}{3}$, any single qubit measurement
has its outcome flipped.
	\item Lastly, with probability $p$, each idle gate location
    is followed by a Pauli error drawn uniformly and
    independently from $\{X, Y, Z\}$.
\end{enumerate}

This noise model is similar to the one used in Refs.~\cite{ChamberlandHeavyHex,ChamberlandColorCode}. However, in this work, we treat each idle location during measurement and reset times of the ancillas as a single idle location failing with probability $p$ (instead of two idling locations each failing with probability $p$).

\section{The effects of throughput and latency on algorithm run-times}
\label{section:AlgoRun}

\begin{figure}
    \centering
    \includegraphics[width=0.8\columnwidth]{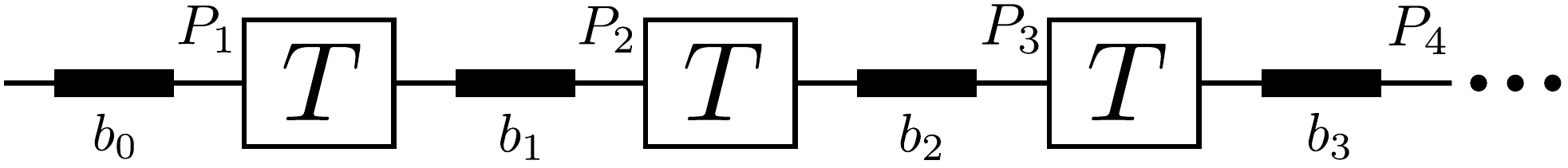}
    \caption{Sequence of $T$ gates separated by buffers $b_j$ (black rectangles). The Pauli operators $P_j$ indicate the Pauli frame immediately prior to implementing the $j$'th $T$ gate. During the buffer time $b_j$, repeated rounds of error correction are performed until the Pauli frame immediately prior to the $j$'th $T$ gate is known. }
    \label{fig:BuffTgates}
\end{figure}

In this section we discuss how latency and decoding times affect the run-time of algorithms. In what follows, we refer to inbound latency as the time it takes for the stabilizer measurement outcomes of an error correcting code to be known to the classical computer which implements the decoding task. By classical computer, we mean the classical device which stores and processes syndrome information arising from stabilizer measurements of an error correcting code in order to compute a correction. We specify ``inbound'' to distinguish this quantity from the ``outbound'' latency, or delay between the arrival of an error syndrome at the decoder and its resolution. We also refer to throughput as the time it takes for the classical computer to compute a correction based on the syndrome measurement outcome. 

We denote the Clifford group as $\mathcal{C}$ which is generated by $\mathcal{C} = \langle H,S, \text{CNOT} \rangle$, with the matrix representation for the Hadamard and phase gates in the computational basis expressed as $H = \frac{1}{\sqrt{2}} \begin{pmatrix} 1 & 1 \\ 1 & -1 \end{pmatrix}$ and $S = \text{diag}(1,i)$. The CNOT gate acts as $\text{CNOT} \ket{a}\ket{b} = \ket{a}\ket{a \oplus b}$.

\begin{figure}
	\centering
	\subfloat[\label{fig:TgateImpv1}]{%
		\includegraphics[width=0.33\textwidth]{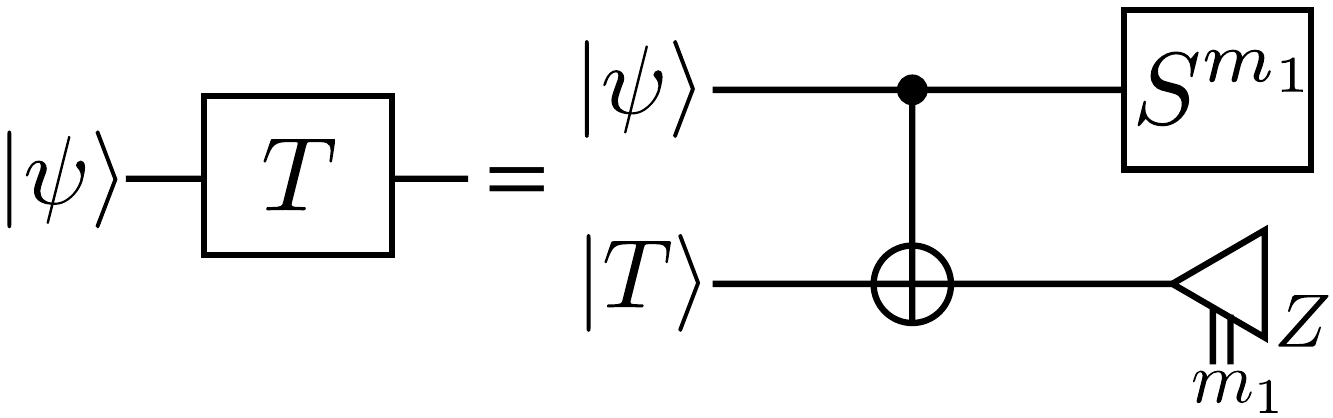}
	}
	\vfill
	\subfloat[\label{fig:TgateImpv2}]{%
		\includegraphics[width=0.39\textwidth]{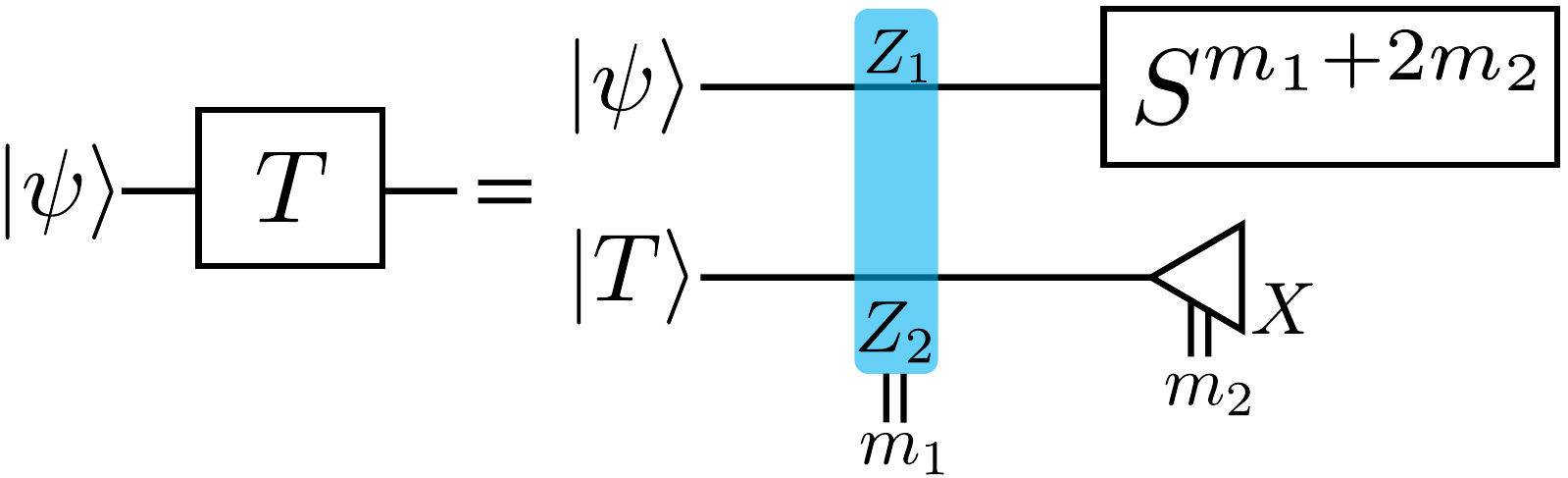}
	}

	\caption{\label{fig:TgateImp} Two equivalent circuits for implementing a $T$ gate. In (a), we show the standard circuit for implementing a $T$ gate using the magic state $\ket{T} = \frac{1}{\sqrt{2}}(\ket{0} + e^{i \pi /4}\ket{1})$ as a resource state. In (b), we provide an equivalent circuit where the logical CNOT gate is replaced by a $Z \otimes Z$ Pauli measurement, which can be implemented via lattice surgery, as discussed for instance in Ref.\cite{CC21}.}
\end{figure}

Consider the sequence of non-parallel $T = \text{diag}(1,e^{i \pi /4})$ gates shown in \cref{fig:BuffTgates}.  Note that $T$ gates are non-Clifford gates, and the set generated by $ \langle H,S, \text{CNOT}, T \rangle$ forms the basis of a universal gate set. We also consider a framework where we keep track of a Pauli frame \cite{Knill05,terhal2015quantum,ChamberlandPauliFrame} throughout the execution of the quantum algorithm. The Pauli frame allows one to keep track of all conditional Pauli's and Pauli corrections arising from error correction (EC) in classical software, thus avoiding the direct implementation in hardware of such gates, which could add additional noise to the device. Since $T P T^{\dagger} \in \mathcal{C}$, when propagating the Pauli frame through a $T$ gate, a Clifford correction may be required in order to restore the Pauli frame. Consequently, buffers are added between the sequence of $T$ gates where repeated rounds of EC are performed until the Pauli frame $P_j$ immediately before applying the $j$'th $T$ gate is known. The buffer immediately after the $j$'th $T$ gate is labeled as $b_j$. We now show how buffer times increase with circuit depth as a function of inbound latency and throughput. 

\begin{figure*}
	\centering
	\subfloat[\label{fig:PlotThroughput}]{%
		\includegraphics[width=0.45\textwidth]{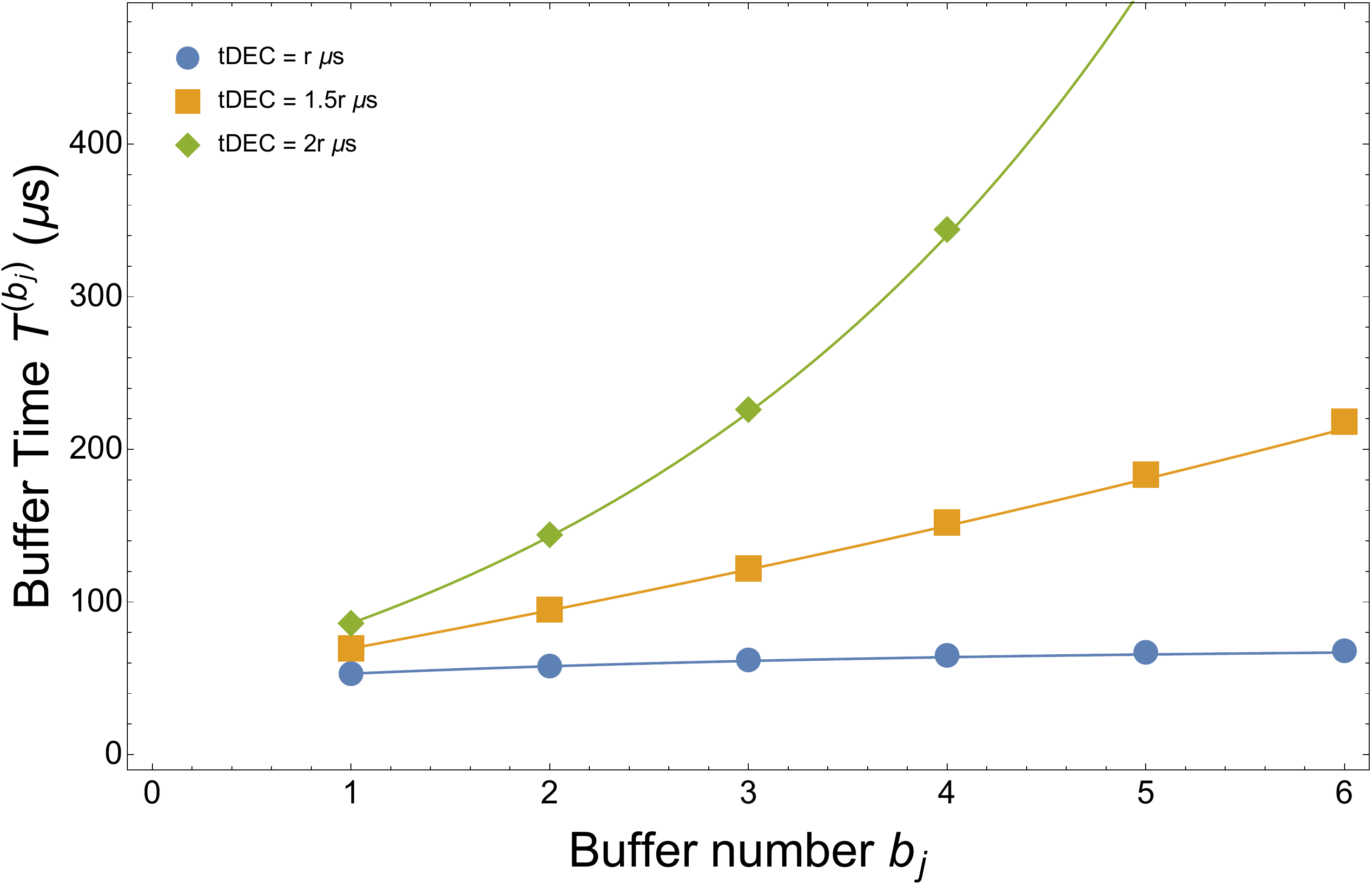}
	}
	\subfloat[\label{fig:PlotLatency}]{%
		\includegraphics[width=0.45\textwidth]{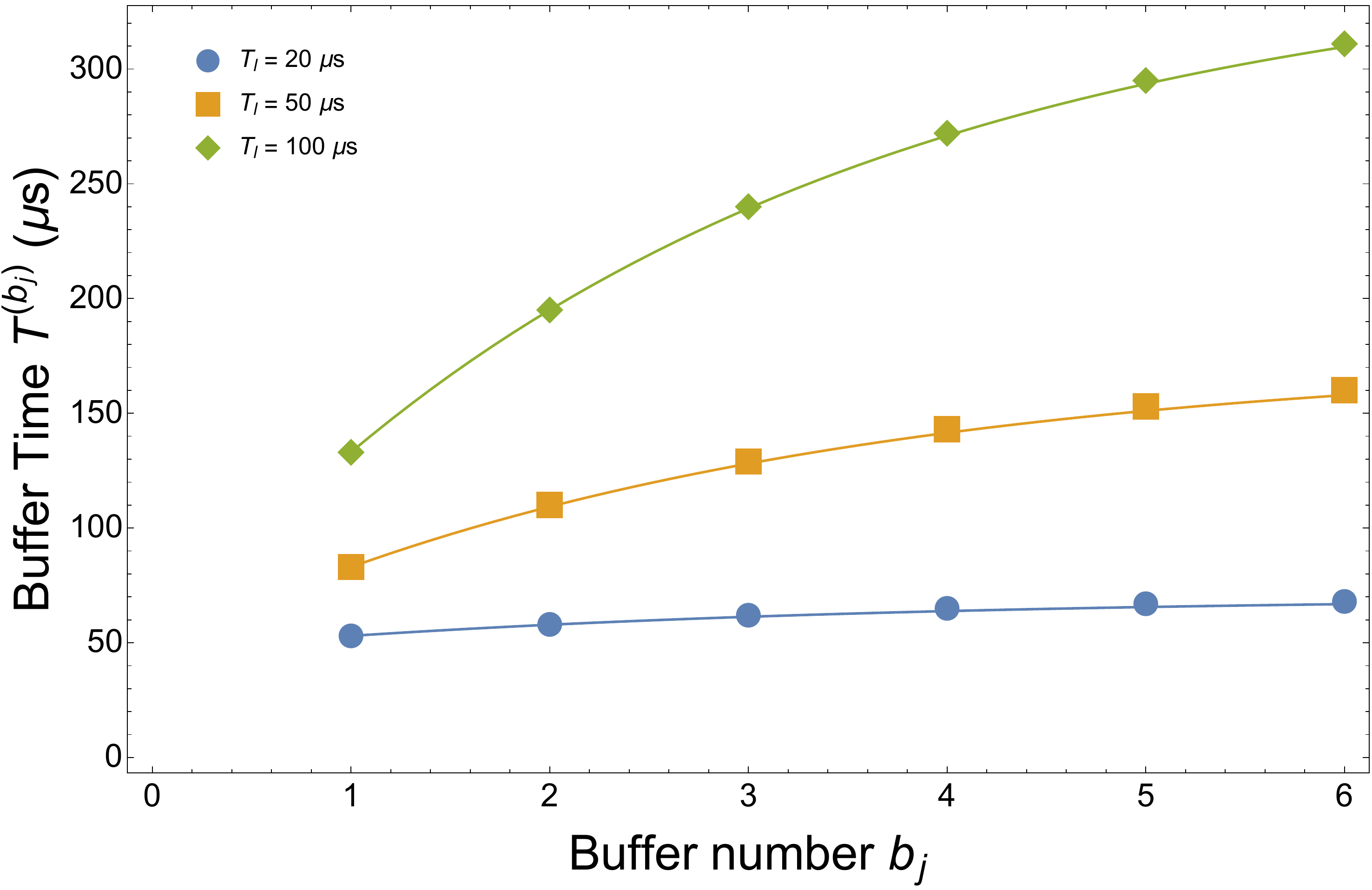}
	}

	\caption{\label{fig:BufferTimePlots} Plots showing the buffer times $T^{b_j}$ as a function of the buffer number $b_j$. We set $r_1 + r_2 = 33$ and consider using the surface code for performing EC. The surface code requires four time steps to implement all CNOT gates used to measure the codes stabilizers, and we assume each CNOT gate takes $100 \text{ns}$. We also assume a measurement plus reset time of the ancillas to be $1 \mu \text{s}$, resulting in a total time $T_s = 1.4 \mu \text{s}$. In (a), we fix the inbound latency to be $T_l = 20 \mu \text{s}$, and assume a decoding time which scales as $T^{(r)}_{\text{DEC}} = cr \mu \text{s}$. Using \cref{eq:Tbuffj}, we plot $T^{b_j}$ for different values of $c$. In (b), we fix $T^{(r)}_{\text{DEC}} = r \mu \text{s}$ and vary the inbound latency $T_l$.}
\end{figure*}

We start with a few definitions. Let $T^{b_j}$ denote the total waiting time during buffer $b_j$, and $T_s$ be the total time it takes to perform one round of stabilizer measurements for a given quantum hardware architecture. Let $T_l$ be the time for the stabilizer measurements of one round of EC to be known to the classical computer. An example circuit using the $\ket{T} = \frac{1}{\sqrt{2}}(\ket{0} + e^{i \pi /4}\ket{1})$ magic state is provided in \cref{fig:TgateImp}. Lastly, we define $T^{(r)}_{\text{DEC}}$ to be the time it takes the classical computer to compute a correction based on syndrome measurement outcomes arising from $r$ rounds of EC. As such, $T^{(r)}_{\text{DEC}}$ corresponds to the throughput for $r$ rounds of EC. 

The $j$'th buffer time $T^{b_j}$ will depend on the particular implementation of the $T$ gate. For many quantum hardware architectures, arbitrary logical CNOT gates must be implemented by lattice surgery \cite{fowler2018low,litinski2018lattice,litinski2019game,CC21,CC22Twist}, which would be equivalent to using the circuit in \cref{fig:TgateImpv2}. In such a case, $T^{b_j}$ will depend not only on the processing of EC rounds during buffer $b_{j-1}$, but also on the processing of the multiple rounds of EC for the $Z \otimes Z$ measurement via lattice surgery since the measurement outcome is needed in order to restore the Pauli frame. We note however that given access to an extra ancilla qubit, the conditional Clifford in \cref{fig:TgateImpv2} can be replaced with a conditional Pauli (see Fig. 17 (b) in Ref.~\cite{litinski2019game}, and for a generalization to CCZ gates, Fig. 4 in Ref.~\cite{GidneyAutoCCZ}). For simplicity, we will use the circuit in \cref{fig:TgateImpv2} as using the circuit in Ref.~\cite{litinski2019game} would simply change the number of syndrome measurement rounds used in our analysis. 

Now, consider the wait time $T^{b_1}$ of the first buffer. Since the Pauli frame $P_1$ and the measurement outcome of the $Z \otimes Z$ measurement must be known to restore the Pauli frame, we have that
\begin{align}
    T^{b_1} = T^{(r_1+r_2)}_{\text{DEC}} + T_l,
    \label{eq:Tbuff1}
\end{align}
where we assume $r_1$ rounds of EC are performed during the waiting time of buffer $b_0$ and $r_2$ rounds of EC are needed for the $Z \otimes Z$ measurement. We also assume that the syndrome measurement outcomes of each EC round have an inbound latency $T_l$, and that the decoder used by the classical computer can begin processing the syndromes after receiving the outcome of the last round. In \cref{appendix:BuffSlideWindow} we discuss how buffer times can be reduced for decoders implemented using sliding windows. However in this section, we consider the case where the decoder takes as input all syndrome measurement rounds until the last round when the data qubits are measured in some basis. 

Now let $n^{(b_j)}_{\text{QEC}}$ denote the total number of QEC rounds needed during the buffer $b_j$. For $b_1$, we have that 
\begin{align}
    n^{(b_1)}_{\text{QEC}} = \lceil T^{b_1}/T_s \rceil,
    \label{eq:TnQEC1}
\end{align}
since each syndrome measurement round takes time $T_s$. Using \cref{eq:TnQEC1}, the buffer time $T^{b_2}$ is then $T^{b_2} = T^{(n^{(b_1)}_{\text{QEC}})}_{\text{DEC}} + T_l$. 

Applying the above arguments recursively, the $j$'th buffer is then 
\begin{align}
    T^{b_j} = T^{(n^{(b_{j-1})}_{\text{QEC}})}_{\text{DEC}} + T_l,
    \label{eq:Tbuffj}
\end{align}
with $n^{(b_j)}_{\text{QEC}} = \lceil T^{b_j}/T_s \rceil$.

If we assume a linear decoding time of the form $T^{(r)}_{\text{DEC}} = cr$ (where the constant $c$ is in microseconds), solving \cref{eq:Tbuff1,eq:TnQEC1,eq:Tbuffj} recursively results in
\begin{align}
    T^{b_j} = \frac{c^jr}{T_s^{j-1}} + T_l\Big[ \frac{T_s^{1-j}(c^j - T_s^j)}{c-T_s} \Big].
    \label{eq:SolBuffLinear}
\end{align}

Plots of \cref{eq:SolBuffLinear} for different values of $c$ and inbound latency times $T_l$ are shown in \cref{fig:BufferTimePlots}. We assume that the surface code is used to perform each round of EC, where the CNOT gates used to measure the stabilizers take four time steps. Each CNOT is assumed to take $100 \text{n}s$, and the measurement and reset time of the ancillas take $1 \mu s$, as is the case for instance in Ref.~\cite{WallrafGateTimes}. Therefore we set $T_s = 1.4 \mu s$. We also assume that the number of syndrome measurement rounds during the buffer $b_0$ and first lattice surgery measurement for $Z \otimes Z$ is $r_1+r_2=33$, which could be the case for the implementation of medium to large size algorithms with a $d \approx 20$ surface code.

As can be seen in \cref{fig:PlotThroughput}, where the inbound latency term $T_l = 20 \mu s$, if $c \lesssim T_s$, then the buffer wait times grow in a manageable way. However for larger values of $c$, there is a large exponential blow-up in the buffer wait times. This can also be seen from the first term in \cref{eq:SolBuffLinear}, which grows linearly if $c \le T_s$. In \cref{fig:PlotLatency}, we consider how changing the inbound latency $T_l$ affects the buffer wait times when keeping $c$ fixed (which we set to $c = 1\mu s$).  As can be seen, increasing inbound latency does not result in an exponential blow-up in buffer wait times. This can also be seen from the second term in \cref{eq:SolBuffLinear} which only depends linearly  on $T_l$. As such, we conclude buffer wait times are much more sensitive to decoding throughput times, and it will thus be very important to have fast EC decoders in order to implement quantum algorithms. 

We conclude this section by remarking that increasing buffer times can also lead to an increase in the code distances $d_x$ and $d_z$ to ensure that logical failure rates remain below the target set by the quantum algorithm. In other words, if the code distance is fixed, buffer times cannot be arbitrarily large. For instance, for a code with full effective code distance (and let $d_x=d_z=d$ as is the case for a depolarizing noise model), the logical $X$ and $Z$ error rates for $d_m$ syndrome measurement rounds scale as
\begin{align}
    p_L(p) = u d d_m (b p)^{(d+1)/2},
\end{align}
for some constants $u$ and $b$ (see for instance Ref.~\cite{CC21}). We must also have $p_L(p) < \delta$ where $\delta$ is the maximum failure rate allowed for a particular algorithm. Hence for a fixed $d$, we must have that $d_m < \delta / ( u d (b p)^{(d+1)/2})$. The parameters $u$ and $b$ depend on the details of the noise model and decoding algorithm used, as discussed in \cref{subsec:Numerics}. The reader may be concerned that a large value of $d_m$ imposed by long buffer wait times may require a large increase in the code distance. In \cref{appendix:SurfDvsDm} we show that the code distance $d$ only grows logarithmically with $d_m$.

\section{Using NN's as local decoders for circuit-level noise}
\label{sec:NNdecsLocal}

In \cref{section:AlgoRun} we motivated the need for fast decoders. In this section, we construct a hierarchical decoding strategy for correcting errors afflicting data qubits encoded in the surface code. Our hierarchical decoder consists of a local decoder which can correct errors of a certain size, and a global decoder which corrects any remaining errors after implementating the local decoder. In this manuscript we use MWPM for the global decoder, though our scheme can easily be adapted to work with other global decoders such as Union Find. We use NN's to train our local decoder arising from the circuit-level noise model described in \cref{sec:SurfRev}. Importantly, the NN decoder is scalable and can be applied to arbitrary sized volumes $(d_x,d_z,d_m)$ where $d_x$ and $d_z$ are the $X$ and $Z$ distances of the surface code, and $d_m$ is the number of syndrome measurement rounds. 

Our local decoder will have an effective distance  $d_{\text{eff}} \le \max{(d_x,d_z)}$ allowing it to remove errors arising from at most $(d_{\text{eff}}-1)/2$ faults. By removing such errors, the goal is to reduce the syndrome density, i.e.\ the number of highlighted vertices in the matching graph $G$ used to implement MWPM, thus resulting in a much faster execution of MWPM. We note that hierarchical decoding strategies have previously been considered in the literature \cite{DelfosseHierarchical,UsmanNN,GermanyNN}. In Ref.~\cite{DelfosseHierarchical}, a subset of highlighted vertices in the matching graph $G$ (which we refer to as \textit{syndrome density}) are removed based on a set of local rules. However, the weight of errors which can be removed by the local rules is limited, and the scheme (analyzed for code capacity noise) requires low physical error rates to see a large reduction in decoding runtimes. The schemes in Refs.~\cite{UsmanNN,GermanyNN} used NN's to train local decoders. In Ref.~\cite{UsmanNN}, a   two-dimensional fully convolutional NN was used to correct errors arising from code capacity noise. However the scheme does not generalize to phenomenological or circuit-level noise, where repeated rounds of syndrome measurements must be performed. In Ref.~\cite{GermanyNN}, fully connected layers were used to train a network based on patches of constant size, and the scheme was adapted to also work with phenomonological noise. However, as will be shown in \cref{subsec:SyndromeCollapse}, the presence of circuit-level noise introduces fault patterns which are distinct from code capacity and phenomenological noise. In particular, we find that for a certain subset of failures, the syndrome density is not reduced even if the local decoder removes the errors afflicting the data qubits (vertical pairs of highlighted vertices arise after the correction performed by the NN decoder). In fact, the use of NN's as local decoders can \textit{increase} the syndrome density if no other operations are performed prior to implementing the global decoder. As such, in \cref{subsec:SyndromeCollapse} we introduce the notion of \textit{syndrome collapse} which not only reduces the syndrome density but also reduces the size of the matching graph $G$, leading to a much faster implementation of MWPM. We also introduce in \cref{subsec:RemoveVerticalPairs} the notion of a \textit{vertical cleanup} which directly removes pairs of highlighted vertices after the application of the local NN decoder, without reducing the size of the matching graph. We also point out that larger NN's are required to correct errors arising from the more complex fault-patterns of circuit-level noise than what was previously considered in the literature. In particular, in \cref{subsec:NNdescription} we describe how three-dimensional fully convolutional NN's can be used to train our local decoder. 

Regarding the implementation of our three-dimensional convolutional NN's, we introduce new encoding strategies for representing the data that not only allows the NN to adapt to different boundaries of a surface code lattice, but also significantly enhances its abilities to correct errors in the bulk. 

Lastly, in \cref{subsec:Numerics} we provide a numerical analysis of our decoding strategy applied to various surface code volumes of size $(d_x,d_z,d_m)$, showing both the logical error rates and syndrome density reductions after the implementation of our local decoder. 

\subsection{Using NN's to train local decoders.}
\label{subsec:NNdescription}

\begin{figure}
    \centering
    \includegraphics[width=0.8\columnwidth]{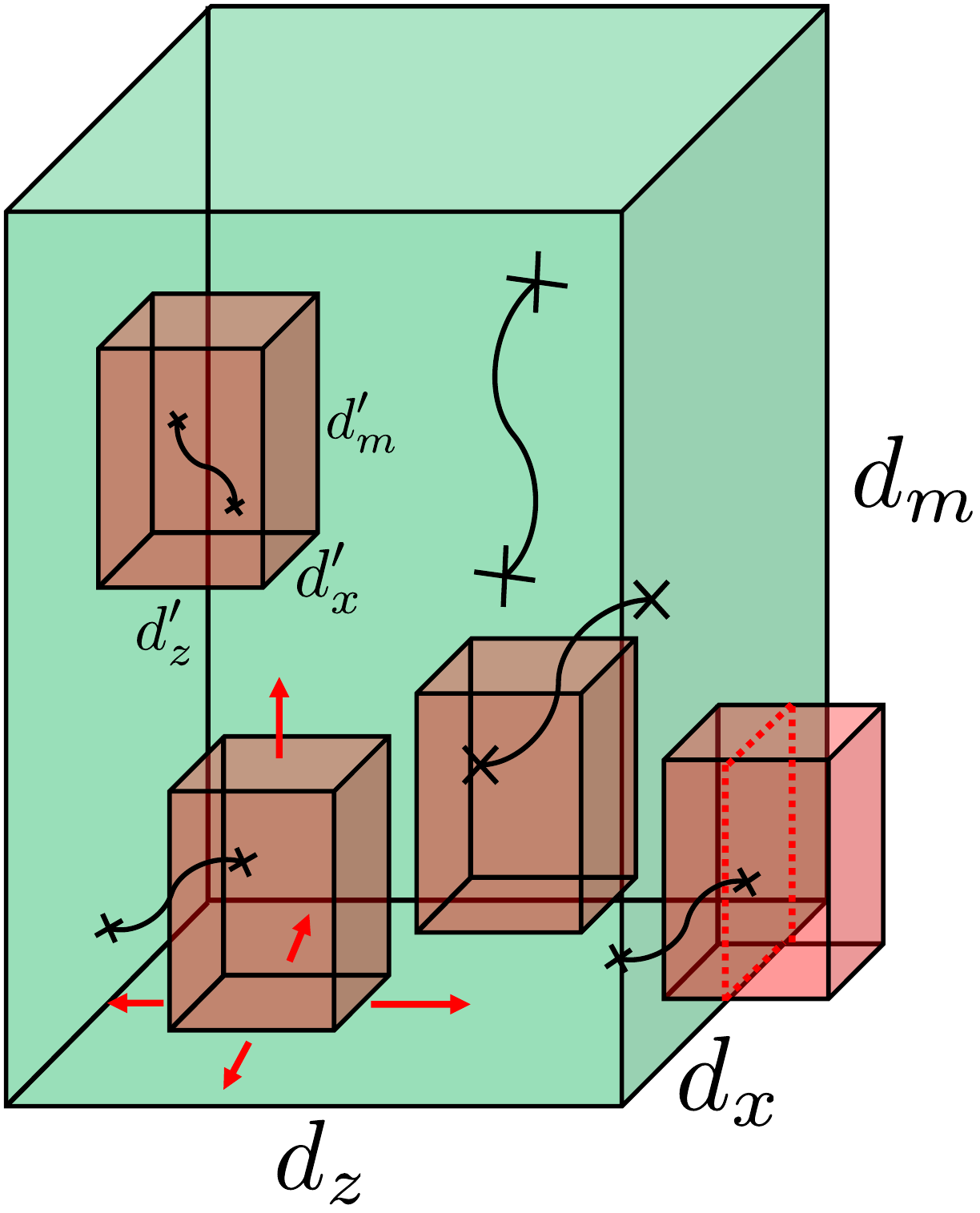}
    \caption{Illustration of a convolution trained on size $(d'_x,d'_z,d'_m)$ applied to a larger surface code volume of size $(d_x,d_z,d_m)$. The network can correct error strings of size at most $(\max{(d'_x,d'_z)} - 1)/2$ by effectively "sweeping" through the larger $(d_x,d_z,d_m)$ volume. }
    \label{fig:slidingwindows}
\end{figure}

Decoding can be considered a pattern recognition task: for each physical data qubit $q_j$ used in the encoding of the surface code, given the syndrome measurements within some local volume $(d'_x,d'_z,d'_m)$ of the lattice, a classifier can predict whether or not there is an error afflicting $q_j$. 

\begin{figure*}
	\centering
	\subfloat[\label{fig:6layer}]{%
		\includegraphics[width=0.7\textwidth]{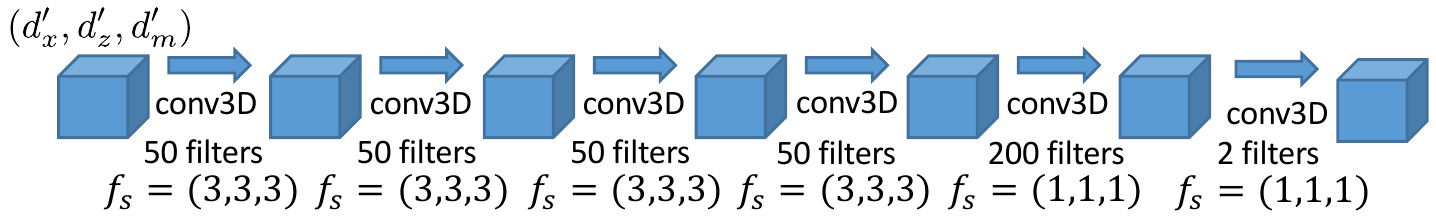}
	}
	\vfill
	\subfloat[\label{fig:11layer}]{%
		\includegraphics[width=0.87\textwidth]{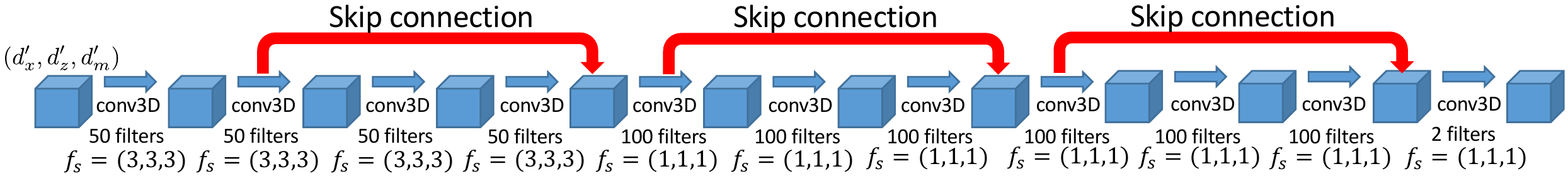}
	}
	\vfill
	\subfloat[\label{fig:SkipConnFig}]{%
		\includegraphics[width=0.55\textwidth]{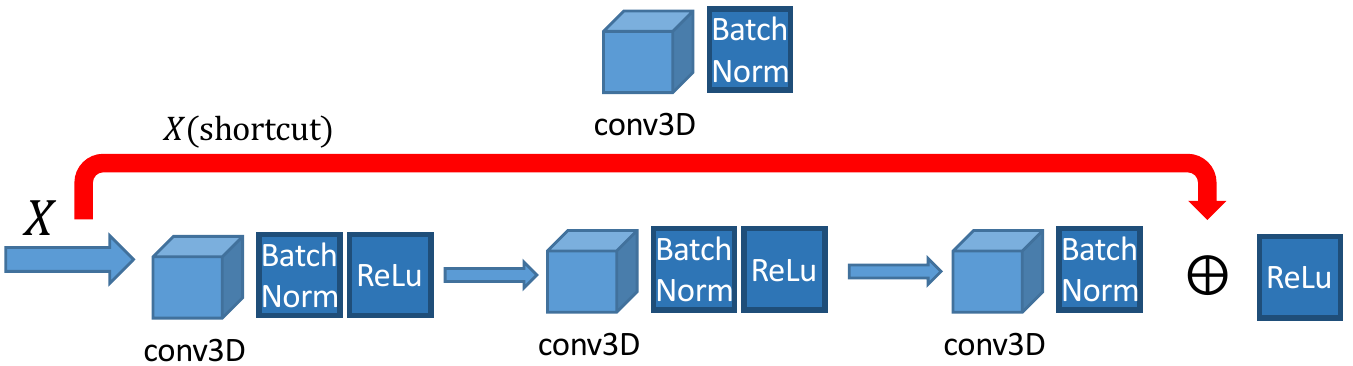}
	}
	\caption{\label{fig:NetworkArchitectures} NN architectures used to train our local decoders. In (a), we consider a network with 6 layers. The first 4 layers have 50 filters of dimension $(3,3,3)$ and serve as feature extractors with a total receptive field of $9 \times 9 \times 9$. The last two layers have filters of dimension $(1,1,1)$, with 200 filters used in the second last layer. The last layer has 2 filters, to predict the $X$ and $Z$ error outputs. The network has a total of $221,660$ parameters. In (b) we use a network with 11 layers. The first 4 layers have 50 filters of dimension $3 \times 3 \times 3$, whereas the next 6 layers have 100 filters of dimension $(1,1,1)$. The last layers use 2 filters of size $(1,1,1)$. The network has a total of $352,210$ parameters. We also use skip connections which becomes more relevant as the number of layers in the network becomes large to avoid exploding/vanishing gradients \cite{HeSkipConn,SkipConn2}. For both networks, we perform batch normalization after each layer. All layers use the ReLu activation function except for the last layer, where we use a sigmoid activation function, to generate predictions for physical qubit errors throughout the lattice. We also use the binary cross-entropy loss function to train our networks. In (c), we provide the details of the implementation of the skip connections. For clarity, we also illustrate the batch normalization step and the implementation of the ReLu activation function.}
	
\end{figure*}

In this work, we design a NN classifier that takes as input a  local volume of size  $(d'_x,d'_z,d'_m)$, and train it to correct data-qubit errors arising from at most $(d'-1)/2$ faults, where $d' = \min{(d'_x,d'_z)}$. To ensure scalability, our NN classifier must be designed in such a way that it corrects errors arising from at most $(d_{\text{eff}}-1)/2$ faults even when applied to larger surface code volumes $(d_x,d_z,d_m)$, where $d_{\text{eff}} \le d'$.

There are many choices for our network architecture. The simplest is a multi-layer perceptron (MLP) with an input layer, hidden layer, and output layer, each of which is a "fully connected" layer where all inputs connect to each neuron in the layer. In this type of network, the $(d'_x,d'_z,d'_m)$ local volume serves as inputs to a set of $N$ neurons in the input layer. The hidden layer takes those $N$ neurons as inputs for a set of $H$ neurons, and finally the $H$ hidden layer neuron outputs are inputs to the final layer neurons that produce the prediction. We implement a network with two outputs, the occurrence of an $X$ error, and the occurrence of a $Z$ error (with $Y$ errors occurring if both $X$ and $Z$ errors are present).

For an efficient computation, we transform (and subsequently enhance) the MLP to be a "fully-convolutional" network, where each layer consists of a set of convolution filters. Convolutions efficiently implement a sliding-window computation\footnote{In this context, the sliding-window computation should not be confused with the sliding window approach of Ref.~\cite{dennis02}, where MWPM is performed in "chuncks" of size $\mathcal{O}(d)$ for a distance $d$ code, with the temporal corrections from the previous window used as input into the MWPM decoder applied to the next window. In our case, the NN takes the entire volume as its input, and performs corrections on each qubit in the volume using only local information.} to produce an output at each location of an input of arbitrary size. For the case of a network with a $(d'_x,d'_z,d'_m)$ local input volume, we use a 3-dimensional convolution of the same size, and so the first layer is a set of $N$ $(d'_x,d'_z,d'_m)$ convolutional filters.  This layer, when applied to a local patch of size $(d'_x,d'_z,d'_m)$, produces $N$ outputs.  The hidden layer, accepting these $N$ inputs for $H$ outputs, can be viewed as a set of $H$ $1 \times 1 \times 1$ convolutional filters. Likewise, the final output layer accepts these $H$ inputs to produce 2 outputs, and can be represented as two $1 \times 1 \times 1$ \texttt{conv3d} filters.

The fully-convolutional network produces a prediction for the data qubit at the center of the local volume it analyzes, as it sweeps through the entire lattice. To allow the network to make predictions right up to the boundary of the lattice, the \texttt{conv} layers are chosen to produce a 'same' output, whereby the input is automatically zero-padded beyond the boundary of the lattice. For example, for a convolution of size 9 to produce an output right at the boundary, the boundary is padded with an additional 4 values. \cref{fig:slidingwindows} illustrates the NN applied throughout the lattice volume, including computing a prediction right at the border of the lattice, in which case some of it's input field lies outside of the lattice volume and receives zero padded values.

To improve the representational power of the network, we can replace the first layer of convolutional filters with multiple layers, taking care to preserve the overall receptive field of the network. For example, if the first layer had filters of size (9,9,9), 4 layers with filters of size (3,3,3) will also have an effective filter size of (9,9,9), since each additional layer increases the effective filter width by 2 from the first layer's width of 3. If each layer were linear, the resulting $N$ outputs in the fourth layer would be mathematically equivalent to a single $9 \times 9 \times 9$ layer with $N$ outputs.  However, since each layer is non-linear, with a nonlinear activation function (ReLu in our case), the two networks are no longer equivalent, and the network with 4 layers of (3,3,3) filters has more representational power, learning nonlinear combinations of features-of-features-of-features. Similarly, we can expand the hidden layer with (1,1,1) filters to become multiple layers of (1,1,1) filters to increase the network's learning capacity. 

In this work we consider two network architectures illustrated in \cref{fig:NetworkArchitectures}. The network in \cref{fig:6layer} has 6 layers, with the first 4 layers having filters of size $(3,3,3)$. The remaining 2 layers have filters of size $(1,1,1)$. In \cref{fig:11layer}, the network has 11 layers, with the first 4 layers having filters of size $(3,3,3)$ and the remaining 7 layers have filters of size $(1,1,1)$. The networks in \cref{fig:6layer,fig:11layer} have a total of $221,600$ and $352,210$ parameters, respectively, with the goal that such networks can learn the complex fault patterns arising from circuit-level noise. Another goal is for the networks to correct errors on timescales similar to those discussed in \cref{section:AlgoRun} using appropriate hardware. More details on the implementation of these networks on FPGA's are discussed in \cref{section:LatencyReduction}. 

\begin{figure}
    \centering
    \includegraphics[width=0.9\columnwidth]{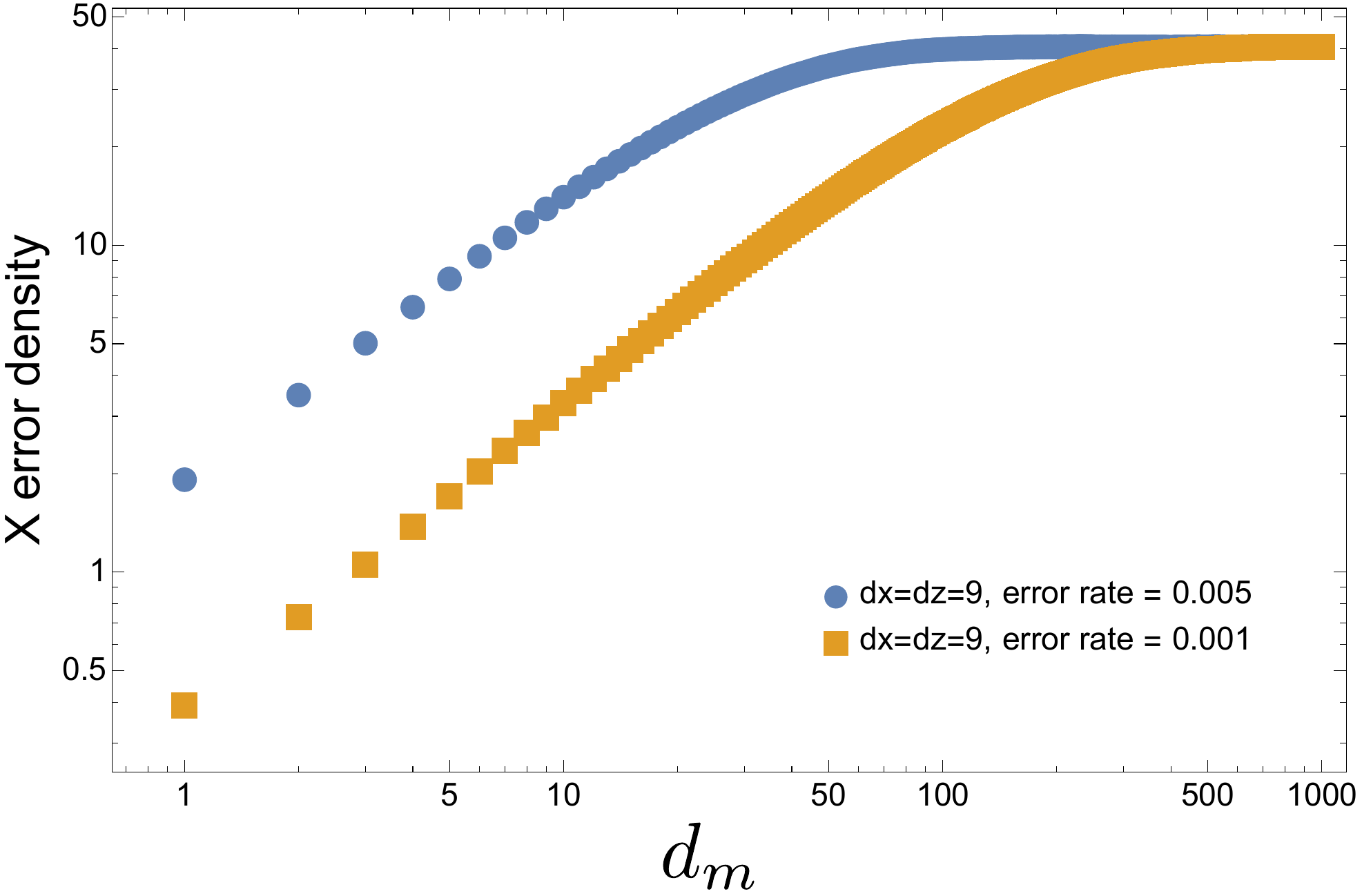}
    \caption{Average number of $X$ errors afflicting data qubits of a $d_x=d_z=9$ rotated surface code lattice as a function of the number of syndrome measurement rounds and the circuit-level noise model described in \cref{sec:SurfRev}. Results are shown for the depolarizing noise parameter $p$ set to $p = 0.001$ and $p = 0.005$. For small noise rates, hundreds of syndrome measurement rounds are required to saturate the average $X$ error density of $50 \%$.}
    \label{fig:ErrorDensities}
\end{figure}

To obtain the data used to train the fully convolutional NN's, we perform $N_{\text{train}}$ Monte Carlo simulations using the circuit-level noise model described in \cref{sec:SurfRev}, with the surface code circuit being used to compute the error syndrome. The training data is then stored using the following format. The input to the network, which we label as \texttt{trainX}, is a tensor of shape $(N_{\text{train}},d_x,d_z,d_m,5)$ for a surface code with $X$ and $Z$ distances $d_x$ and $d_z$, with $d_m$ syndrome measurement rounds. Following \cref{defXZsyndiff}, the first two inputs to \texttt{trainX} contain the syndrome \textit{differences} $s^{\text{diff}}_X(d_m)$ and $s^{\text{diff}}_Z(d_m)$ obtained for $d_{m}-1$ rounds of noisy syndrome measurements, followed by one round of perfect error correction. Tracking changes in syndrome measurement outcomes between consecutive rounds ensures that the average syndrome density remains constant across different syndrome measurement rounds. The next two inputs of \texttt{trainX} contain spatial information used to enable the network to associate syndrome measurement outcomes with data qubits in both the bulk and along boundaries that can influence the observed outcome. The data is represented as $d_x$ by $d_z$ binary matrices labelled $\texttt{enc}(X)$ and $\texttt{enc}(Z)$, where 1 values are inserted following a particular mapping between the position of the ancillas (grey vertices in \cref{fig:SurfaceCoded5}) and data qubits (yellow vertices in \cref{fig:SurfaceCoded5}) which interact with the ancillas. The details of our mapping is described in \cref{app:DataRep}. We note that the matrices $\texttt{enc}(X)$ and $\texttt{enc}(Z)$ are provided for each syndrome measurement round, and are identical in each round unless the lattice changes shape between consecutive syndrome measurement rounds, as would be the case during a lattice surgery protocol \cite{fowler2018low,litinski2018lattice,litinski2019game,CC21,CC22Twist}. Further, the syndrome differences stored in the first two inputs of \texttt{trainX} also follow the same mapping used in $\texttt{enc}(X)$ and $\texttt{enc}(Z)$ between stabilizers and entries in the matrix representations, except that a 1 is only inserted for non-zero values of $s^{\text{diff}}_X(d_m)$ and $s^{\text{diff}}_Z(d_m)$ (more details are provided in \cref{app:DataRep}). Finally, the fifth input of \texttt{trainX} contains the temporal boundaries, which specify the first and last syndrome measurement round. Since the last syndrome measurement round is a round of perfect error correction\footnote{A round of perfect error correction is a syndrome measurement round where no new errors are introduced, and arises when the data qubits are measured directly in some basis at the end of the computation. A measurement error which occurs when the data qubits are measured directly is equivalent to an error on such data qubits in the prior round. See for instance Appendix I in Ref.~\cite{chamberland2020building}.}, the syndrome measurement outcome will always be compatible with the errors afflicting the data qubits arising from the second last round. As such, since the last syndrome measurement round behaves differently than the other rounds, it is important to specify its location (as well as the location of the first round) in \texttt{trainX} so that the trained network can generalize to volumes with arbitrary $d_m$ values. More details for how the data is represented in \texttt{trainX} and the mappings discussed in this paragraph are provided in \cref{app:DataRep}.

Next, the output targets that the NN will attempt to predict (i.e. the locations of $X$ and $Z$ data qubit errors) are stored in a tensor \texttt{trainY} of shape $(N_{\text{train}},d_x,d_z,d_m,2)$. In particular, \texttt{trainY} contains the $X$ and $Z$ data errors afflicting the data qubits for syndrome measurement rounds 1 to $d_m$. In order for the data stored in \texttt{trainY} to be compatible with \texttt{trainX}, we only track changes in data qubit errors between consecutive syndrome measurement rounds, since \texttt{trainX} tracks changes in syndrome measurement outcomes between consecutive rounds. Tracking changes in data qubit errors also ensures that the average error densities are independent of the number of syndrome measurement rounds. Otherwise, one would need to train the network over a very large number of syndrome measurement rounds in order for the networks to generalize well to arbitratry values of $d_m$. An illustration showing the increase in the average data qubit error densities with the number of syndrome measurement rounds is shown in \cref{fig:ErrorDensities}. 

When performing the Monte Carlo simulations to collect the training data, there are many cases where two errors $E_1$ and $E_2$ can have the same syndrome ($s(E_1) = s(E_2)$) with $E_1E_2 = g$ where $g$ is in the stabilizer group of the surface code. We say that such errors are homologically equivalent. In training the NN's, we found that choosing a particular convention for representing homologically equivalent errors in \texttt{trainY} leads to significant performance improvements, as was also remarked in Ref.~\cite{UsmanNN}. A detailed description for how we represent homologically equivalent errors in \texttt{trainY} is provided in \cref{appendix:HomEquivConv}. 

We conclude this section by remarking that the performance of the networks not only depend on the network architecture and how data is represented in \texttt{trainX} and \texttt{trainY}, but also on the depolarizing error rate $p$ used to generate the training data, and the size of the input volume $(d'_x,d'_z,d'_m)$. For instance, since the local receptive field of the networks in \cref{fig:6layer,fig:11layer} is 9x9x9, we used input volumes of size $(13,13,18)$ to allow the network to see spatial and temporal data located purely in the bulk of the volume (i.e. without being influenced by boundary effects). We also trained our networks at error rates $p = 0.005$ and $p = 0.001$, and found that training networks at higher physical error rates did not always lead to superior performance relative to networks trained at low physical error rates. More details are provided in \cref{subsec:Numerics}.

\subsection{Performing a syndrome collapse by sheets.}
\label{subsec:SyndromeCollapse}

\begin{figure}
	\centering
	\subfloat[\label{fig:CNOTfailSynColl}]{%
		\includegraphics[width=0.46\textwidth]{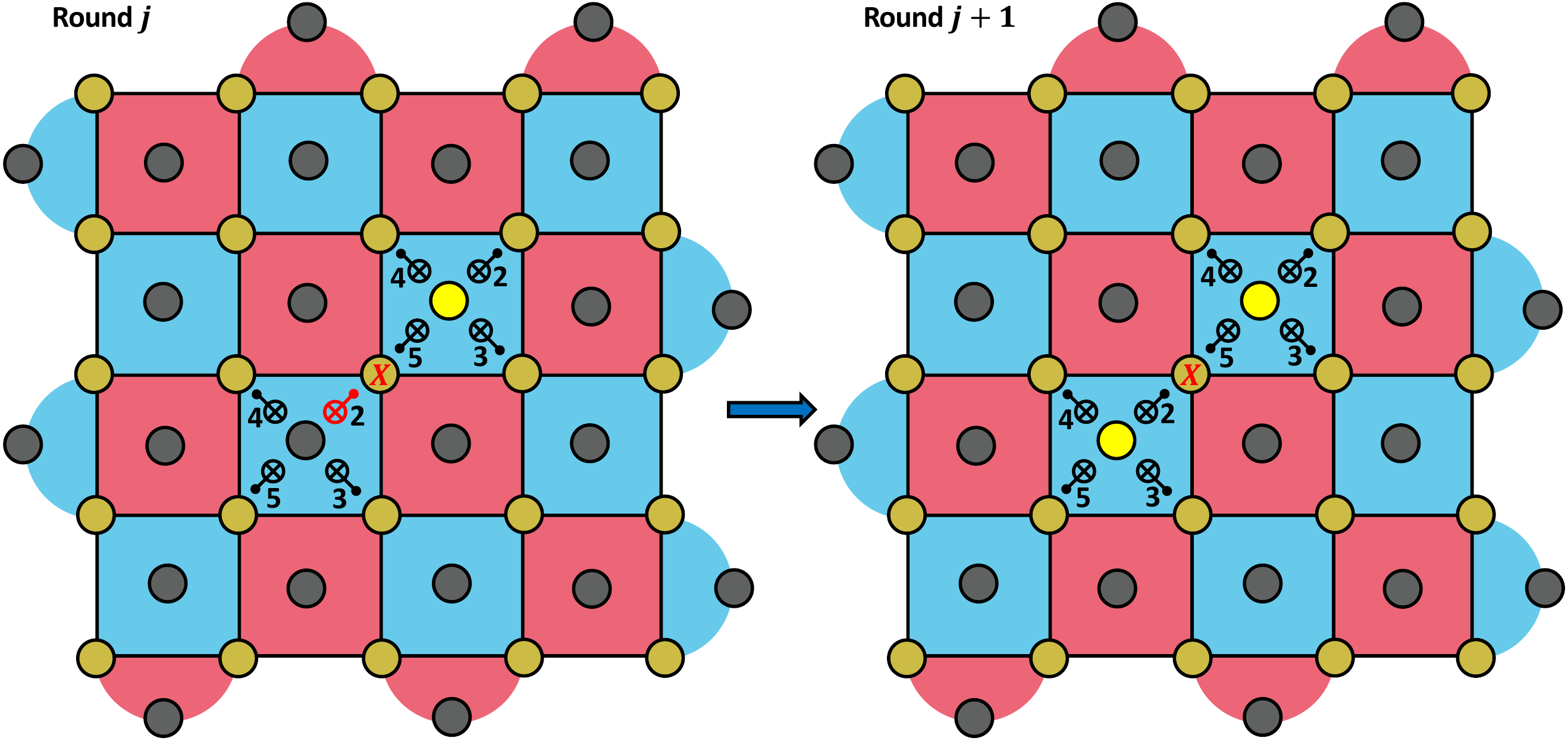}
	}
	\vfill
	\subfloat[\label{fig:SpacetimeGraphv1}]{%
		\includegraphics[width=0.45\textwidth]{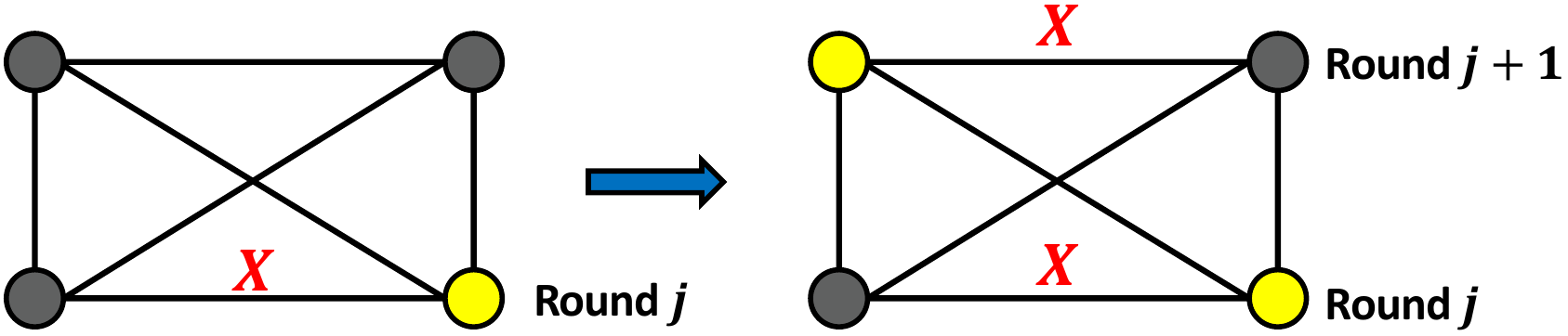}
	}
	\vfill
	\subfloat[\label{fig:SpacetimeGraphv2}]{%
		\includegraphics[width=0.25\textwidth]{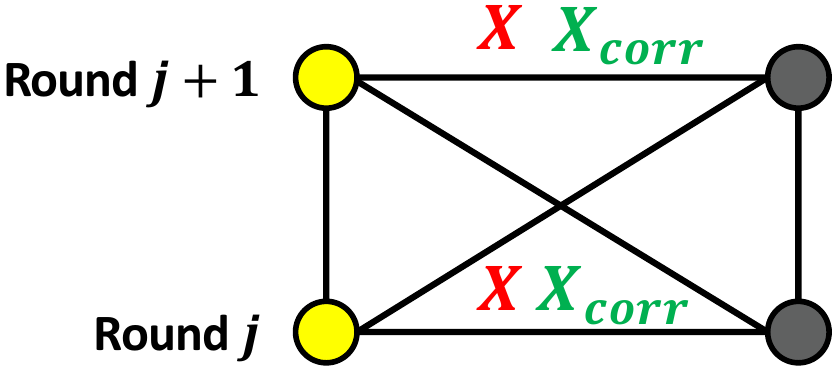}
	}
	\caption{\label{fig:SynCollCNOTfail} (a) CNOT failure (shown in red) resulting in a $X$ data qubit error in the $j$'th syndrome measurement round (we only show CNOT gates which are part of the stabilizers used in this example). Due to the time steps in which the CNOT gates are implemented, only a single $Z$-type stabilizer detects the error in round $j$, with two stabilizers detecting the error in round $j+1$. (b) Subset of the matching graph $G$ associated with the $d_x=d_z=5$ surface code shown in (a). The vertices in $G$ are highlighted (shown in yellow) when changes in syndrome measurement outcomes are detected between consecutive syndrome measurement rounds. (c) Transformation of $G$ after the local decoder applies a correction removing the $X$ error. Even though the local decoder removes the error, the correction creates a vertical pair of highlighted vertices.}
\end{figure}

\begin{figure}
	\centering
	\subfloat[\label{fig:SynCollGraphv1}]{%
		\includegraphics[width=0.46\textwidth]{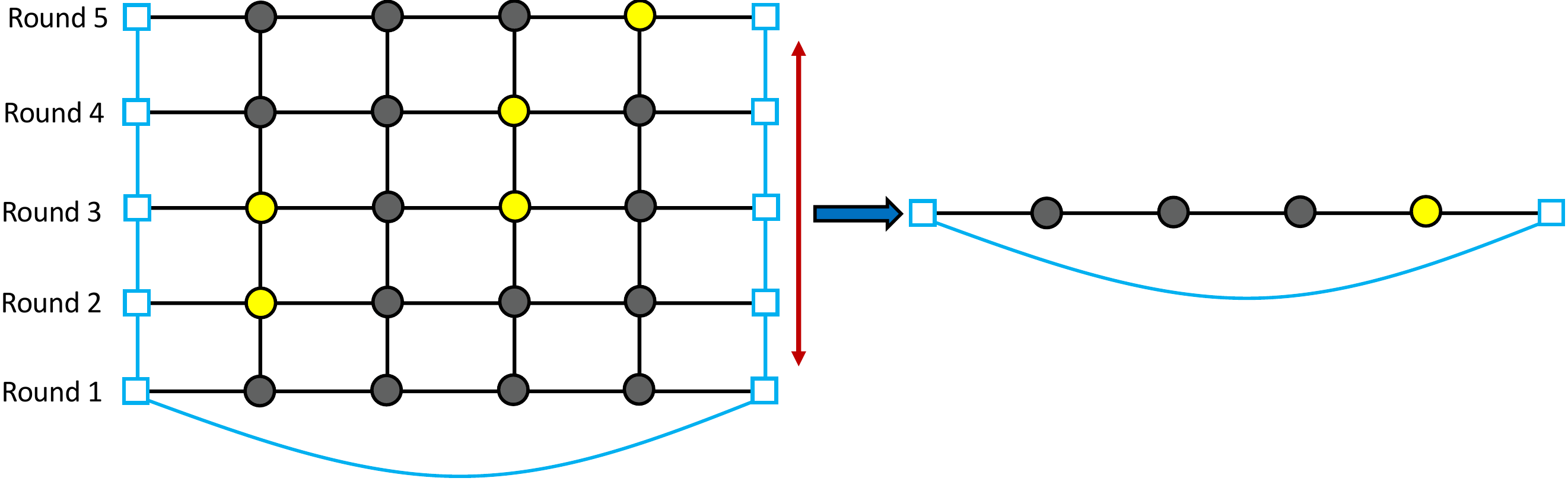}
	}
	\vfill
	\subfloat[\label{fig:SynCollGraphv2}]{%
		\includegraphics[width=0.45\textwidth]{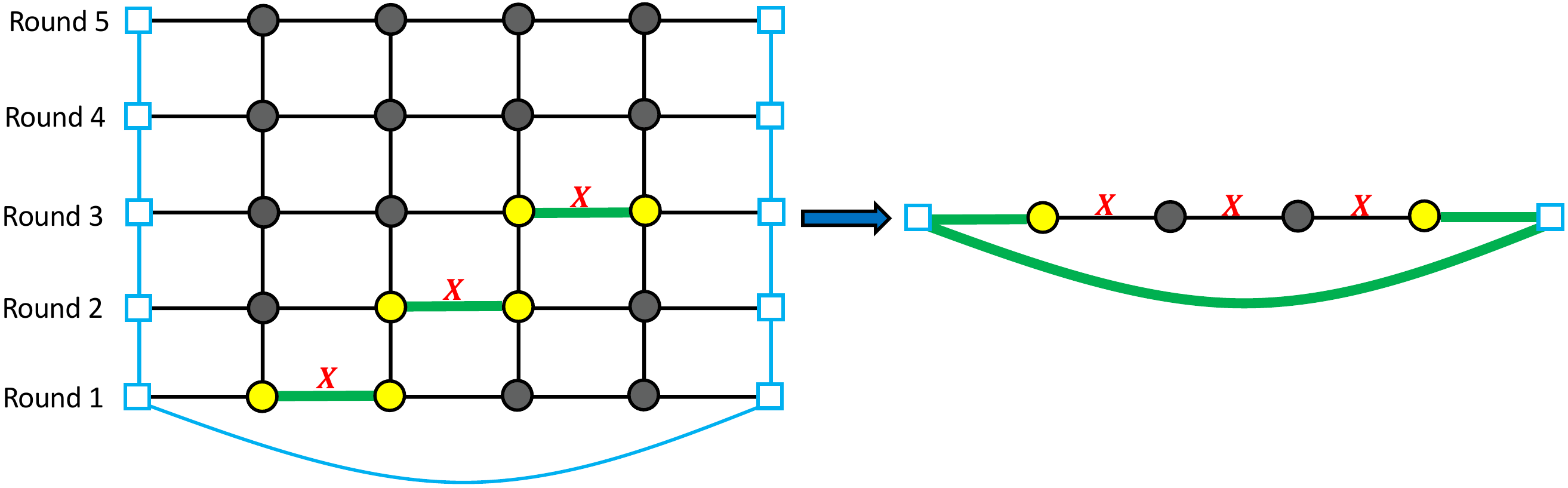}
	}
	\caption{\label{fig:SynCollGraph} (a) On the left is a two-dimensional slice of a subset of the surface code matching graph for 5 rounds of stabilizer measurements. Horizontal edges correspond to data qubits, vertices correspond to stabilizer measurement outcomes, and the blue squares are boundary vertices connected by blue edges of zero weight. The graph has two vertical pairs of highlighted vertices. On the right of the figure is the graph obtained after performing the syndrome collapse. Both vertical pairs vanish after performing the syndrome collapse. (b) On the left of the figure is a sequence of $X$ data qubit errors which are temporally separated, i.e. they occur in different syndrome measurement rounds. The thick green edges show the minimum-weight path which pairs all highlighted vertices (we assume all edges in the graph have unit weight) effectively correcting the errors. On the right of the figure is the graph obtained after performing the syndrome collapse, along with the minimum-weight path pairing the highlighted vertices. The correction thus results in a logical $X$ error.}
\end{figure}

\begin{figure*}
    \centering
    \includegraphics[width=1.7\columnwidth]{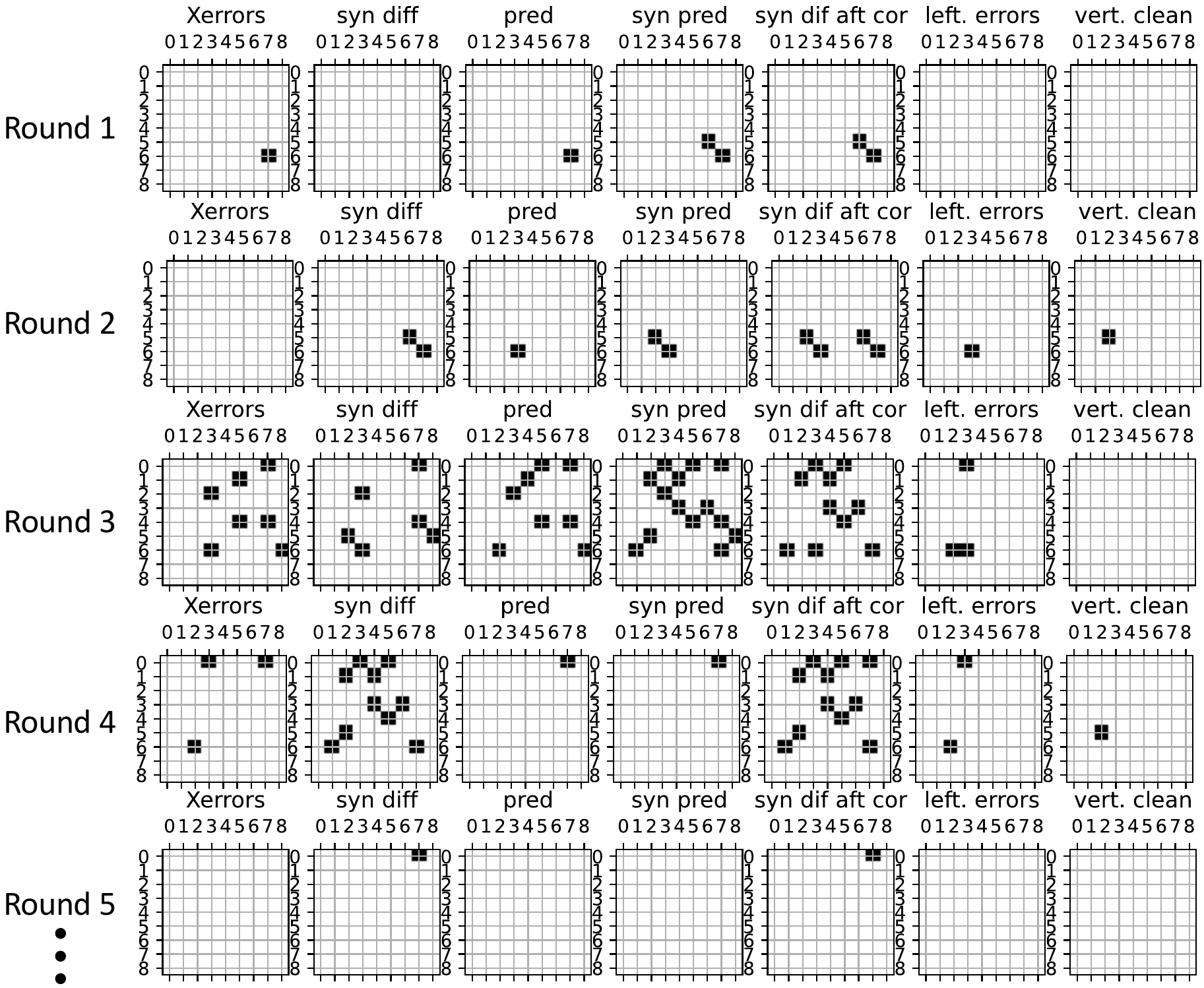}
        \caption{Illustration of $X$-type Pauli errors occurring in a $d_x=d_z=9$ surface code in consecutive syndrome measurement rounds (where we only track changes in errors between consecutive rounds) along with the syndrome differences observed in each round. Note that syndrome differences are mapped to a $d \times d$ grid following the mapping described in \cref{app:DataRep}. We also show the correction applied by the local NN decoder, and resulting homologically equivalent errors in the column \texttt{left. errors} (see \cref{appendix:HomEquivConv}) and syndrome differences after the correction is applied. The plots in the last column labelled \texttt{vert clean} shows the remaining syndrome differences after all pairs of vertical highlighted vertices have been removed. As can be seen, the vast majority of highlighted vertices after the application of the local NN decoder results in vertical pairs. Further, since the NN sees syndrome differences in both the future and the past given the size of its receptive field, in some cases it performs a correction on a data qubit in a round before the error actually occurs, leading to the creation of a vertical pair of highlighted vertices.}
    \label{fig:VertCleanExamp}
\end{figure*}

Consider a CNOT failure during a $Z$-type stabilizer measurement resulting in an $X \otimes I$ error in the $j$'th syndrome measurement round, as shown in \cref{fig:CNOTfailSynColl}. The failure results in an $X$ error on a data qubit. However, given the ordering of the CNOT gates, only a single $Z$-type stabilizer detects the error in round $j$, with two stabilizers detecting the error in round $j+1$. We refer to such failure mechanisms as space-time correlated errors. In \cref{fig:SpacetimeGraphv1} we illustrate the resulting highlighted vertices in a subset of the matching graph $G$ which is used to implement MWPM. As explained in \cref{sec:SurfRev}, a vertex in $G$ associated with the stabilizer $g_k$ is highlighted in round $j$ if the measurement outcome of $g_k$ changes from rounds $j-1$ to $j$. Now, suppose a local decoder correctly identifies the observed fault pattern, and removes the $X$ error on the afflicted data qubit. \cref{fig:SpacetimeGraphv2} shows how $G$ transforms after applying the correction. Importantly, even though the error is removed, a vertical pair of highlighted vertices is created in $G$. We also note that the creation of vertical pairs arising from a correction performed by the local decoder due to a two-qubit gate failure is intrinsic to circuit-level noise and would not be observed for code capacity or phenomenological noise models. In fact, we observe numerically that the average number of highlighted vertices in $G$ after the corrections applied by the local decoder will \textit{increase} rather than decrease. However, as the example of \cref{fig:SynCollCNOTfail} illustrates, many of the highlighted vertices in $G$ will be due to the creation of vertical pairs induced by the corrections arising from the local decoder (see also \cref{fig:VertCleanExamp} in \cref{subsec:RemoveVerticalPairs}). 

One way to reduce the number of vertical pairs after the correction is applied by the local decoder is to perform what we call a \textit{syndrome collapse by sheets}. More specifically, consider the syndrome difference  $s^{\text{diff}}_X(d_m) = (e^{(1)}_{X} \tilde{e}^{(2)}_{X}  \cdots \tilde{e}^{(d_m)}_{X})$ as defined in \cref{defXZsyndiff} and let us assume for simplicity that $d_m = \gamma d'_m$ for some integer $\gamma$. We can partition $s^{\text{diff}}_X(d_m)$ as
\begin{align}
    s^{\text{diff}}_X(d_m) &= (e^{(1)}_{X} \tilde{e}^{(2)}_{X} \cdots \tilde{e}^{(d'_m)}_{X} | \tilde{e}^{(d'_m+1)}_{X} \cdots \tilde{e}^{(2d'_m)}_{X} | \cdots |  \nonumber \\
    &\tilde{e}^{(d_m - d'_m + 1)}_{X}  \cdots \tilde{e}^{(d_m)}_{X}).
    \label{eq:SynDiffBitSt}
\end{align}
A syndrome collapse by sheets of size $d'_m$ transforms $s^{\text{diff}}_X(d_m)$ as
\begin{align}
    \overline{s}^{\text{diff}}_X(d_m) =  (\overline{e}^{(1)}_{X} \overline{e}^{(2)}_{X} \cdots \overline{e}^{(\gamma)}_{X}),
    \label{eq:SynColBitSt}
\end{align}
where 
\begin{align}
    \overline{e}^{(j)}_{X} = \bigoplus^{d'_m}_{i = 1} \tilde{e}^{((j-1)d'_m + i)}_{X},
    \label{eq:sumOplus}
\end{align}
with the sum being performed modulo 2 (if $j=1$, the first term in \cref{eq:sumOplus} is $e^{(1)}_X$, without the tilde). Note that if $d_m$ is not a multiple of $d'_m$, there will be $\lceil d_m / d'_m \rceil$ sheets with the last sheet having size $d_m - \beta d'_m$ where $\beta = \lfloor d_m/d'_m \rfloor$. The above steps can also be performed analogously for syndromes corresponding to $Z$ errors. 

Performing a syndrome collapse by sheets reduces the size of the original matching graph $G$ since $G$ contained $d_m$ sheets prior to performing the collapse. We label $G_{\text{sc}}$ as the graph resulting from performing the syndrome collapse on the original graph $G$. An illustration of how the syndrome collapse removes vertical pairs is shown in \cref{fig:SynCollGraphv1}. Note that without the presence of a local decoder, one would not perform a syndrome collapse using a MWPM decoder since such an operation would remove the decoders ability to correct errors which are temporally separated. An example is shown in \cref{fig:SynCollGraphv2}. However by performing a syndrome collapse on a surface code of distance $d$ \textit{after} the application of the local decoder with $d'_m  = \mathcal{O}(d_{\text{eff}})$ where $d_{\text{eff}}$ is the effective distance of the local decoder (which depends on the local receptive field and size of the volume the network was trained on), we expect such an operation to result in a global effective distance which is equal or close to $d$. The reason is that errors contained within each sheet arising from less than or equal to $(d_{\text{eff}}-1)/2$ faults should be removed by the local decoder. We say ``should'' because local NN decoders are not necessarily guaranteed to correct any error arising from $(d_{\text{eff}}-1)/2$ faults. Since NN decoders offer no fault-tolerance guarantees, we cannot provide a proof giving the effective distance of the surface code decoded using the local NN, followed by a syndrome collapse and application of a global decoder. However, we observed numerically that using larger networks (i.e. a network with more layers and filters per layers) resulted in increased slopes of the logical error rate curves. In \cref{subsec:Numerics} we present numerical results showing the effective distances of various surface code lattices when performing a syndrome collapse after the application of local decoders implemented by NN's.

\begin{figure}
	\centering
	\subfloat[\label{fig:VertCleanCase1}]{%
		\includegraphics[width=0.3\textwidth]{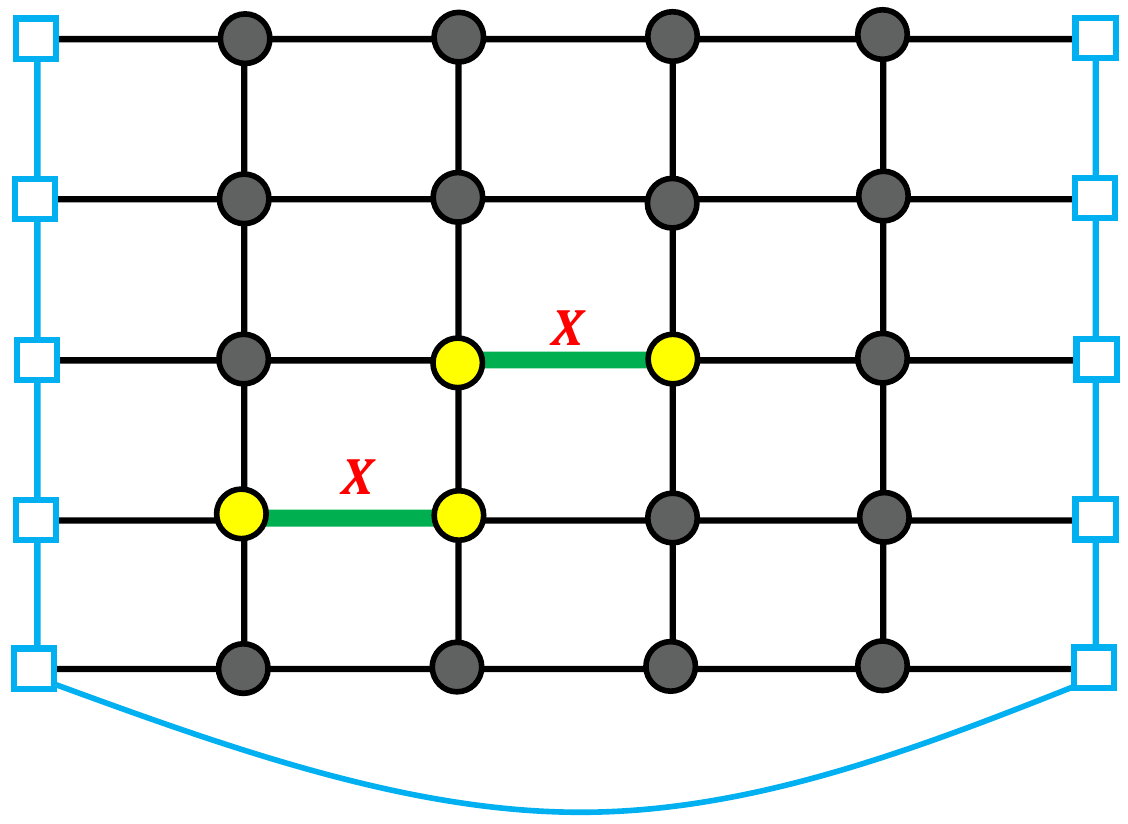}
	}
	\vfill
	\subfloat[\label{fig:VertCleanCase2}]{%
		\includegraphics[width=0.3\textwidth]{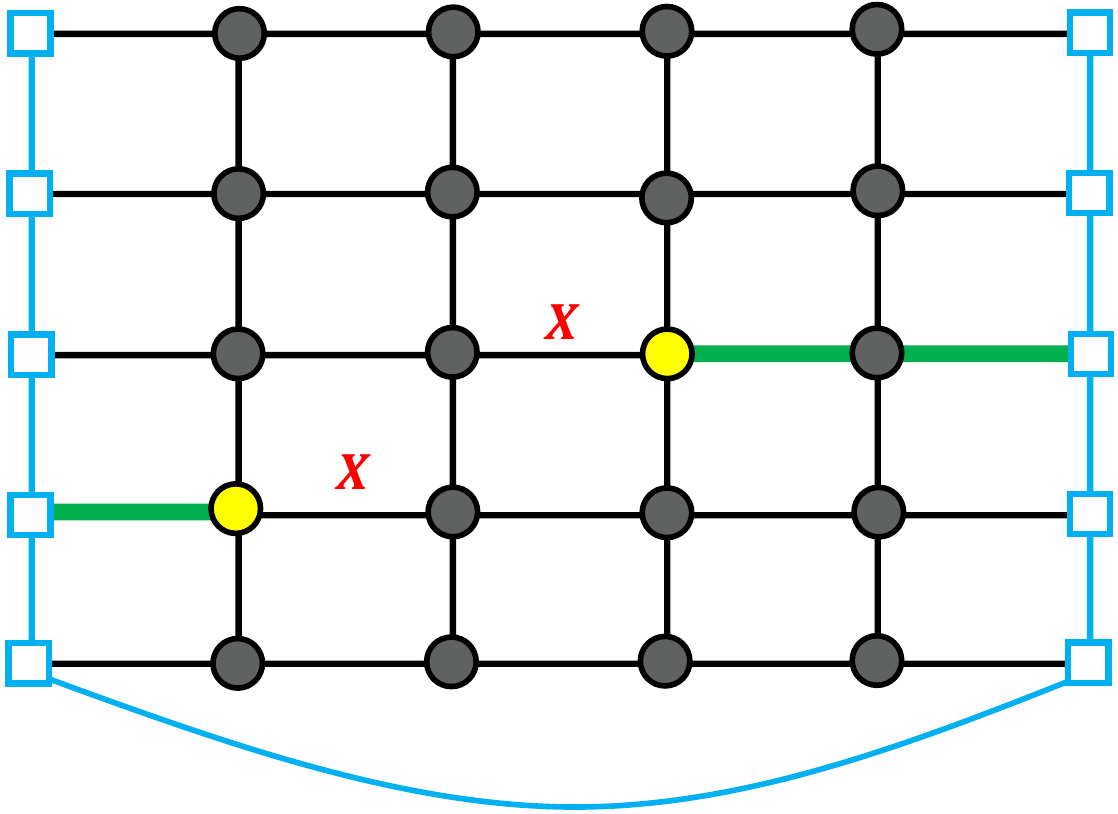}
	}
	\caption{\label{fig:VertCleanBadCase} (a) Two $X$-type errors temporally separated by one syndrome measurement round, along with the highlighted vertices in a two-dimensional strip of a subset of a $d=5$ surface code decoding graph $G_X$ used to correct $X$-type Pauli errors. All black edges are taken to have unit weight. The green shaded edges correspond to the minimum-weight correction, which correctly removes the errors.  (b) Resulting graph after performing the vertical cleanup. The green shaded edges correspond to a  minimum-weight correction, which results in a logical fault.}
\end{figure}

We now give an important remark regarding performing a syndrome collapse during a parity measurement implemented via lattice surgery. As discussed in detail in Ref.~\cite{CC21}, when performing a parity measurement via lattice surgery, there is a third code distance related to timelike failures, where the wrong parity measurement would be obtained. The timelike distance is given by the number of syndrome measurement rounds which are performed when the surface code patches are merged. If a syndrome collapse were to be performed in the region of the merged surface code patch (see for instance Fig. 7 in Ref.~\cite{CC21}), the timelike distance would be reduced and would result in timelike failures which would be too large. As such, a syndrome collapse should not be implemented when performing a parity measurement via lattice surgery unless additional syndrome measurement rounds are performed on the merged surface code patches to compensate for the loss in timelike distance. However, the timelike distance can still potentially be made small using a temporal encoding of lattice surgery protocol (TELS) as described in Ref.~\cite{CC21}. Alternatively, the vertical cleanup protocol described below in \cref{subsec:RemoveVerticalPairs} (which can also significantly reduce the syndrome density) could be used (see also \cref{appendix:VertCleanLatticeSurgery} regarding the required number of syndrome measurement rounds to maintain the timelike distance).

Lastly, we conclude by remarking that a NN architecture that performs a correction by identifying edges in the matching and flipping the vertices incident to such edges could potentially avoid creating vertical pairs after performing its corrections. In such settings, a syndrome collapse or a vertical cleanup as described in \cref{subsec:RemoveVerticalPairs} may not be required. 

\subsection{Performing a vertical cleanup}
\label{subsec:RemoveVerticalPairs}

In \cref{fig:VertCleanExamp} we show an example of the application of the 11-layer NN decoder (trained on an $(13,13,18)$ input volume at $p = 0.005$) to test set data of size $(9,9,9)$ generated at $p = 0.005$. In the figure, each row containing a series of plots corresponds to a syndrome measurement round. For a given row, the first plot labelled \texttt{Xerrors} shows changes in $X$ data qubit errors from the previous round, and the second plot labelled \texttt{syn diff} shows changes in observed syndromes from the previous round (see \cref{app:DataRep} for how changes in syndrome measurement outcomes are represented as $d_x \times d_z$ binary matrices). The third plot labelled \texttt{pred} gives the correction applied by the NN decoder, and the fourth plot labelled \texttt{syn pred} corresponds to the syndrome compatible with the applied correction. The fifth plot labelled \texttt{syn dif aft cor} shows the remaining syndromes after the correction has been applied, and the sixth plot labelled \texttt{left errors} gives any remaining $X$ data qubit errors after the correction has been applied. The last plot labelled \texttt{vert clean} shows the remaining syndromes after all vertical pairs of highlighted vertices have been removed. Vertical pairs are formed when the vertex associated with the measurement of a stabilizer $g_i$ is highlighted in two consecutive syndrome measurement rounds. 

Comparing the fifth and seventh plot in any given row, it can be seen that the vast majority of remaining syndromes after the NN decoder has been applied consists of vertical pairs, since removing vertical pairs eliminates nearly all highlighted vertices. In \cref{subsec:SyndromeCollapse} we described our protocol for performing a syndrome collapse by sheets, which removes any vertical pairs of highlighted vertices within a given sheet, but not vertical pairs between sheets. As the plots in the last column of \cref{fig:VertCleanExamp} suggest, another strategy which can significantly reduce the density of highlighted vertices is to remove all vertical pairs of highlighted vertices which are present \textit{after} the local NN decoder has been applied. More specifically, for the syndrome difference $s^{\text{diff}}_X(d_m) = (e^{(1)}_{X} \tilde{e}^{(2)}_{X}  \cdots \tilde{e}^{(d_m)}_{X})$, we start with the syndrome in the first round $e^{(1)}_{X}$. If $e^{(1)}_{X}(j) = 1$ and $\tilde{e}^{(2)}_{X}(j) = 1$ for some $j \in \{1, \cdots, r_2\}$, we set $e^{(1)}_{X}(j) = \tilde{e}^{(2)}_{X}(j) = 0$. Such a process is repeated by comparing $\tilde{e}^{(m)}_{X}(j)$ and $\tilde{e}^{(m+1)}_{X}(j)$ for $m \in \{2, \cdots, d_m \}$ and for all $j \in \{1, \cdots, r_2 \}$, and setting them to zero if $\tilde{e}^{(m)}_{X}(j) = \tilde{e}^{(m+1)}_{X}(j) = 1$. An identical step is performed for the syndrome differences $s^{\text{diff}}_Z(d_m)$. Note that when performing parity measurements via lattice surgery, there is  a preferred direction in which a vertical cleanup should be performed (i.e. staring from the  first round  and moving upwards to the last vs starting from last round and  moving downwards to the first). The particular direction depends on the syndrome densities above and below some reference point, and is used to maintain a higher effective distance for protecting against temporal errors. More details are provided in \cref{appendix:VertCleanLatticeSurgery}.

We remark that performing a vertical cleanup without an accompanying local decoder can result in a correctable error no longer being correctable by the global decoder. In \cref{fig:VertCleanBadCase}, we show two $X$-type errors which are temporally separated by one syndrome measurement round, along with the corresponding highlighted vertices in a two-dimensional strip of a $d=5$ surface code decoding graph $G_X$, with the subscript $X$ indicating it is a graph for correcting $X$ errors. We assume that all black edges in $G_X$ have unit weight. In \cref{fig:VertCleanCase1}, the green shaded edges correspond to the minimum-weight correction which removes the $X$ errors. In \cref{fig:VertCleanCase2}, we show the resulting highlighted vertices in $G_X$ after performing a vertical cleanup. In this case, one possible minimum-weight correction results in a logical fault as shown by the green shaded edges. 

\begin{figure*}
	\centering
	\subfloat[\label{fig:Plot6layerLogX}]{%
		\includegraphics[width=0.46\textwidth]{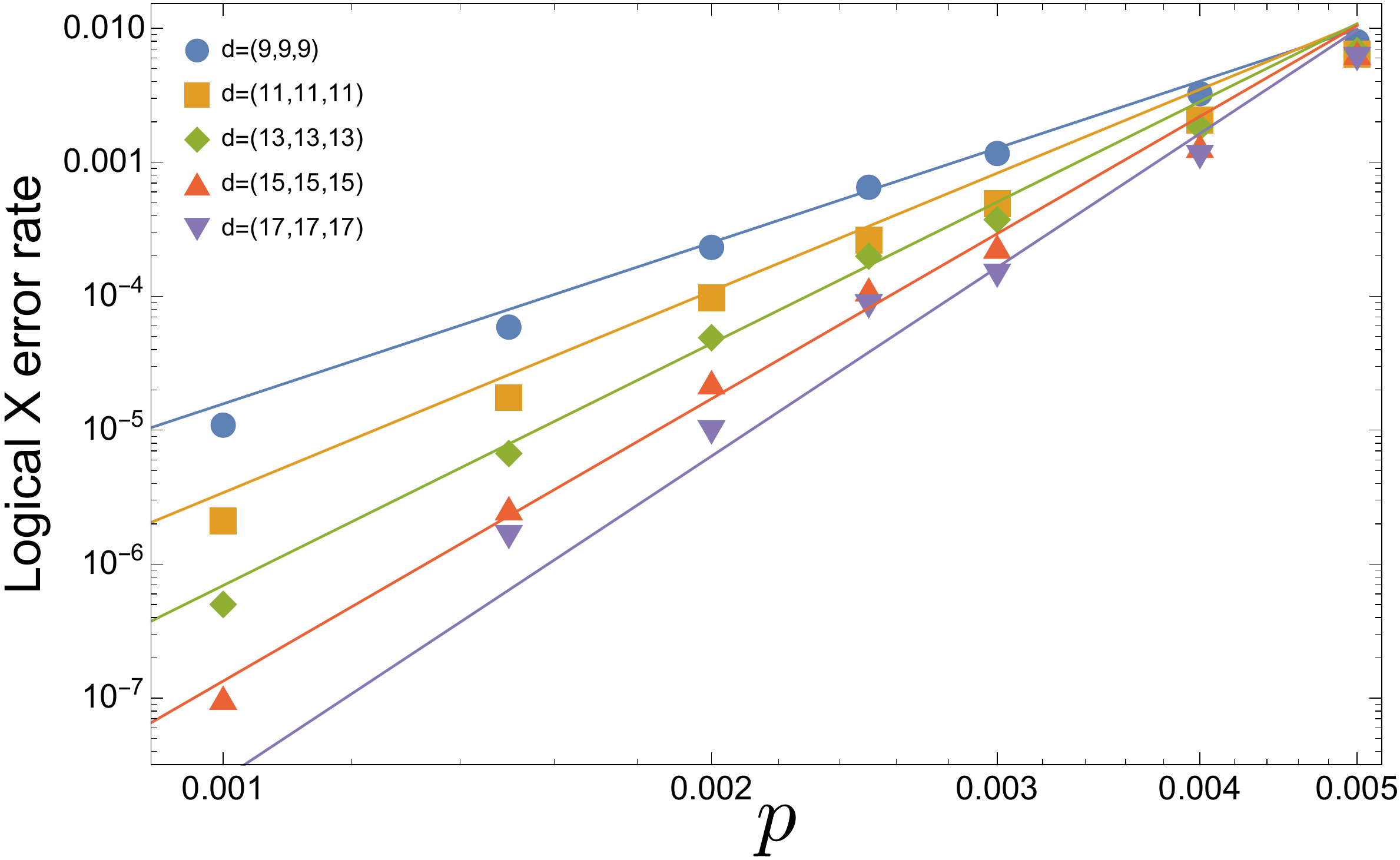}
	}
	\subfloat[\label{fig:Plot11layerLogX}]{%
		\includegraphics[width=0.46\textwidth]{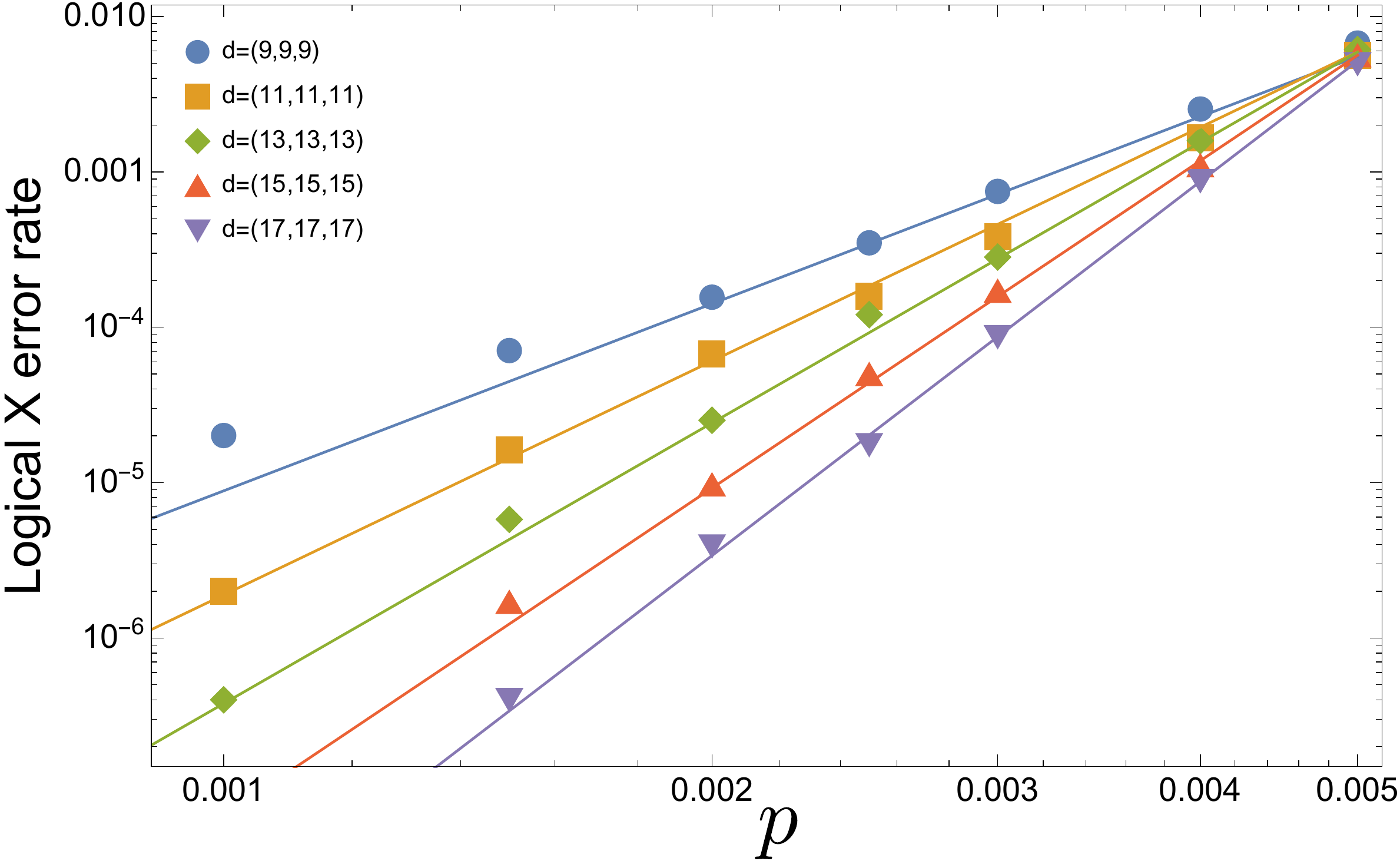}
	}
	\caption{\label{fig:logXErrorRates} Logical $X$ error rates for surface code volumes ranging between $(9,9,9)$ and $(17,17,17)$ after the application of the local NN decoder, followed by a syndrome collapse (with the input volumes partitioned into sheets of temporal height $d'_m = 6$) and MWPM to correct any remaining errors. In (a) the results are for the 6-layer network whereas in (b) the results are for the 11-layer network.}
\end{figure*}

\begin{figure*}
	\centering
	\subfloat[\label{fig:Ratio6Layer}]{%
		\includegraphics[width=0.5\textwidth]{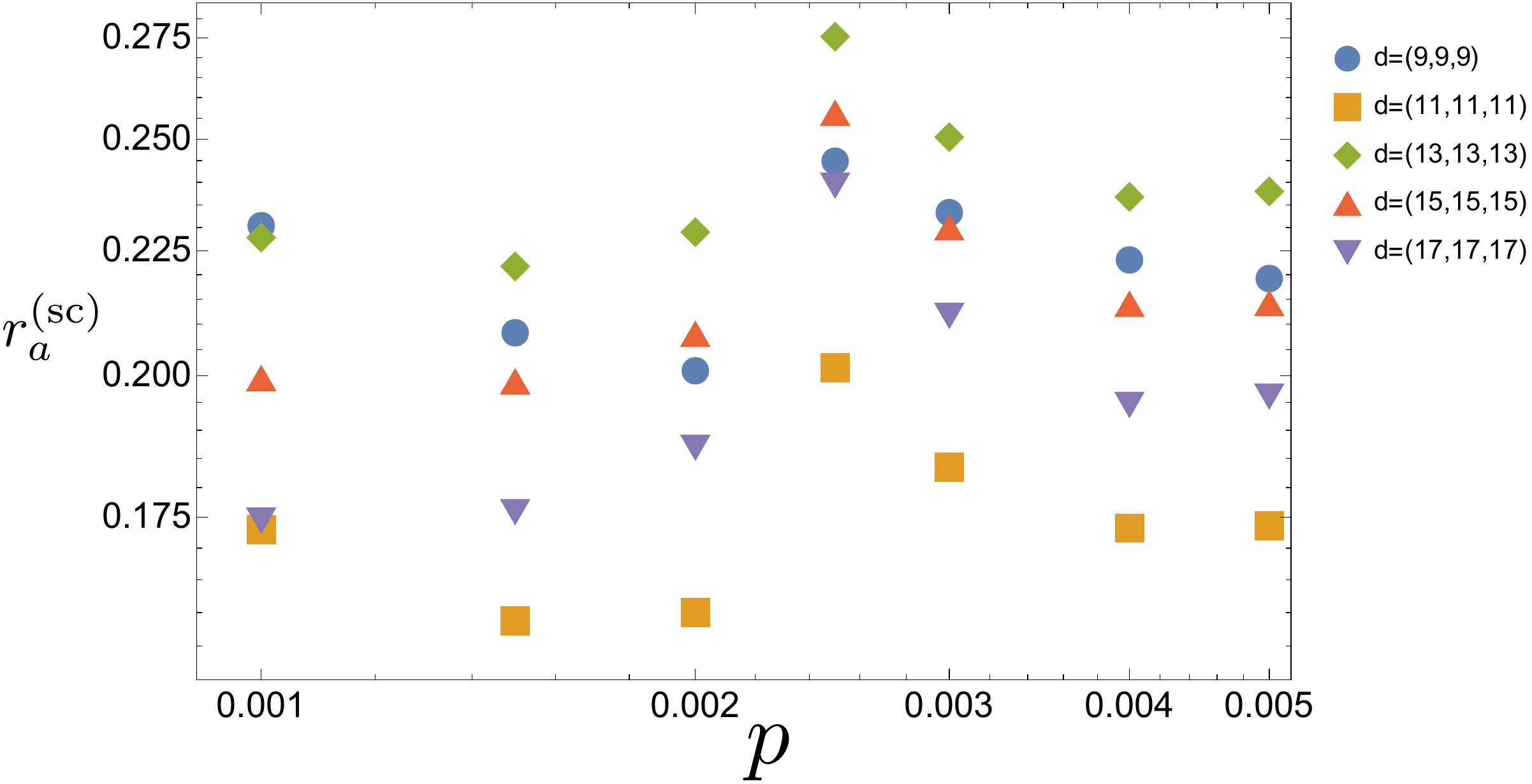}
	}
	\subfloat[\label{fig:Ratio11Layer}]{%
		\includegraphics[width=0.5\textwidth]{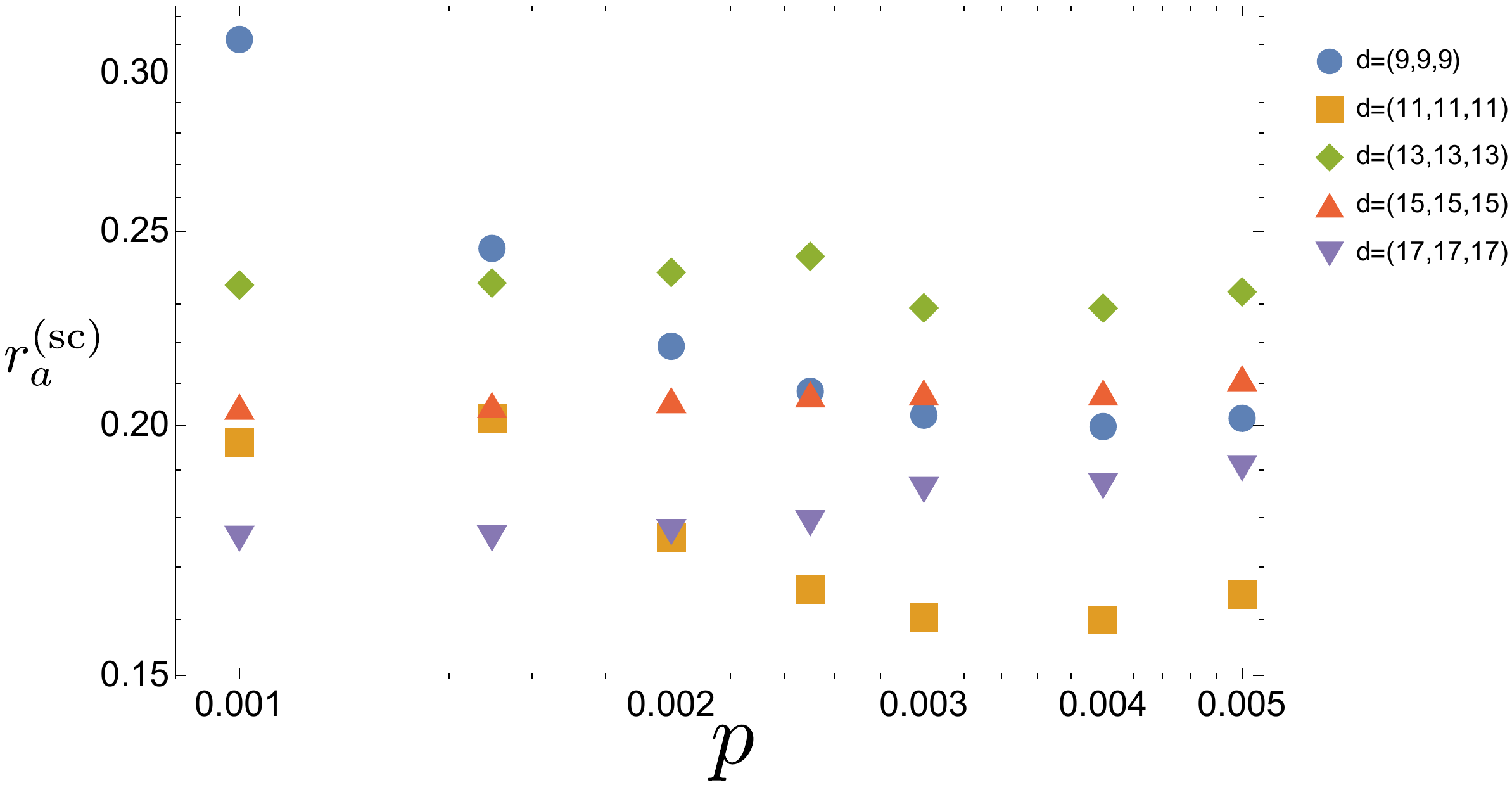}
	}
	\caption{\label{fig:SyndromDensityPlots} After applying corrections from the local NN decoder, we plot the ratio $r^{(\text{sc})}_a$ (see the main text) between the average number of highlighted vertices in the matching graph $G_{\text{sc}}^{(N)}$ where a syndrome collapse has been performed (using sheets of size $d'_m = 6$) to the average number of highlighted vertices in the original matching graph $G$ prior to the application of the local NN decoder followed by a syndrome collapse. In (a), the results are shown for the 6-layer network whereas in (b) the results are shown for the 11-layer network. the relationship of $r^{(\text{sc})}_a$ as a function of the code distance is less intuitive here compared to the results obtained in \cref{fig:SyndromDensityPlotsVertClean} for the vertical cleanup. The reason is that the number of two-dimensional sheets in the matching graph depend on the surface code distance, and there can be a jump of one sheet when the distances increase, as is the case for example with the $d=11$ and $d=13$ graphs.  }
	\label{fig:Ratio}
\end{figure*}

If a local NN decoder with effective distance $d_{\text{eff}}=5$ was applied prior to performing a vertical cleanup, such $X$-type errors would be removed and no logical failures would occur. However, we generally caution that a vertical cleanup could in fact reduce the effective code distance of the surface code if the local NN decoder has an effective distance smaller than the volume to which it is applied. Nonetheless, as shown in \cref{subsec:Numerics} below, low logical error rates and near optimal effective code distance are indeed achievable with our local NN decoders and vertical cleanup. 

Lastly, at the end of \cref{subsec:SyndromeCollapse}, we explained how a syndrome collapse reduces the timelike distance of a lattice surgery protocol. Performing a vertical cleanup does not have the same effect on the timelike distance, and can be applied during a lattice surgery protocol. More details are provided in \cref{appendix:VertCleanLatticeSurgery}.

\subsection{Numerical results.}
\label{subsec:Numerics}

\begin{table*}
\centering
\begin{tabular*}{\textwidth}{@{\extracolsep{\fill}}rrrrccccccc@{}}
 \toprule
 & & & & \multicolumn{7}{c}{best $p_{\textrm{train}} \in \{\CC{1.0 \times 10^{-3}}, \LG{2.5 \times 10^{-3}}, \SG{5.0 \times 10^{-3}}\}$} \\
 \cmidrule{5-11}
 layers & $d_x$ & $d_z$ & $d_m$ & at $p = 1.0 \times 10^{-3}$ & $1.5 \times 10^{-3}$ & $2.0 \times 10^{-3}$ & $2.5 \times 10^{-3}$ & $3.0 \times 10^{-3}$ & $4.0 \times 10^{-3}$ & $5.0 \times 10^{-3}$  \\
 \midrule
 6 &  9 &  9 &  9 & $\CC{1.0 \times 10^{-3}}$ & $\CC{1.0 \times 10^{-3}}$ & $\CC{1.0 \times 10^{-3}}$ & $\SG{5.0 \times 10^{-3}}$ & $\SG{5.0 \times 10^{-3}}$ & $\SG{5.0 \times 10^{-3}}$ & $\SG{5.0 \times 10^{-3}}$ \\ 
   & 11 & 11 & 11 & $\CC{1.0 \times 10^{-3}}$ & $\CC{1.0 \times 10^{-3}}$ & $\CC{1.0 \times 10^{-3}}$ & $\SG{5.0 \times 10^{-3}}$ & $\SG{5.0 \times 10^{-3}}$ & $\SG{5.0 \times 10^{-3}}$ & $\SG{5.0 \times 10^{-3}}$ \\
   & 13 & 13 & 13 & $\CC{1.0 \times 10^{-3}}$ & $\CC{1.0 \times 10^{-3}}$ & $\CC{1.0 \times 10^{-3}}$ & $\SG{5.0 \times 10^{-3}}$ & $\SG{5.0 \times 10^{-3}}$ & $\SG{5.0 \times 10^{-3}}$ & $\SG{5.0 \times 10^{-3}}$ \\
   & 15 & 15 & 15 & $\CC{1.0 \times 10^{-3}}$ & $\CC{1.0 \times 10^{-3}}$ & $\CC{1.0 \times 10^{-3}}$ & $\SG{5.0 \times 10^{-3}}$ & $\SG{5.0 \times 10^{-3}}$ & $\SG{5.0 \times 10^{-3}}$ & $\SG{5.0 \times 10^{-3}}$ \\
   & 17 & 17 & 17 & $\CC{1.0 \times 10^{-3}}$ & $\CC{1.0 \times 10^{-3}}$ & $\CC{1.0 \times 10^{-3}}$ & $\SG{5.0 \times 10^{-3}}$ & $\SG{5.0 \times 10^{-3}}$ & $\SG{5.0 \times 10^{-3}}$ & $\SG{5.0 \times 10^{-3}}$ \\
 \addlinespace
 11 &  9 &  9 &  9 & $\LG{2.5 \times 10^{-3}}$ & $\SG{5.0 \times 10^{-3}}$ & $\SG{5.0 \times 10^{-3}}$ & $\SG{5.0 \times 10^{-3}}$ & $\SG{5.0 \times 10^{-3}}$ & $\SG{5.0 \times 10^{-3}}$ & $\SG{5.0 \times 10^{-3}}$ \\ 
    & 11 & 11 & 11 & $\LG{2.5 \times 10^{-3}}$ & $\LG{2.5 \times 10^{-3}}$ & $\SG{5.0 \times 10^{-3}}$ & $\SG{5.0 \times 10^{-3}}$ & $\SG{5.0 \times 10^{-3}}$ & $\SG{5.0 \times 10^{-3}}$ & $\SG{5.0 \times 10^{-3}}$ \\
    & 13 & 13 & 13 & $\LG{2.5 \times 10^{-3}}$ & $\LG{2.5 \times 10^{-3}}$ & $\LG{2.5 \times 10^{-3}}$ & $\LG{2.5 \times 10^{-3}}$ & $\SG{5.0 \times 10^{-3}}$ & $\SG{5.0 \times 10^{-3}}$ & $\SG{5.0 \times 10^{-3}}$ \\
    & 15 & 15 & 15 & $\LG{2.5 \times 10^{-3}}$ & $\LG{2.5 \times 10^{-3}}$ & $\LG{2.5 \times 10^{-3}}$ & $\LG{2.5 \times 10^{-3}}$ & $\SG{5.0 \times 10^{-3}}$ & $\SG{5.0 \times 10^{-3}}$ & $\SG{5.0 \times 10^{-3}}$ \\
    & 17 & 17 & 17 & $\LG{2.5 \times 10^{-3}}$ & $\LG{2.5 \times 10^{-3}}$ & $\LG{2.5 \times 10^{-3}}$ & $\LG{2.5 \times 10^{-3}}$ & $\SG{5.0 \times 10^{-3}}$ & $\SG{5.0 \times 10^{-3}}$ & $\SG{5.0 \times 10^{-3}}$ \\
 \bottomrule
\end{tabular*}
\caption{Table showing the error rates at which the 6 and 11-layer NN were trained to give the lowest \textit{total} logical $X+Z$ error rate when applied to test set data of volume $(d_x,d_z,d_m)$ and physical error rate $p$. The first column gives the input volume of the test set data. Subsequent columns give the error rates used to train the best performing NN model when applied to the physical error rates used to generate the test set data given in the top row.}
\label{tab:6layerTrainedModel}
\end{table*}

In this section, we show the logical error rates and and syndrome density reductions achieved by the 6 and 11-layer NN's described in \cref{subsec:NNdescription} (see \cref{fig:NetworkArchitectures}). We obtain our numerical results by first applying the trained NN decoder to the input volume $(d_x,d_z,d_m)$, followed by either performing a syndrome collapse (as described in \cref{subsec:SyndromeCollapse}) or a vertical cleanup (as described in \cref{subsec:RemoveVerticalPairs}). After the syndrome collapse or vertical cleanup, any remaining errors are removed by performing MWPM on the resulting graph. We set edges to have unit weights since the error distributions change after applying the local NN decoder. 

In what follows, we define $G$ to be the matching graph with highlighted vertices \textit{prior} to applying the local NN decoder. Since we consider a symmetric noise model, we focus only on correcting $X$-type Pauli errors, as $Z$-type errors are corrected analogously using the same network. To optimize speed, the global decoder uses separate graphs $G_X$ and $G_Z$ for correcting $X$ and $Z$-type Pauli errors. However since we focus on results for $X$-type Paulis, to simplify the discussion we set $G=G_X$. The graph obtained after the application of the NN decoder is labelled $G^{(N)}$ (which will in general have different highlighted vertices than $G$), and the reduced graph obtained by performing the syndrome collapse on $G^{(N)}$ is labelled $G_{\text{sc}}^{(N)}$. Lastly, the graph obtained after applying the local NN decoder followed by a vertical cleanup is labeled $G_{\text{vc}}^{(N)}$.

We trained the 6 and 11-layer networks on data consisting of input volumes of size $(13,13,18)$. The data was generated for physical depolarizing error rates of $p = 10^{-3}$, $p = 2.5 \times 10^{-3}$ and $p = 5 \times 10^{-3}$, resulting in a total of six models. For each of the physical error rates mentioned above, we generated $10^7$ training examples by performing Monte Carlo simulations using the noise model described in \cref{sec:SurfRev}. Both the 6 and 11-layer networks were trained for 40 epochs when $p = 10^{-3}$, and for 80 epochs when $p = 2.5 \times 10^{-3}$ and $p = 5 \times 10^{-3}$. The networks were then applied to test set data generated at physical error rates in the range $10^{-3} \le p \le 5 \times 10^{-3}$ (see \cref{tab:6layerTrainedModel} which describes which models gave the best results for a given physical error rate used in the test set data). The networks described in \cref{fig:NetworkArchitectures} have a receptive field of size $9 \times 9 \times 9$, and thus have a maximal effective local distance of $d_{\text{eff}} \le 9$. Recall that in the last layer we use a sigmoid activation function (instead of ReLu) to ensure that the two output tensors describing $X$ and $Z$ data qubit corrections in each of the $d_m$ syndrome measurement rounds consists of numbers between zero and one. If this output is greater than $0.5$ we apply a correction to a given qubit, otherwise we do nothing. We found numerically that a decision threshold of $0.5$ gave the best results. In other words, the outputs consist of $d_m$ matrices of size $d_x \times d_z$ for $X$ corrections, and $d_m$ matrices of size $d_x \times d_z$ for $Z$ corrections. If the $(i,j)$ coordinate of the matrix for $X$ ($Z$) Pauli corrections in round $k$ is greater than $0.5$, we apply an $X$ ($Z$) Pauli correction on the data qubit at the $(i,j)$ coordinate of the surface code lattice in the $k$'th syndrome measurement round. 

\subsubsection{Numerical analysis when performing a syndrome collapse.}
\label{subsubsec:NumericsSynColl}

When performing a syndrome collapse, we considered sheets of size $d'_m \in \{4,5,6\}$. We found numerically that using sheets of size $d'_m = 4$ resulted in worse performance compared to sheets of size $d'_m = 5$ and $d'_m = 6$. Using sheets of size $d'_m = 5$ and $d'_m = 6$ resulted in nearly identical performance. However since sheets of size $d'_m = 6$ results in a smaller graph $G^{(N)}_{\text{sc}}$ compared to using sheets of size $d'_m = 5$, in what follows we give numerical results using sheets of size $d'_m = 6$.

\begin{figure*}
	\centering
	\subfloat[\label{fig:Plot6LayerVertClean}]{%
		\includegraphics[width=0.46\textwidth]{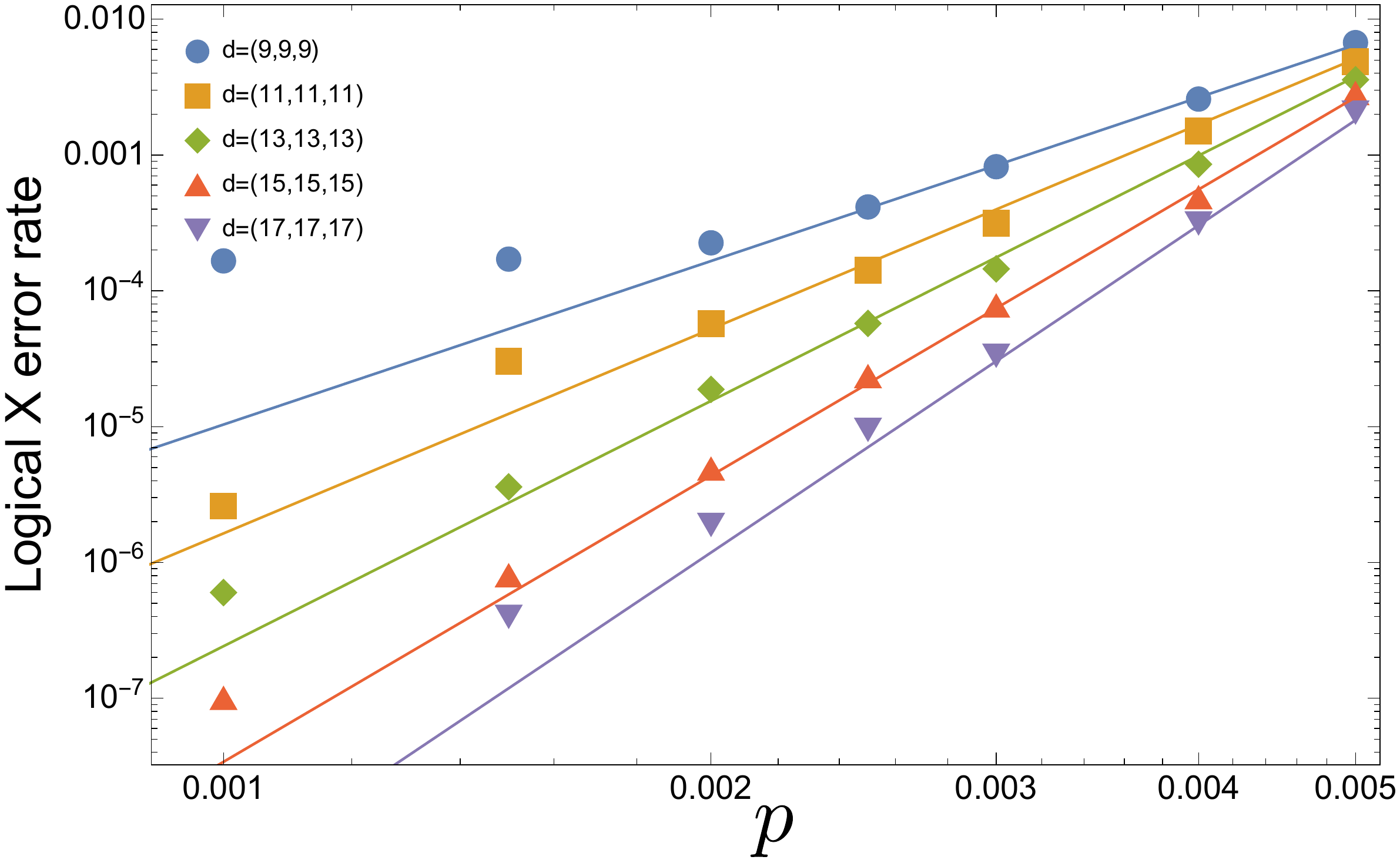}
	}
	\subfloat[\label{fig:Plot11LayerVertClean}]{%
		\includegraphics[width=0.46\textwidth]{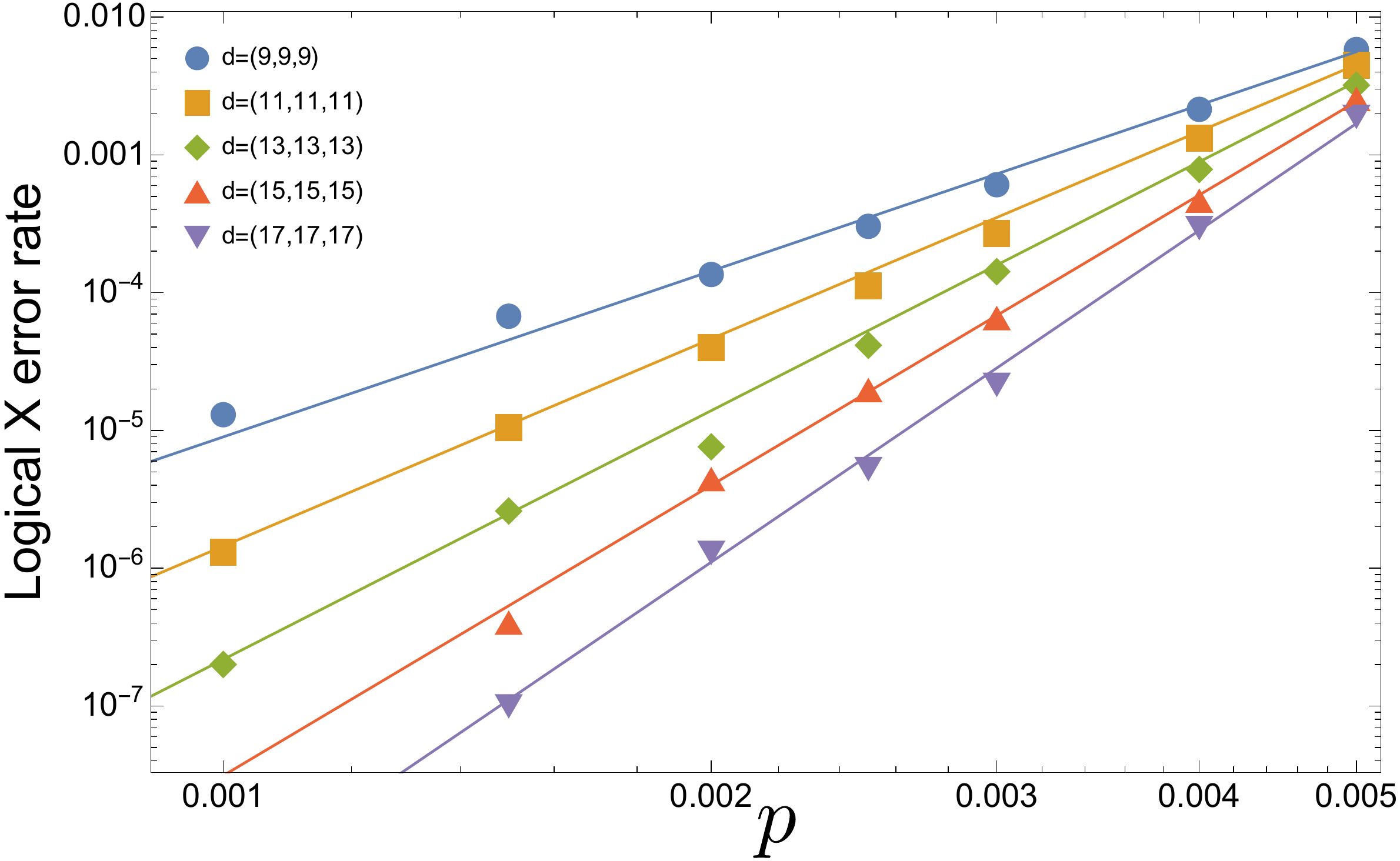}
	}
	\caption{\label{fig:logXErrorRatesVertClean} Logical $X$ error rates for surface code volumes ranging between $(9,9,9)$ and $(17,17,17)$ after the application of the local NN decoder, followed by a vertical cleanup and MWPM to correct any remaining errors. In (a) the results are for the 6-layer network whereas in (b) the results are for the 11-layer network.}
\end{figure*}

\begin{figure*}
	\centering
	\subfloat[\label{fig:Ratio6LayerVertClean}]{%
		\includegraphics[width=0.46\textwidth]{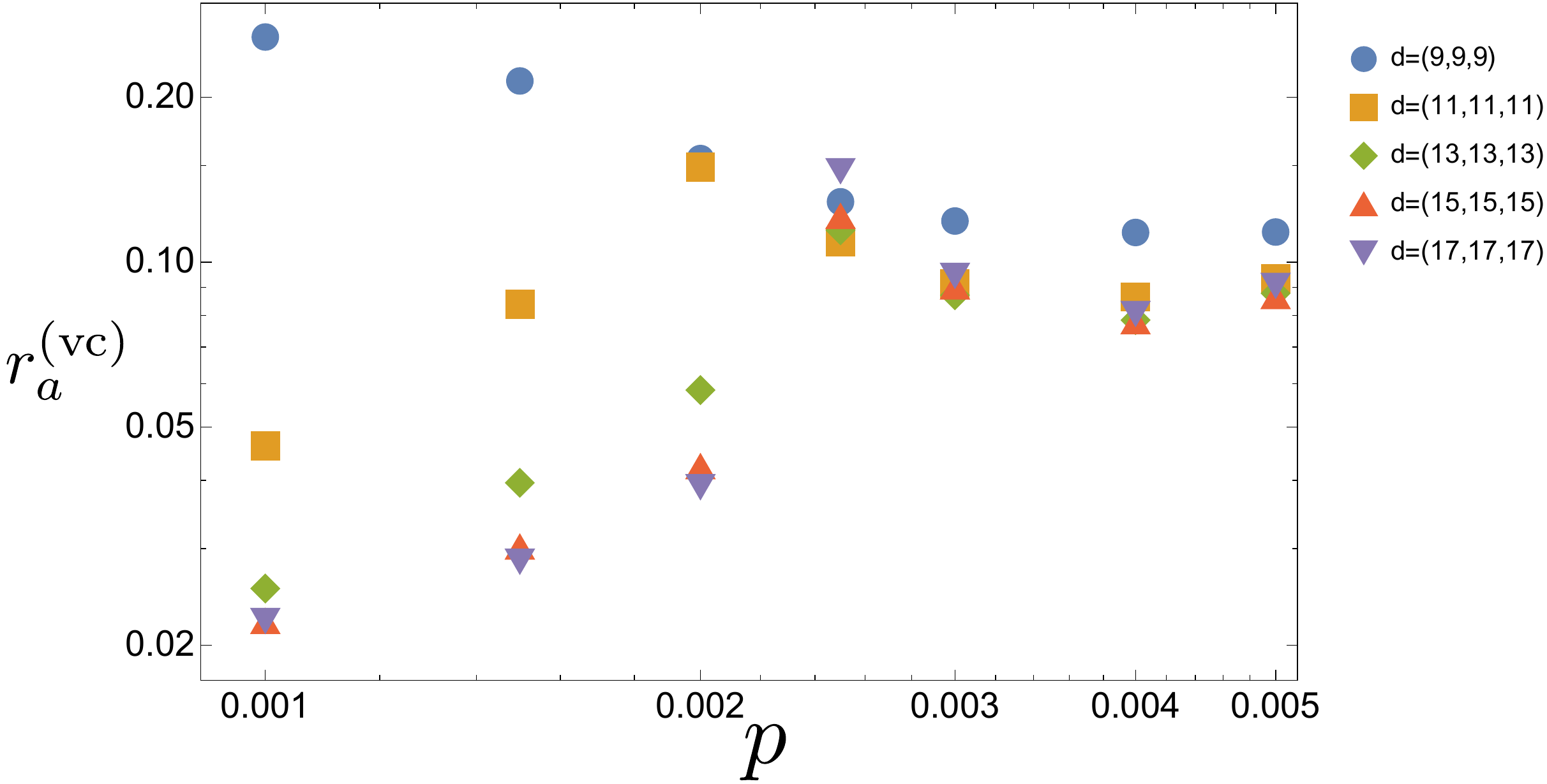}
	}
	\subfloat[\label{fig:Ratio11LayerVertClean}]{%
		\includegraphics[width=0.46\textwidth]{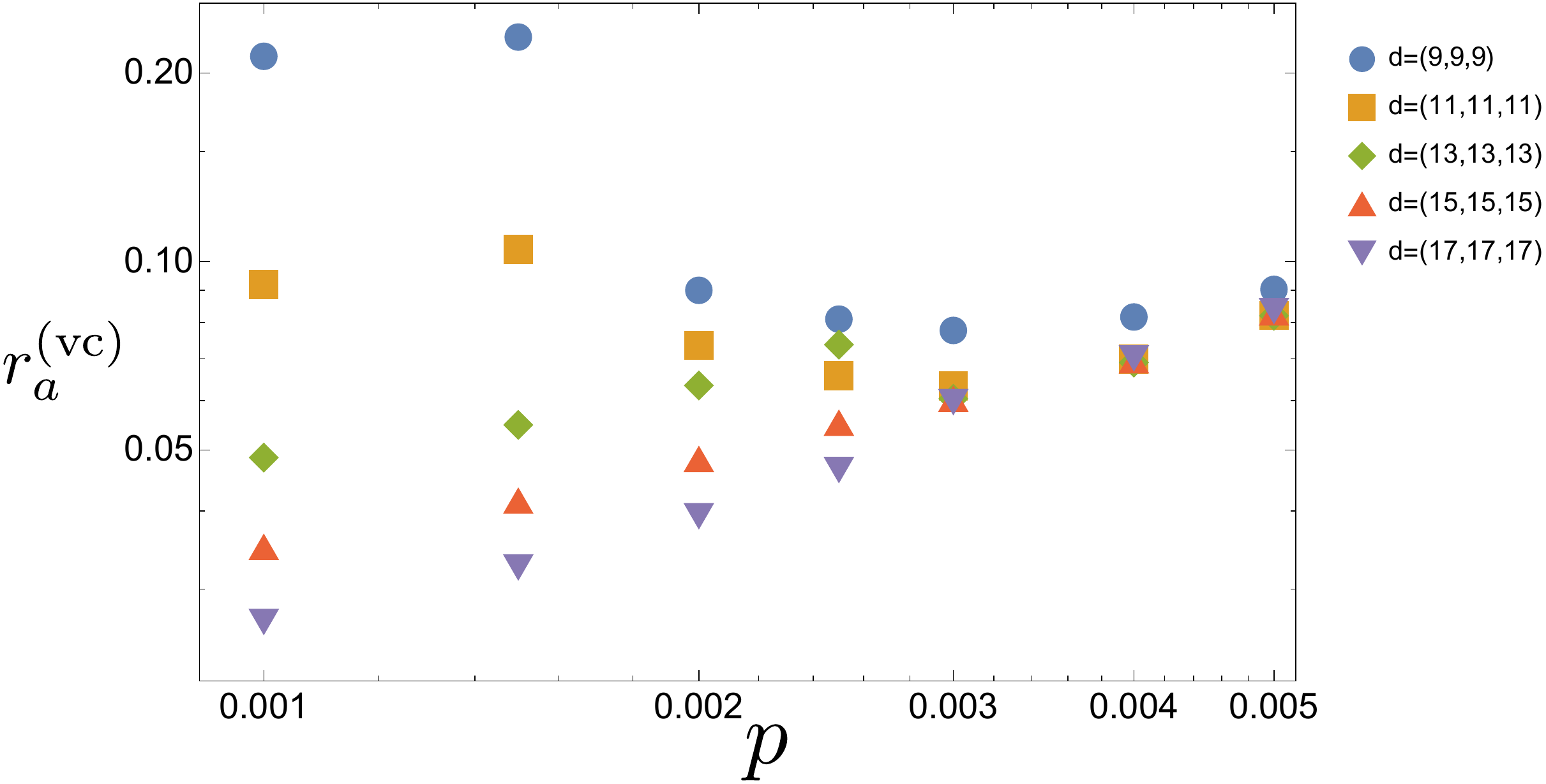}
	}
	\caption{\label{fig:SyndromDensityPlotsVertClean} After applying corrections from the local NN decoder, we plot the ratio $r^{(\text{vc})}_a$ (see the main text) between the average number of highlighted vertices in the matching graph $G_{\text{vc}}^{(N)}$ where a vertical cleanup has been performed to the average number of highlighted vertices in the original matching graph $G$ prior to the application of the local NN decoder followed by a vertical cleanup. In (a), the results are shown for the 6-layer network whereas in (b) the results are shown for the 11-layer network.}
\end{figure*}

The logical $X$ error rate curves for the 6 and 11-layer networks are shown in \cref{fig:Plot6layerLogX,fig:Plot11layerLogX}. As a first remark, we point out that networks trained at high physical error rates don't necessarily perform better when applied to test set data obtained at lower error rates (which is in contrast to what was observed in previous works such as Ref.~\cite{CR18NN}). In \cref{tab:6layerTrainedModel} it can be seen that the 6-layer network trained at $p = 0.005$ outperforms the model trained at $p = 0.0025$ and $p=0.001$ when applied to test set data generated at $p \ge 0.0025$. However, for test set data generated in the range $0.001 \le p \le 0.002$, the model trained at $p = 0.001$ achieves lower total logical error rates. For the 11-layer network, the model trained at $p = 0.0025$ always outperform the model trained at $p=0.001$ for all the sampled physical error rates. The window of out-performance also depends on the surface code volume. For instance, the 11-layer network trained at $p = 0.005$ outperforms the model trained at $p=0.0025$ for $p > 0.001$ when applied to a $(9,9,9)$ surface code volume. However, when applied to a $(17,17,17)$ surface code volume, the model trained at $p=0.0025$ outperforms the model trained at $p=0.005$ for $p \le 0.0025$. More details comparing models trained at different physical error rates are discussed in \cref{appendix:CompareModels} and \cref{fig:CompareModelsPlot}.

Note that to achieve better results, one can train a network for each physical error rate used in the test set data. However, in generating our results, one goal was to see how well a network trained at a particular error rate would perform when applied to data generated at a \textit{different} physical error rate. In realistic settings, it is often difficult to fully characterize the noise model, and circuit-level failure rates can also fluctuate across different regions of the hardware. As such, it is important that our networks trained at a particular value of $p$ perform well when applied to other values of $p$. An alternative to using models trained at different values of $p$ would be to train a single network with data generated at different values of $p$. However, doing so might reduce the network's accuracy at any particular error rate. Since in practice one would have some estimate of the error rate, it would be more favorable to train the network near such error rates.

In general, we expect the logical $X$ error rate polynomial as a function of the code distance and physical error rate $p$ to scale as (assuming $d_x=d_z=d$)
\begin{align}
    p^{(X)}_L(p) = u d d_m (bp)^{(cd+w)},
    \label{eq:AnsatzLogError}
\end{align}
for some parameters $u$, $b$, $c$ and $w$, and where $d_m$ is the number of syndrome measurement rounds. Using the data from \cref{fig:Plot11layerLogX}, we find that the 11-layer network has a logical $X$ error rate polynomial 
\begin{align}
    p^{(X;\text{sc})}_{L;\text{11l}}(p) = 0.000260 d^2(143.084p)^{(d-1)/2},
    \label{eq:11layerPolyDMP6}
\end{align}
and the 6-layer network has a logical error rate polynomial
\begin{align}
    p^{(X;\text{sc})}_{L;\text{7l}}(p) = 0.000436 d^2(145.277p)^{(d-1)/2},
    \label{eq:6layerPolyDMP6}
\end{align}
where $d_m=d$ for all of our simulations. As such, for a distance $d_x=d_z=d$ surface code, applying both the 6 and 11-layer local NN decoder followed by a syndrome collapse with each sheet having height $d'_m=6$ results in an effective code distance $d_{\text{eff}} \approx d-2$. The plots in \cref{fig:logXErrorRates} also show a threshold of $p_{\text{th}} \approx 5 \times 10^{-3}$. Note that in \cref{eq:11layerPolyDMP6,eq:6layerPolyDMP6} we added labels to distinguish the polynomials arising from the 11 and 6-layer networks, and to indicate that the results are obtained from performing a syndrome collapse. 

In \cref{fig:Ratio}, we give the ratio $r^{(\text{sc})}_a = A_{\text{syn}}(G_{\text{sc}}^{(N)}) / A_{\text{syn}}(G)$ where $A_{\text{syn}}(G)$ corresponds to the average number of ``raw'' syndrome differences appearing in a given spacetime volume and $A_{\text{syn}}(G_{\text{sc}}^{(N)})$ corresponds to the average number syndrome differences after the application of the local NN decoder and syndrome collapse. As a side note, we remark that due to the possible creation of vertical pairs of highlighted vertices after the NN has been applied, $A_{\text{syn}}(G^{(N)})$ (i.e. the average number of syndrome differences after the application of the NN decoder but before performing a syndrome collapse) may have more highlighted vertices than what would be obtained if no local corrections were performed. 

A small $r^{(\text{sc})}_a$ ratio indicates that a large number of highlighted vertices vanish after applying the local NN decoder and performing a syndrome collapse, and results in a faster implementation of MWPM or Union Find. More details on how the ratio $r^{(\text{sc})}_a$ affects the throughput performance of a decoder are discussed in \cref{subsec:DownstreamPerfReqs}.

The reader may remark that there are discontinuities in the plots of \cref{fig:Ratio6Layer,fig:Ratio11Layer}, as well as the logical error rate plots in \cref{fig:logXErrorRates}. There are two reasons contributing to the discontinuities. The first is because the models were trained at different physical error rates; at each error rate $p$, we chooe the model that performs best as outlined \cref{tab:6layerTrainedModel}. However, upon careful inspection the discontinuities are more pronounced for surface code volumes of size $(9,9,9)$ and $(11,11,11)$. This is because the NN models were trained on a $(13,13,18)$ volume in order for the network to see data which is purely in the bulk (since the local receptive field of our models is $9 \times 9 \times 9$). We do not expect a model trained on a volume where the receptive field sees data purely in the bulk to generalize well to smaller surface code volumes given the network's  local receptive field will always see data containing boundaries in these scenarios. As such, to achieve better performance on volumes with $d_x=d_z < 13$, one should train a network on a volume of that size.

\subsubsection{Numerical analysis when performing a vertical cleanup.}
\label{subsubsec:NumericsVertClean}

The logical $X$ error rates when performing a vertical cleanup after applying the 6 and 11-layer local NN decoders are shown in \cref{fig:Plot6LayerVertClean,fig:Plot11LayerVertClean}. The models trained at $p=0.001$, $p=0.0025$ and $p=0.005$ were applied to the test set data following \cref{tab:6layerTrainedModel}. The discontinuities in the logical error rate curves occur for the same reasons as outlined above for the syndrome collapse protocol, and are particularly apparent for the 6-layer network applied to test set data generated on a $(9,9,9)$ volume as shown in \cref{fig:Plot6LayerVertClean}. Comparing the logical $X$ error rate curves in \cref{fig:Plot6LayerVertClean} and \cref{fig:Plot11LayerVertClean} also shows the performance improvement that is gained by using a larger network (however for $d \ge 13$, only a small performance gain is observed from using the 11-layer network). The logical error rate polynomial for the 11-layer network is 
\begin{align}
    p^{(X;\text{vc})}_{L;\text{11l}}(p) = 0.0008198 d^2(107.803p)^{(d-1)/2},
    \label{eq:11layerPolyVC}
\end{align}
and for the 6-layer network is
\begin{align}
    p^{(X;\text{vc})}_{L;\text{7l}}(p) = 0.001022 d^2(105.752p)^{(d-1)/2}.
    \label{eq:6layerPolyVC}
\end{align}
As with the syndrome collapse, applying the local NN decoders followed by a vertical cleanup results in an effective distance $d_{\text{eff}} \approx d-2$. It can also be observed that at $p = 0.005$, the logical error rate decreases when increasing the code distance $d$, indicating a threshold $p_{\text{th}} > 0.005$ when applying the local NN decoder followed by a vertical cleanup. Note that we did not generate data for $p > 0.005$ since we are primarily concerned with the error rate regime where low logical error rates can be achieved while simultaneously being able to implement our decoders on the fast time scales required by quantum algorithms. 

In \cref{fig:Ratio6LayerVertClean,fig:Ratio11LayerVertClean} we show the ratio's $r^{(\text{vc})}_a = A_{\text{syn}}(G_{\text{vc}}^{(N)}) / A_{\text{syn}}(G)$ which is identical to $r^{(\text{sc})}_a$, but where a vertical cleanup is performed instead of a syndrome collapse. For $p = 0.001$ and the distance $d=17$ surface code, we see a reduction in the average number of highlighted vertices by nearly two orders of magnitude. Further, comparing with the ratio's $r^{(\text{sc})}_a$ obtained in \cref{fig:SyndromDensityPlots}, we see that performing a vertical cleanup results in fewer highlighted vertices compared to performing a syndrome collapse by sheets. Such a result is primarily due to the fact that vertical pairs of highlighted vertices between sheets do not vanish after performing a syndrome collapse. Lastly we observe an interesting phenomena for the 11-layer networks trained at $p=0.001$ and $p=0.0025$ when applied to test set data generate near $p=0.001$. Although the 11-layer trained at $p=0.0025$ achieves a lower total logical failure rate (see \cref{tab:6layerTrainedModel}), the network trained at $p=0.001$ results in smaller ratio $r^{(\text{vc})}_a$. This can be seen for instance by comparing the results in \cref{fig:Ratio6LayerVertClean,fig:Ratio11LayerVertClean}, where although the 6-layer network is outperformed by the 11-layer network, a smaller $r^{(\text{vc})}_a$ is achieved at $p=0.001$ since the the 6-layer network trained at $p=0.001$ was applied to the test set data, compared to the 11-layer network which was trained at $p=0.0025$. 

\section{Hardware implementation of our NN's}
\label{section:LatencyReduction}

Let us now consider possible suitable embodiment's of NN decoders on classical hardware.
One of the appealing features of NN evaluation is that it involves very little conditional logic.
In theory, this greatly helps in lowering NN evaluation strategies to specialized hardware, where one can discard the bulk of a programmable processor as irrelevant and one can make maximal use of pipelined data pathways.
In practice, such lowering comes with significant costs, among them slow design iteration, custom manufacturing, bounded size, and a great many concerns around integration with existing electronics.
In this section we consider some candidate technologies which occupy compromise positions among these costs.

\subsection{FPGA implementation performance}

One option for specialized hardware is a Field-Programmable Gate Array (FPGA).
A typical FPGA consists of a fixed set of components, including flip-flops, look-up tables (LUTs), block RAM (BRAM), configurable logic blocks (CLBs), and digital signal processing (DSP) slices, all of whose inputs can be selectively routed into one another to perform elaborate computations ranging from fixed high-performance arithmetic circuits to entire programmable processors.
FPGAs have been used for NN evaluation in a variety of real-time applications; one use case particularly close to ours is the recognition of nontrivial events at the Large Hadron Collider.
That working group has produced an associated software package \textsf{hls4ml}~\cite{hls4mlArticle} which produces a High-Level Synthesis (HLS) description of an evaluation scheme for a given initialized NN, and one can then compile that description into a high-throughput and low-latency FPGA embodiment.
The tool \textsf{hls4ml} itself has several tunable parameters which trade between resource occupation on the target FPGA and performance in throughput and latency, e.g.: re-use of DSP slices to perform serial multiply-and-add operations rather than parallel operations; ``quantization'' of intermediate results to a specified bit width; and so on.

At the time of this writing, \textsf{hls4ml} does not support 3D convolutional layers.
Rather than surmount this ourselves, we explored the realization through \textsf{hls4ml} of 1D and 2D convolutional networks of a similar overall structure and parameter count to the models considered in \cref{subsec:NNdescription} under the assumption that the generalization to 3D will not wildly change the inferred requirements.%
\footnote{%
As evidence, we verified that 1D and 2D convolutional networks of similar shape and size occupy similar FPGA resources under \textsf{hls4ml}.
See also the argument in Section 2.2 of Ref.~\cite{LiuEtAlConv3D}.
}
We report one such experiment in \cref{fig:DSPUsage}, which includes both the details of the analogous model and the resulting FPGA resource usage; other networks and other \textsf{hls4ml} settings are broadly similar.

\begin{figure*}
\centering
\begin{tabular}{@{}rrrrrrrrrrr@{}} \toprule
\textbf{Layers}  & BRAM &     (\%) &     LUT &    (\%) & DSP &    (\%) &      FF &    (\%) &         latency & rate \\ \midrule
\textbf{11}      &  656 & (24.4\%) & 139~378 & (8.1\%) &   1 &   (0\%) & 131~383 & (3.8\%) & 2~785.55 $\mu$s & 359 Hz \\
one-layer $\max$ &   35 &  (1.3\%) &  12~528 & (0.7\%) &  58 & (1.4\%) &  17~483 & (0.5\%) & 1~566.72 $\mu$s & 638 Hz \\
\bottomrule
\end{tabular}
\caption{%
FPGA resource costs for an \textsf{hls4ml} embodiment of a NN composed of 2D convolutional layers, each with $3 \times 3$ kernels and $60$ output channels, taking an initial $32 \times 32$ trichannel image, for a total of $360,180$ trainable parameters and a per-layer maximum of $32,580$ trainable parameters.
This model is chosen so as to limit ourselves to the functionality provided in \textsf{hls4ml}, while maintaining structural similarity to the models of direct interest given in \Cref{subsec:NNdescription}.
Relative percentages reported are taken against the resources available on a Virtex Ultrascale+ FPGA (XCU250-FIGD2104-2L-E).
Note that our strong quantization settings often caused \textsf{hls4ml} to trade DSPs for LUTs to use as multipliers.
}
\label{fig:DSPUsage}
\end{figure*}

One way to improve model throughput is by \emph{inter-layer pipelining}, i.e., deploying its individual layers to different hardware components and connecting those components along communication channels which mimic the structure of the original network.
Whereas the throughput of a conventional system is reciprocal to the total time between when input arrives and when output is produced (i.e., the computation latency), the throughput of a pipelined system is reciprocal only to the computation latency of its slowest constituent component.
Accordingly, we also report the FPGA resource usage for the largest layer in the network, so as to calculate pipelined throughput.

Out of the synthesis details, we highlight the re-use parameter $R$: the set of available such parameter values is discrete and increasingly sparse for large $R$; latency scales linearly with choice of large values of $R$ and synthesis will not converge for small values of $R$; and the size of our model necessitated choosing the rather large re-use parameter $R = 540$ to achieve synthesis.
In fact, even just synthesizing one layer required the same setting of $R = 540$, which results in rather meager throughput savings achieved by pipelining FPGAs, one per layer.
Unfortunately, we conclude these models are nontrivial to realize within the constraints of contemporary FPGA hardware.

A promising avenue to close this gap may be networks that reduce computational cost by encoding parameters in at most a few bits, while incurring some small loss in accuracy.
For instance, authors in Ref.~\cite{Geng2019} used an optimized Binary Convolution NN on a Xilinx KCU1500 FPGA with order 100 $\mu$s inference latencies on networks with millions of parameters (e.g., AlexNeT, VGGNet, and ResNet).

\subsection{ASIC performance and Groq}

The programmability of FPGAs makes them popular for a wide variety of tasks, and hence they appear as components on a wide variety of commodity hardware.
However, flexibility is double-edged: FPGAs' general utility means they are likely to be under-optimized for any specific task.
Application-Specific Integrated Circuits (ASICs) form an alternative class of devices which are further tailored to a specific application domain, typically at the cost of general programmability.
Groq~\cite{GroqArticle} is an example of a programmable ASIC which is also a strong candidate target: it is tailored toward low-latency, high-throughput evaluation of NN's, but without prescribing at manufacturing time a specific NN to be evaluated.

We applied Groq's public software tooling to synthesize binaries suitable for execution on their devices.
In \Cref{fig:GroqUsage}, we report the synthesis statistics for the 11-layer network of \Cref{subsec:NNdescription} and for the single largest layer of the network, both as embodied on a single Groq chip.
Otherwise, we left all synthesis settings at their default, without exploring optimizations.
Even with these default settings, the reported throughput when performing per-layer pipelining is within 6--10$\times$ of the target value of $\approx 40$ kHz.
We believe that further tuning, perhaps entirely at the software level, could close this gap, amounting to one path to hardware feasibility.
Such tunable features include pruning near-zero weights, quantizing the intermediate arithmetic to some lossier format, intra-layer distributed evaluation (i.e., evaluating the outputs of a given convolutional layer in parallel over several chips), instruction timing patterns, and so on.

\begin{figure}
\begin{tabular}{@{}rrr@{}} \toprule
\textbf{Layers} & latency & rate \\ \midrule
\textbf{11} & 672.53 $\mu$s & 1~487 Hz \\
one-layer $\max$ & 168.71 $\mu$s & 5~927 Hz \\
\bottomrule
\end{tabular}
\caption{Resource costs for single Groq chip embodiments of the $11$-layer NN model given in \Cref{subsec:NNdescription}.}
\label{fig:GroqUsage}
\end{figure}

\subsection{Effect on global decoders}
\label{subsec:DownstreamPerfReqs}

In \cref{fig:Ratio,fig:SyndromDensityPlotsVertClean}, we reported a multiplicative relationship between the number of ``raw'' syndromes  $A_{\text{syn}}(G)$ appearing in a given spacetime volume to the number of syndromes $A_{\text{syn}}(G_{\text{sc}}^{(N)})$ remaining after the application of the local NN decoder and syndrome collapse $A_{\text{syn}}(G_{\text{sc}}^{(N)}) = r^{(\text{sc})}_a \cdot A_{\text{syn}}(G)$ or the application of the local NN decoder followed by a vertical cleanup $A_{\text{syn}}(G_{\text{vc}}^{(N)}) = r^{(\text{vc})}_a \cdot A_{\text{syn}}(G)$, where $r^{(\text{sc})}_a, r^{(\text{vc})}_a < 1$.
In what follows, the reader is to interpret $r_a$ to mean either $r_a^{\text{(sc)}}$ or $r_a^{\text{(vc)}}$, according to whether they are applying syndrome collapse or vertical clean-up respectively.

This value $r_a$ has significant implications for the hardware performance requirements of global decoders, which arise from the same need described in \Cref{section:AlgoRun} to meet overall throughput.
For example, the UF decoder is a serial algorithm whose runtime is nearly linear in its inbound syndrome count (see \cref{sec:SurfRev}), from which it follows that preceding a UF decoder by a NN preprocessing relaxes its performance requirements by the same factor $r_a$ needed meet the same throughput deadline.
One can make a similar argument for more elaborate distributed decoders, such as the Blossom variant proposed by Fowler~\cite{FowlerOhOne}: if the rate at which a given worker encounters highlighted syndromes is reduced by a factor of $r_a$, then the amount of time it can spend processing a given syndrome is scaled up by a factor of $1/r_a$, so that minimum performance requirements in turn are scaled up by $1/r_a$.

In fact, for the syndrome collapse protocol, these improvements are quite pessimistic.
A decoder could take advantage of the simpler edge structure of $G_{\text{sc}}^{(N)}$ relative to $G$ given that the syndrome collapse shrinks the size of the graph.
In particular, the number of vertices and edges in $G$ is reduced by a factor of at least $d'_m$, with $d'_m$ being the size of the sheets in a syndrome collapse.
For instance, the complete implementation of a serial MWPM decoder can be decomposed into two steps.
The first is the construction of the syndrome graph using Dijkstra’s algorithm which finds the shortest path between a source highlighted vertex and all other highlighted vertices.
The second is the implementation of the serial Blossom algorithm on such graphs.
Following Ref.~\cite{PyMatching}, the syndrome graph using Dijkstra’s algorithm has time complexity $\mathcal{O}(h (N \log(N)+M))$ where $h$ is the number of highlighted vertices in the matching graph (in our case $G_{\text{sc}}^{(N)}$ for the syndrome collapse protocol) with $N$ vertices and $M$ edges.
The application of the local NN decoder followed by a syndrome collapse with sheets of size $d'_m$ reduces $h$ by a factor of $r_a$ and $N$ by a factor of $d'_m$. $M$ is reduced by a factor greater than $d'_m$ because not only are there edges incident to vertices for a given syndrome measurement rounds, but there are also vertical and space-time correlated edges incident to vertices in consecutive syndrome measurement rounds.
A serial Blossom algorithm when applied to a matching graph with $h$ highlighted vertices has complexity $\mathcal{O}(h^3 \log h)$.
As such, the runtime of the serial blossom algorithm is reduced by a factor of $\mathcal{O}(1/(r_a^3))$. 

These improvements in speed come algorithmically cheap: the procedures of syndrome collapse and vertical cleanup are both trivially spatially parallelizable, adding $O(d_m')$ operations of preprocessing before applying the global decoder.

\section{Conclusion}
\label{sec:Conclusion}

In this work we developed local NN decoders using fully three-dimensional convolutions, and which can be applied to arbitrary sized $(d_x,d_z,d_m)$ surface code volumes. We discussed more efficient ways of representing the training data for our networks adapted to circuit-level noise, and discussed how vertical pairs of highlighted vertices are created when applying local NN decoders. We showed how applying our local NN decoders paired with a syndrome collapse or vertical cleanup can significantly reduce the average number of highlighted vertices seen by a global decoder, thus allowing for a much faster implementation of such decoders. Performing a syndrome collapse also reduces the size of the matching graph used by the global NN decoder, providing even further runtime improvements. For some code distances and physical error rates, the syndrome densities were reduced by almost two orders of magnitude, and we expect even larger reductions when applying our methods to larger code distances than what was considered in this work. Further, our numerical results showed competitive logical error rates and a threshold of $p_{\text{th}} \approx 5 \times 10^{-3}$ for the syndrome collapse scheme and $p_{\text{th}} > 5 \times 10^{-3}$ for the vertical cleanup scheme. A trade-off between throughput and performance may be required in order to run algorithms with reasonable hardware overheads while still having fast enough decoders to avoid exponential backlogs during the implementation of algorithms. Although a more direct implementation of our local NN decoders on FPGA's appears challenging, encoding the NN parameters using fewer bits may satisfy the throughput requirements discussed in \cref{section:AlgoRun}. Using application-specific integrated circuits (ASICs) may also allow the implementation of our NN's on time scales sufficient for running algorithms. 

\begin{figure*}
	\centering
	\subfloat[\label{fig:SynXPrep}]{%
		\includegraphics[width=0.7\textwidth]{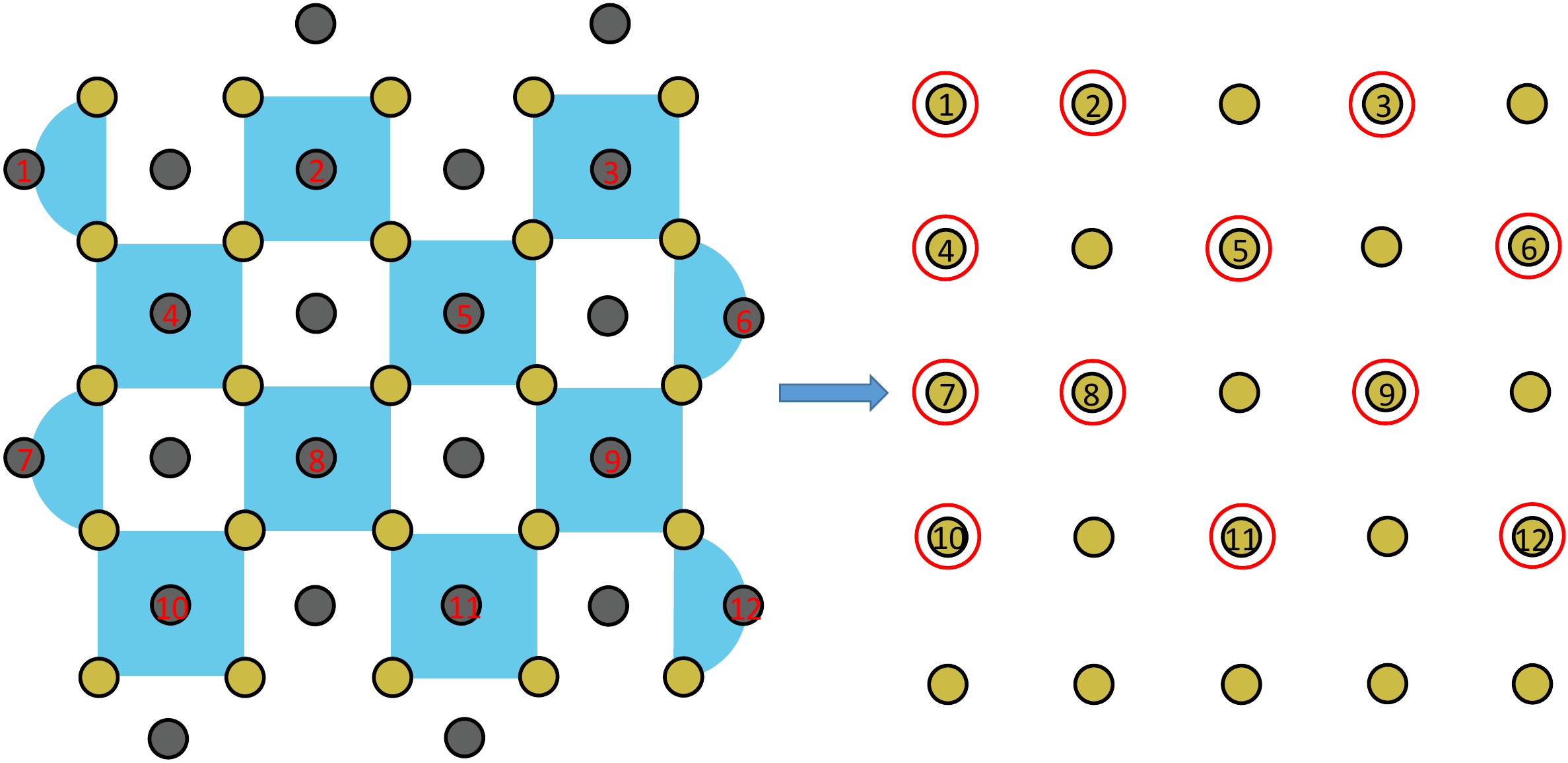}
	}
	\vfill
	\subfloat[\label{fig:SynZPrep}]{%
		\includegraphics[width=0.7\textwidth]{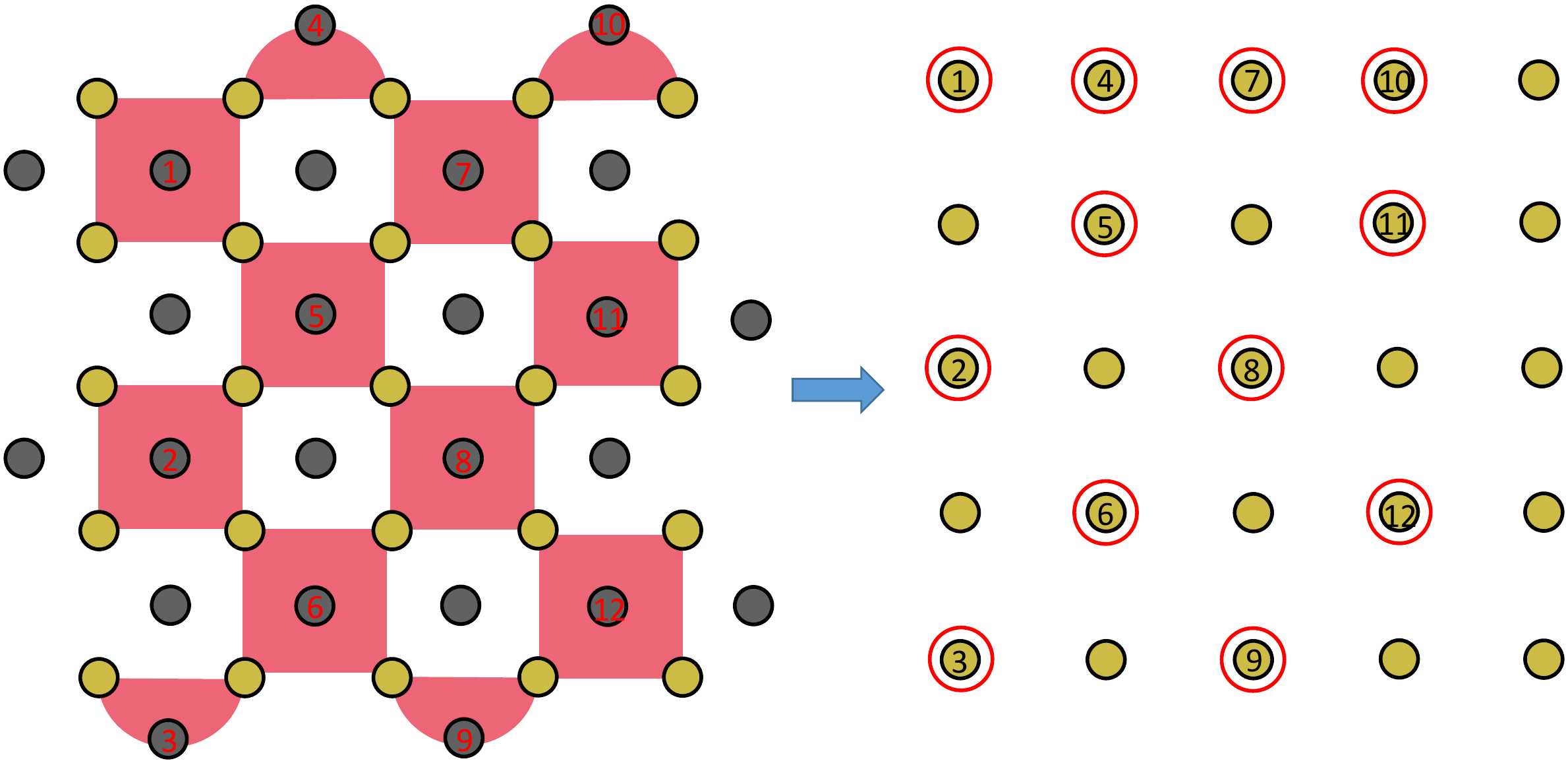}
	}

	\caption{\label{fig:SyndromeRepresentation} (a) Mapping of the $Z$-type stabilizer measurement outcomes for a $d=5$ surface code lattice to the matrix $M_{syn_X}$. For each stabilizer, which we label from 1 to $(d^2-1)/2$ going from left to right, top to bottom, the corresponding bit $b^{(X)}_k \in \{0,1\}$ where $1 \le k \le (d^2-1)/2$ (which is one if the stabilizer is measured non-trivially and zero otherwise) is mapped to a data qubit located at the top left corner of the square if the stabilizer is weight-4, or if it is a weight-2 stabilizer along the right boundary of the lattice. For weight-2 stabilizers along the left boundary of the lattice, the bit is mapped to the top right data qubit. The final binary matrix $M_{syn_X}$ has $d$ rows and $d$ columns, with ones at the circled regions in red if the corresponding stabilizer is measured non-trivially, otherwise the entry is zero. We also label each stabilizer numerically, starting at 1 and increasing by 1 left to right, top to bottom. The corresponding entries in $M_{syn_X}$ are given the same label. (b) Similar to (a), but for $Z$ error syndromes. The $X$-type red stabilizers map $b^{(Z)}_k$ to the top-left data qubit, except for weight-2 stabilizers on the top boundary of the lattice, which map $b^{(Z)}_k$ to the bottom-left data qubit.}
\end{figure*}

There are several avenues of future work. Firstly, adapting our NN decoding protocol to be compatible with sliding windows may lead to improved throughput times, as shown in \cref{appendix:BuffSlideWindow}. A broader NN architecture search may lead to networks with fewer parameters that still achieve low logical failure rates with modest hardware resource overhead requirements. For instance, graph based convolutional NN's \cite{GraphBased} appear to be promising in this regard. We can also design a network architecture which removes edges from the matching graph as part of its correction, rather than applying a data qubit correction followed by an error syndrome updated based on the correction. Such an architecture could make the syndrome collapse or vertical cleanup step unnecessary since for instance vertices incident to diagonal edges arising from space-time correlated errors would be flipped. By not performing a syndrome collapse or vertical cleanup, we anticipate that such networks could achieve lower logical error rates. Another important avenue would be to show how local NN architectures can be adapted to lattice surgery settings, where surface code patches change shape through time, and where new fault patterns which are unique to lattice surgery settings can occur \cite{CC22Twist}.  

Given the size of the NN's, we only considered performing one pass of the NN prior to implementing MWPM. However, performing additional passes may lead to sparser syndromes, which could be a worthwhile trade-off depending on how quickly the NN's can be implemented in classical hardware.

The training data also has a large asymmetry between the number of ones and zeros for the error syndromes and data qubit errors, with zeros being much more prevalent than ones. It may be possible to exploit such asymmetries by asymmetrically weighting the two cases.  

Lastly, other classical hardware approaches for implementing local NN decoders, such as ASICs, should be considered.

\section{Acknowledgements}
\label{sec:acknowledg}
C.C. would like to thank Aleksander Kubica, Nicola Pancotti, Connor Hann, Arne Grimsmo and Oskar Painter for useful discussions.

\appendix

\section{Data representation for training the NN's}
\label{app:DataRep}

In this appendix we describe how we represent the data used to train our convolutional NN's. In what follows, we refer to \texttt{trainX} as the input data to the NN used during training and \texttt{trainY} as the output targets. 

\begin{figure*}
	\centering
	\subfloat[\label{fig:BoundaryX}]{%
		\includegraphics[width=0.35\textwidth]{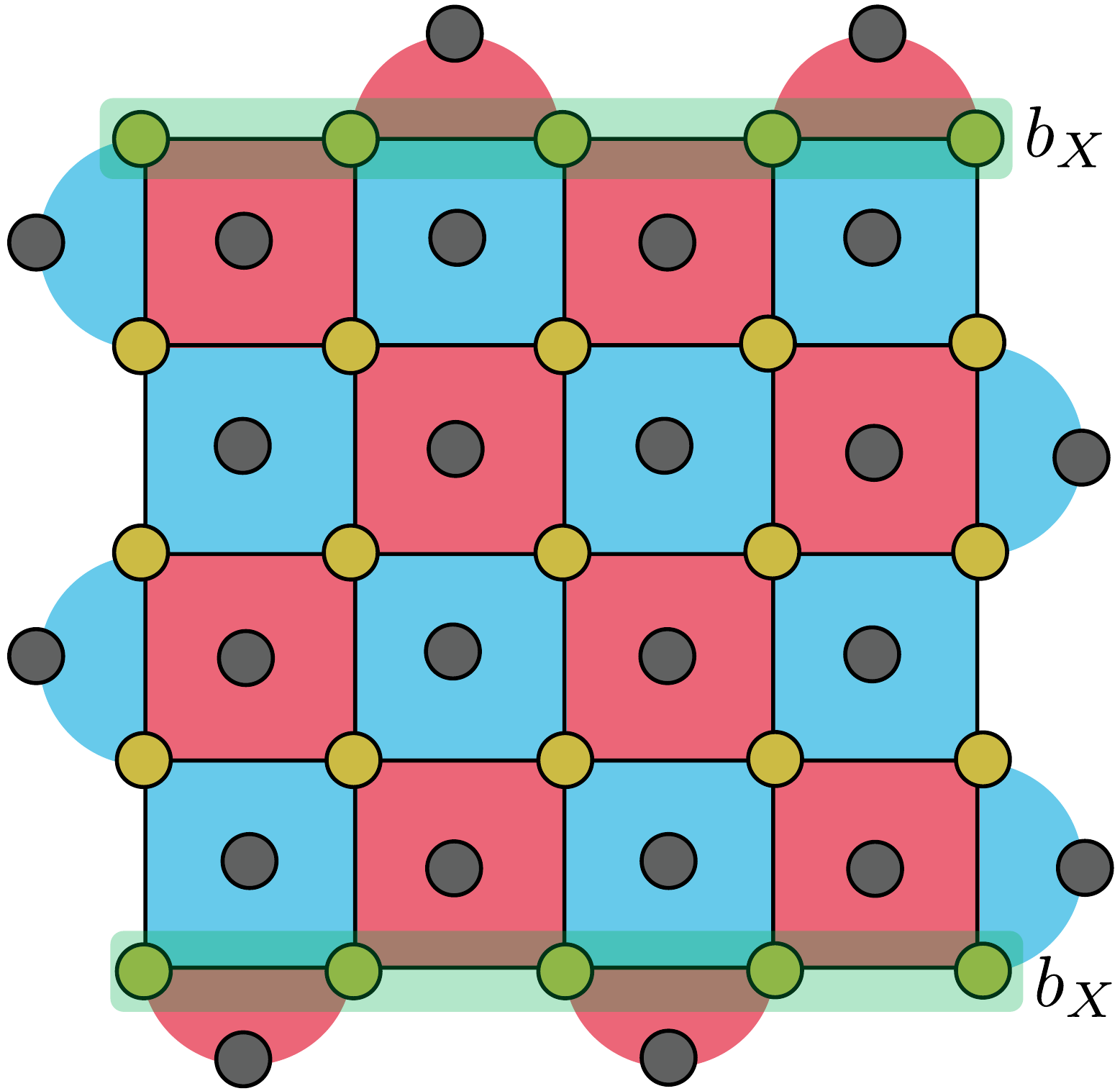}
	}
	\subfloat[\label{fig:BoundaryZ}]{%
		\includegraphics[width=0.35\textwidth]{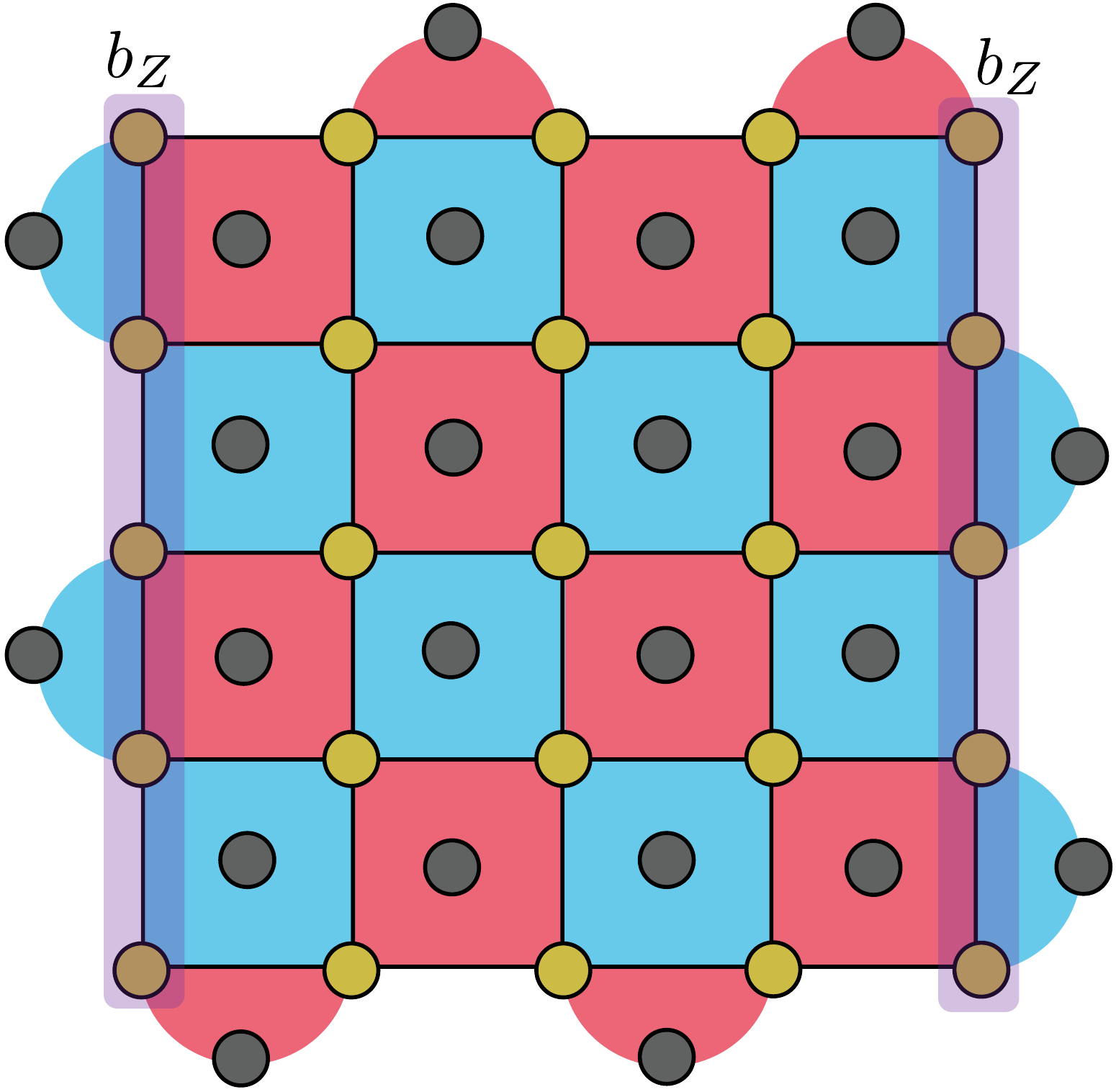}
	}

	\caption{\label{fig:BoundaryQubits} (a) Boundary $X$ qubits, highlighted in green, are located along the horizontal top and bottom boundaries of the lattice. (b) Boundary $Z$ qubits, highlighted in purple, are located along the vertical left and right boundaries of the lattice. }
\end{figure*}

As mentioned in \cref{subsec:NNdescription}, \texttt{trainX} is a tensor of shape $(N_{\text{train}},d_x,d_z,d_m,5)$, where $N_{\text{train}}$ is the number of training examples, $d_x$ and $d_z$ correspond to the size of the vertical and horizontal boundaries of the lattice, and $d_m$ corresponds to the number of syndrome measurement rounds, with the last round being a round of perfect error correction where the data qubits are measured in some basis. We also set $d_x=d_z=d$. 

The first two input channels to \texttt{trainX} correspond to the syndrome difference history $s^{\text{diff}}_X(d_m)$ and $s^{\text{diff}}_Z(d_m)$ defined in \cref{defXZsyndiff} where we only track changes in syndromes between consecutive rounds. Further, in order to make it easier for the NN to associate syndrome measurement outcomes with the corresponding data qubit errors resulting in that measured syndrome, syndrome measurement outcomes for the $j$'th round are converted to two-dimensional $d \times d$ binary matrices labelled $M_{syn_X}(j)$ and $M_{syn_Z}(j)$ following the rules shown in \cref{fig:SyndromeRepresentation}. Note however that the rules described in \cref{fig:SyndromeRepresentation} show how to construct the $M_{syn_X}(j)$ and $M_{syn_Z}(j)$ matrices based on the measurement outcomes of each stabilizer of the surface code in round $j$. To get the final representation for $s^{\text{diff}}_X(d_m)$ and $s^{\text{diff}}_Z(d_m)$, we compute the matrices $\tilde{M}_{syn_X}(j) = M_{syn_X}(j) \oplus M_{syn_X}(j-1)$ and $\tilde{M}_{syn_Z}(j) = M_{syn_Z}(j) \oplus M_{syn_Z}(j-1)$ for $j \ge 2$, with $\tilde{M}_{syn_X}(1) = M_{syn_X}(1)$ and $\tilde{M}_{syn_Z}(1) = M_{syn_Z}(1)$.

\begin{figure*}
	\centering
	\subfloat[\label{fig:HomologicalX}]{%
		\includegraphics[width=0.35\textwidth]{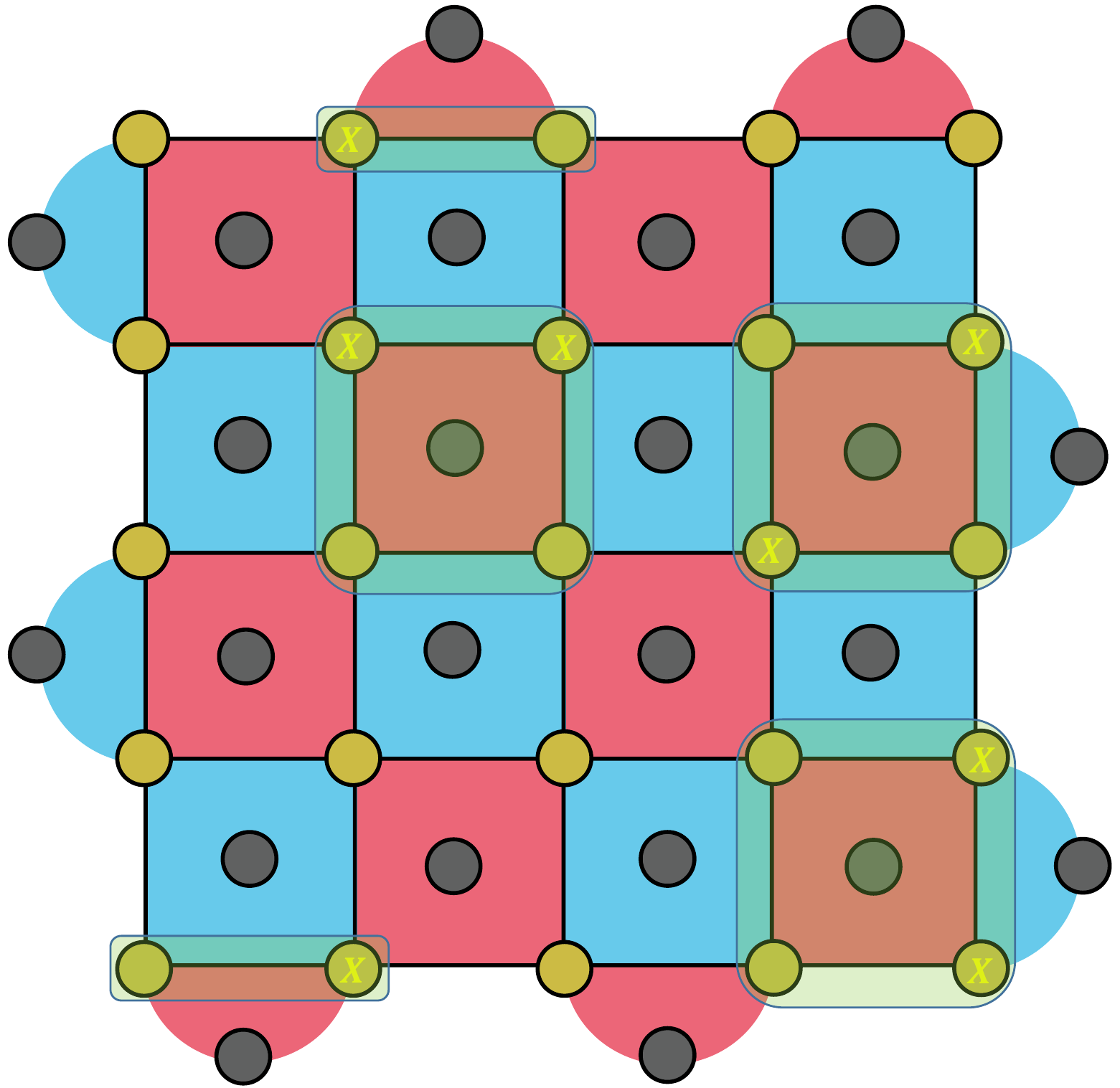}
	}
	\subfloat[\label{fig:HomologicalZ}]{%
		\includegraphics[width=0.35\textwidth]{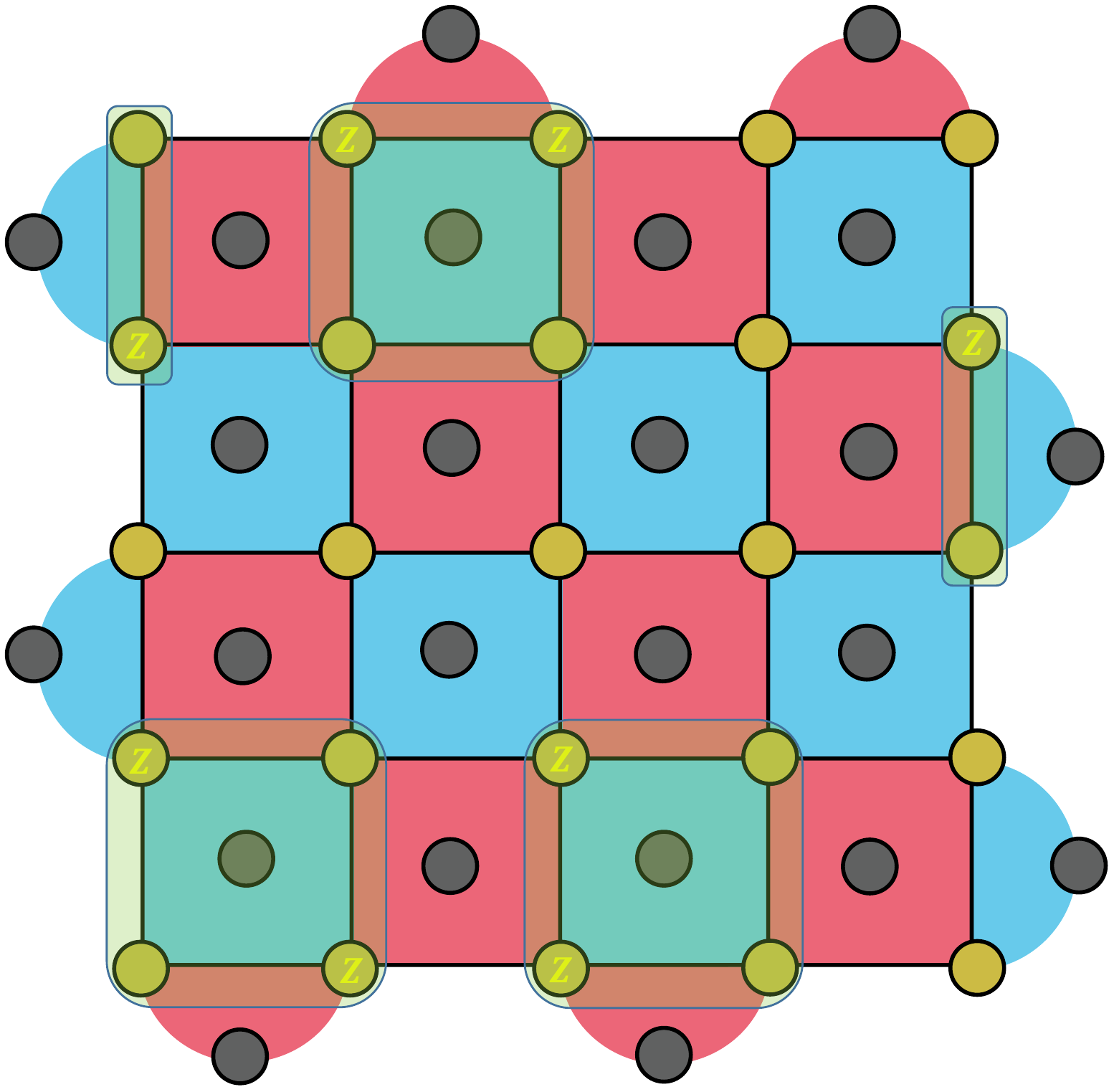}
	}

	\caption{\label{fig:HomologicalConvention} Homological equivalence convention as shown on a $d=5$ surface code lattice. (a) $X$ error configurations which are invariant under the transformations of the functions \texttt{weightReductionX} and \texttt{fixEquivalenceX}. (b) $Z$ error configurations which are invariant under the transformations of the functions \texttt{weightReductionZ} and \texttt{fixEquivalenceZ}.   }
\end{figure*}

As discussed in \cref{subsec:NNdescription}, the next two channels to \texttt{trainX} correspond to the matrices $\texttt{enc}(X)$ and $\texttt{enc}(Z)$ which are identical in each syndrome measurement round unless the surface code lattice changes shape, as would be the case when performing a parity measurement via lattice surgery. The matrices $\texttt{enc}(X)$ and $\texttt{enc}(Z)$ are encoded using the same rules as the encoding of the matrices $M_{syn_X}$ and $M_{syn_Z}$, except that a 1 is always inserted \textit{regardless} of whether a stabilizer is measured non-trivially or not. For instance, for a $d=5$ surface code, the matrices $\texttt{enc}(X)$ and $\texttt{enc}(Z)$ (of shape 5x5) would have 1's at all red circular regions in \cref{fig:SyndromeRepresentation} and 0 for all other positions. So, assuming a surface code patch which doesn't change shape through time, for this $d=5$ example we have
\begin{align}
    \texttt{enc}(X)_j = \begin{pmatrix} 1 & 1 & 0 & 1 & 0 
    \\ 1 & 0 & 1 & 0 & 1
    \\ 1 & 1 & 0 & 1 & 0 
    \\ 1 & 0 & 1 & 0 & 1
    \\ 0 & 0 & 0 & 0 & 0
    \end{pmatrix},
\end{align}
\begin{align}
    \texttt{enc}(Z)_j = \begin{pmatrix} 
    1 & 1 & 1 & 1 & 0
    \\ 0 & 1 & 0 & 1 & 0
    \\ 1 & 0 & 1 & 0 & 0 
    \\ 0 & 1 & 0 & 1 & 0
    \\ 1 & 0 & 1 & 0 & 0
    \end{pmatrix},
\end{align}
where $j \in \{ 1, \cdots, d_m \}$.

When the NN is in the bulk of the lattice, it can be seen from \cref{fig:SyndromeRepresentation} that syndromes associated with a particular data qubit changes shape depending on which data qubit is observed. For instance, on the second row of the lattice in \cref{fig:SynXPrep}, compare the vertices in red surrounding the qubit in the second column versus those surrounding the qubit in the third column. Since the matrices $\texttt{enc}(X)$ and $\texttt{enc}(Z)$ encode this information, providing such inputs to \texttt{trainX} helps the network distinguish between the different types of data qubits when the network's receptive field only sees qubits in the bulk. Similarly, $\texttt{enc}(X)$ and $\texttt{enc}(Z)$ allow the network to identify data qubits along the boundaries of the lattice. At the boundary, the pattern of 1's and 0's in $\texttt{enc}(X)$ and $\texttt{enc}(Z)$ is different than in the bulk. By using the encoding described by $\texttt{enc}(X)$ and $\texttt{enc}(Z)$, we observed significant performance improvements compared to an encoding which only specifies the location of the boundary $X$ and $Z$ data qubits, which are shown in \cref{fig:BoundaryX,fig:BoundaryZ} for the $d=5$ surface code. By boundary $X$ ($Z$) qubits, we refer to data qubits that result in a single non-trivial stabilizer measurement outcome when afflicted by an $X$ ($Z$) error.

Lastly, since the last round of error correction is a round of perfect error correction where the data qubits are measured in some basis, it is also important to specify the \textit{temporal} boundaries of the lattice. Specifying temporal boundaries allows the network to generalize to arbitrary syndrome measurement rounds. As such, the last channel of \texttt{trainX} contains the temporal boundaries, represented using $d_x \times d_z$ binary matrices for each syndrome measurement round. We choose an encoding where the matrices are filled with ones for rounds 1 and $d_m$, and filled with zeros for all other rounds.

\section{Homological equivalence convention for representing data qubit errors}
\label{appendix:HomEquivConv}

Let $E_1$ and $E_2$ be two data qubit errors. We say that $E_1$ and $E_2$ are homologically equivalent for a code $\mathcal{C}$ if $s(E_1) = s(E_2)$, and $E_1E_2 \in \mathcal{S}$ where $\mathcal{S}$ is the stabilizer group of $\mathcal{C}$. In other words, $E_1$ and $E_2$ are homologically equivalent for a code $\mathcal{C}$ if they have the same error syndrome, and are identical up to products of stabilizers. 

In Ref.~\cite{UsmanNN}, it was shown that training a NN where the data qubit errors were represented using a fixed choice of homological equivalence resulted in better decoding performance. In this appendix, we describe our choice of homological equivalence for representing the data qubit errors in \texttt{trainY} which resulted in improved decoding performance. 

Recall that \texttt{trainY} is a tensor of shape $(N_{\text{train}},d_x,d_z,d_m,2)$ where $N_{\text{train}}$ is the number of training examples. For a given training example, the first channel consists of $d_m$ binary $d \times d$ matrices $M^{(X_{(\alpha,\beta)})}_e(j)$, with $1 \le j \le d_m$ being the label for a particular syndrome measurement round, and $\alpha,\beta \in \{1,2, \cdots, d\}$ labelling the data qubit coordinates in the surface code lattice. Since \texttt{trainY} tracks changes in data qubit errors between consecutive syndrome measurement rounds, $M^{(X_{(\alpha,\beta)})}_e(j) = 1$ if the data qubit at coordinate $(\alpha,\beta)$ has a change in an $X$ or $Y$ error between rounds $j-1$ and $j$, and is zero otherwise. Similarly, the second channel of \texttt{trainY} consists of $d_m$ binary $d \times d$ matrices $M^{(Z_{(\alpha,\beta)})}_e(j)$ which tracks changes of $Z$ or $Y$ data qubit errors between consecutive syndrome measurement rounds. 

Now, consider a weight-4 $X$-type stabilizer $g^{(X)}_k$ represented by a red plaquette in \cref{fig:HomologicalConvention} (with $1 \le k \le (d^2-1)/2$), and let $(\alpha, \beta)$ be the data qubit coordinate at the top left corner of $g^{(X)}_k$. Any weight-3 $X$ error, with support on $g^{(X)}_k$ can be reduced to a weight-one error by multiplying the error by $g^{(X)}_k$. Similarly, a weight-4 $X$ error with support on $g^{(X)}_k$ is equal to $g^{(X)}_k$ and can thus be removed entirely. We define the function \texttt{weightReductionX} which applies the weight-reduction transformations described above to each stabilizer. Similarly, \texttt{weightReductionX} also removes weight-2 $X$ errors at weight-2 $X$-type stabilizers along the top and bottom boundaries of the lattice. 

Let $E_x$ be a weight-2 $X$ error with support on a weight-4 stabilizer $g^{(X)}_k$, where the top left qubit has coordinates $(\alpha, \beta)$. We define the function \texttt{fixEquivalenceX} as follows:
\begin{enumerate}
\item Suppose $E_x$ has support at the coordinates $(\alpha+1,\beta)$ and $(\alpha+1,\beta+1)$. Then \texttt{fixEquivalenceX} maps $E_x$ to a weight-2 error at coordinates $(\alpha,\beta)$ and $(\alpha,\beta+1)$. Thus horizontal $X$ errors at the bottom of $g^{(X)}_k$ are mapped to horizontal $X$ errors at the top of $g^{(X)}_k$.

\item Suppose $E_x$ has support at the coordinates $(\alpha,\beta)$ and $(\alpha+1,\beta)$.  Then \texttt{fixEquivalenceX} maps $E_x$ to a weight-2 error at coordinates $(\alpha,\beta+1)$ and $(\alpha+1,\beta+1)$. Thus vertical $X$ errors at the left of $g^{(X)}_k$ are mapped to vertical $X$ errors at the right of $g^{(X)}_k$.

\item Suppose $E_x$ has support at the coordinates $(\alpha,\beta)$ and $(\alpha+1,\beta+1)$.  Then \texttt{fixEquivalenceX} maps $E_x$ to a weight-2 error at coordinates $(\alpha,\beta+1)$ and $(\alpha+1,\beta)$. Thus diagonal $X$ errors from the top left to bottom right of $g^{(X)}_k$ are mapped to diagonal $X$ errors at the top right to bottom left of $g^{(X)}_k$.
\end{enumerate}

Next, let $g^{(X)}_k$ be a weight-2 $X$-type stabilizer along the top of the surface code lattice, with the left-most qubit in its support having coordinates $(\alpha, \beta)$. If $E_x$ is a weight-1 error at coordinates $(\alpha, \beta+1)$, \texttt{fixEquivalenceX} maps $E_x$ to a weight-1 error at coordinates $(\alpha, \beta)$. On the other hand, if $g^{(X)}_k$ is a weight-2 $X$-type stabilizer along the bottom of the surface code lattice with  with the left-most qubit in its support having coordinates $(\alpha, \beta)$, and $E_x$ is a weight-1 error at coordinates $(\alpha, \beta)$, \texttt{fixEquivalenceX} maps $E_x$ to a weight-1 error at coordinates $(\alpha, \beta+1)$.

Next let \texttt{simplifyX} be a function which applies \texttt{weightReductionX} and \texttt{fixEquivalenceX} to all $X$-type stabilizers of the surface code lattice in each syndrome measurement round (with \texttt{weightReductionX} being applied first), with $E_x$ errors in round $1 \le j \le d_m$ described by the binary matrix $M^{(X_{(\alpha,\beta)})}_e(j)$ for all $(\alpha, \beta)$ data-qubit coordinates. Thus \texttt{simplifyX} maps matrices $M^{(X_{(\alpha,\beta)})}_e(j)$ to homologically equivalent matrices $\tilde{M}^{(X_{(\alpha,\beta)})}_e(j)$ using the transformations described above. Our homological equivalence convention for $X$ data qubit errors is implemented by repeatedly calling the function \texttt{simplifyX} until all matrices $M^{(X_{(\alpha,\beta)})}_e(j)$ satisfy the condition $\texttt{simplifyX}(M^{(X_{(\alpha,\beta)})}_e(j)) = M^{(X_{(\alpha,\beta)})}_e(j)$ for all syndrome measurement rounds $j$ and data qubit coordinates $(\alpha, \beta)$.

For $Z$-type data qubit errors, we similarly have a \texttt{weightReductionZ} function which reduces the weights of $Z$ errors at each $Z$-type stabilizer. The function \texttt{fixEquivalenceZ} is chosen to be rotationally symmetric to the function \texttt{fixEquivalenceX} under a $90^{\circ}$ rotation of the surface code lattice. We then define a \texttt{simplifyZ} function in an identical way as \texttt{simplifyX}, but which calls the functions \texttt{weightReductionZ} and \texttt{fixEquivalenceZ}. Errors which are invariant under the transformations \texttt{simplifyX} and \texttt{simplifyZ} are shown in \cref{fig:HomologicalConvention}.

\begin{figure}
    \centering
    \includegraphics[width=0.95\columnwidth]{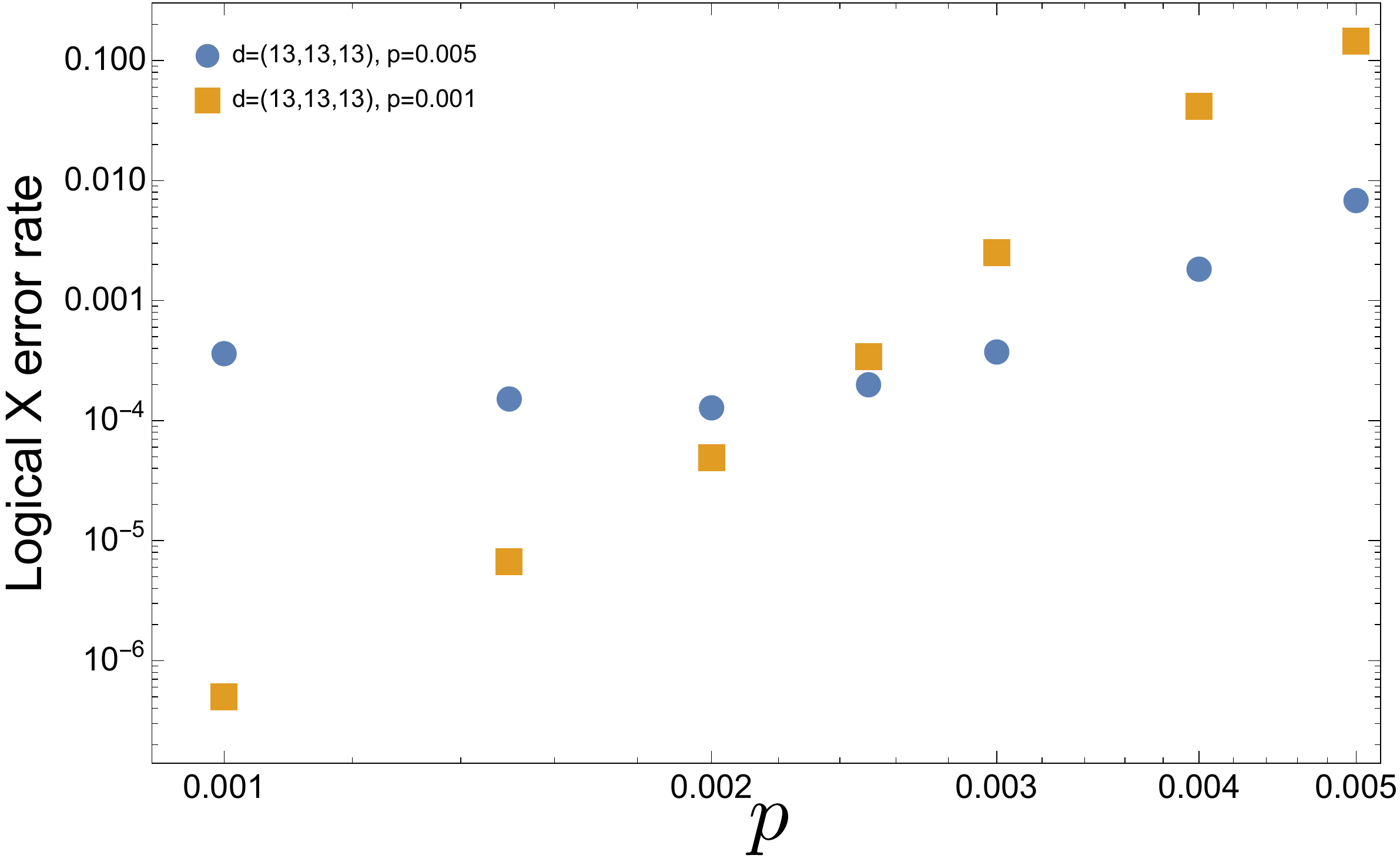}
    \caption{Comparison of the logical $X$ error rate of the 6-layer model trained at $p = 0.001$ (orange squares) and $p = 0.005$ (blue circles) applied to all physical error rates of the test data generated with a volume of size $(13,13,13)$. After the application of the local NN decoder, we perform a syndrome collapse with sheets of size $d'_m=6$ followed by MWPM to correct any remaining errors.}
    \label{fig:CompareModelsPlot}
\end{figure}

\begin{figure*}
	\centering
	\subfloat[\label{fig:SlideWind1}]{%
		\includegraphics[width=0.45\textwidth]{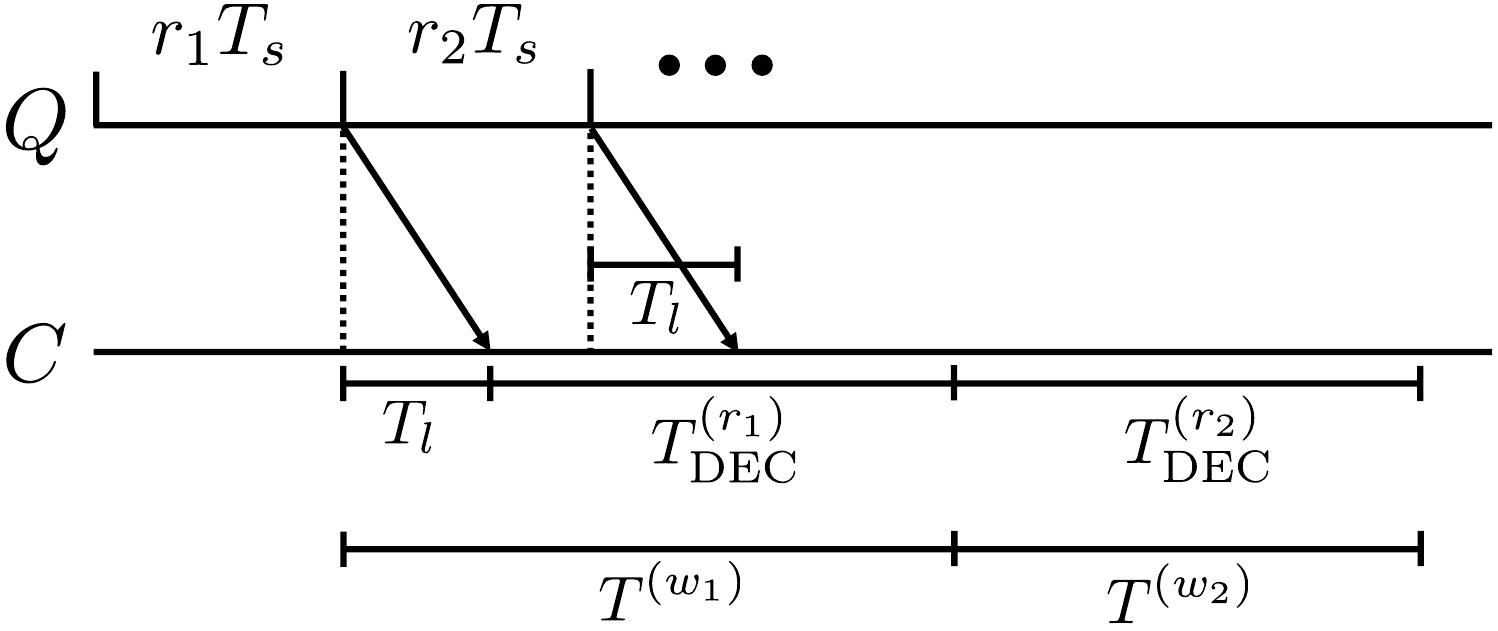}
	}
	\subfloat[\label{fig:SlideWind2}]{%
		\includegraphics[width=0.45\textwidth]{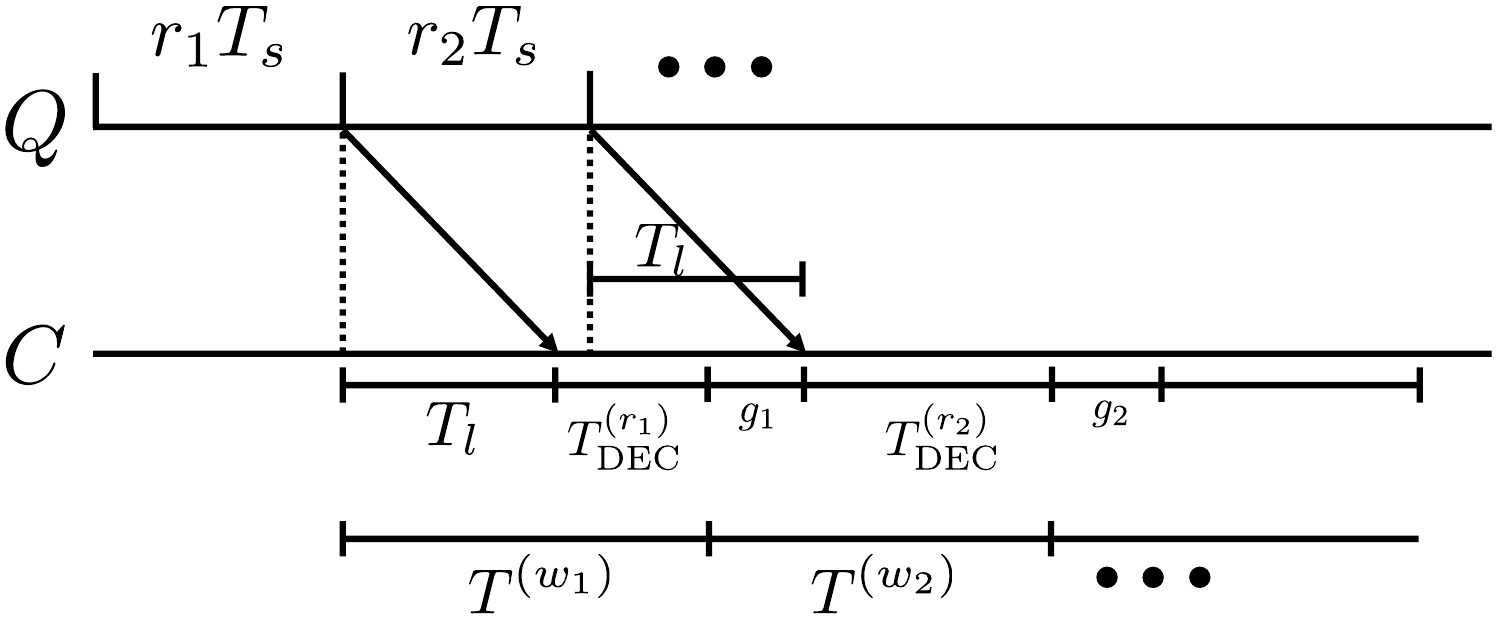}
	}

	\caption{\label{fig:SlideWind} Dividing the number of syndrome measurement rounds into windows, with the $j$'th window containing $r_j$ rounds. In (a), we consider the "slow" case where decoding $r_j$ rounds takes longer than performing $r_j$ syndrome measurement rounds so that $r_j T_s < T^{(r_j)}_{\text{DEC}}$. In (b), we consider the "fast" case where $r_j T_s > T^{(r_j)}_{\text{DEC}}$. Here the $Q$ axis indicates operations performed on the quantum computer, and the $C$ axis are operations performed on the classical computer, with $T_l$ being the latency time.}
\end{figure*}

\section{Comparing models trained at different error rates}
\label{appendix:CompareModels}
In this appendix we discuss in more detail the effects of applying a network trained at different physical error rates to the test set data. 

In \cref{fig:CompareModelsPlot}, we show the logical $X$ error rate curves of the 6-layer network in \cref{fig:NetworkArchitectures} trained at $p = 0.005$ and $p = 0.001$ on training data of size $(13,13,18)$ when applied to test set data generated with a volume of size $(13,13,13)$. The application of the local NN decoder is followed by a syndrome collapse with sheets of size $d'_m=6$ and MWPM to correct any remaining errors. As can be seen in the plot, for $p \ge 0.0025$, the network trained at $p=0.005$ outperforms the network trained at $p=0.001$. However, when we apply the network trained at $p=0.005$ to test set data generated for $p \le 0.002$, not only does the model under-perform the one trained at $p=0.001$, but the logical failure rate \textit{increases} with decreasing $p$. Such a result suggests that the model trained at $p=0.005$ is over-fitting to the data generated at higher physical error rates which has denser syndromes. Consequently, the model does not generalize well to data containing sparser syndromes observed at lower physical error rates.  

The above results show the importance of training models at different physical error rates when applying such models to the test set data. 

\section{Effects on buffer times using sliding windows}
\label{appendix:BuffSlideWindow}

In this appendix we show how the buffer times, as described in \cref{section:AlgoRun}, can be improved by decoding using sliding windows instead of decoding over all syndrome measurement rounds of the full syndrome measurement volume. In particular, we focus on showing how the expression for $T^{b_1}$ in \cref{eq:Tbuff1} is modified in the when using sliding windows.

Suppose we perform $r$ syndrome measurement rounds. We divide all syndrome measurement rounds into $n_w$ windows $\{ w_1, w_2, \cdots, w_{n_w} \}$ with window $w_j$ containing $r_j$ syndrome measurement rounds. In our analysis we consider two cases. The "slow" case is when decoding $r_j$ rounds takes longer than performing $r_j$ syndrome measurement rounds as shown in \cref{fig:SlideWind1}. In this case we have $r_j T_s < T^{(r_j)}_{\text{DEC}}$. The "fast" case is the opposite where decoding $r_j$ rounds takes a shorter amount of time than performing $r_j$ syndrome measurement rounds, so that $r_j T_s > T^{(r_j)}_{\text{DEC}}$. An illustration of the fast case is shown in \cref{fig:SlideWind2}. In what follows, we define $T^{(w_j)}$ as the time it takes to perform all $r_j$ syndrome measurement rounds and decode them for the window $w_j$. Thus we have that 
\begin{align}
    T^{b_1} = \sum_{j = 1}^{n_w} T^{(w_j)}.
    \label{eq:AppTb1}
\end{align}
We also assume that $T_l < r_j T_s$ for all $1 \le j \le n_w$.

For both the fast and slow cases, we have that $T^{(w_1)} = T_l + T^{(r_1)}_{\text{DEC}}$ since the classical computer must wait for a time $T_l$ from the last syndrome measurement round in the first window before it can begin decoding the $r_1$ syndrome measurement rounds, which takes time $T^{(r_1)}_{\text{DEC}}$. For the second window, if $r_2T_s < T^{(r_1)}_{\text{DEC}}$, then the signal from the last syndrome measurement round in the second window will arrive to the classical computer while it is still decoding syndromes from the first window, so that $T^{(w_2)} = T^{(r_2)}_{\text{DEC}}$. On the other hand, if $r_2T_s > T^{(r_1)}_{\text{DEC}}$, then decoding errors in the first window will complete before the syndrome information from the second window is available to the classical computer. As such, the total time to process syndrome in the second window will be $T^{(w_2)} = g_1 + T^{(r_2)}_{\text{DEC}}$ where $g_1$ is the time it takes for the syndrome information from the second window to be made available to the classical computer after decoding syndromes from the first window. From \cref{fig:SlideWind2} we see that $g_1 = r_2 T_s - T^{(r_1)}_{\text{DEC}}$. Summarizing, we have that
\begin{align}
T^{(w_2)} = 
    \begin{cases} 
      T^{(r_2)}_{\text{DEC}} &  r_2 T_s < T^{(r_1)}_{\text{DEC}} \\
      T^{(r_2)}_{\text{DEC}} + r_2 T_s - T^{(r_1)}_{\text{DEC}} & r_2 T_s > T^{(r_1)}_{\text{DEC}}
    \end{cases}
\end{align}
Implementing the above steps recursively, we find that
\begin{align}
    T^{b_1} = 
    \begin{cases} 
       T_l + \sum_{i=1}^{n_w}T^{(r_i)}_{\text{DEC}} &  r_i T_s < T^{(r_{i-1})}_{\text{DEC}} \\
      T_l + \sum_{i=2}^{n_w}r_i T_s & r_i T_s > T^{(r_{i-1})}_{\text{DEC}}
    \end{cases}
    \label{eq:AppTb1Final}
\end{align}
We see that if $r_i T_s < T^{(r_{i-1})}_{\text{DEC}}$ for all $i \in \{1,  \cdots ,  n_w. \}$, then the analysis leading to \cref{eq:AppTb1Final} shows that the buffer times will satisfy \cref{eq:SolBuffLinear} in the case where $T^{(r_j)}_{\text{DEC}} = cr_j$ for all $j$. However for decoding times expressed as a polynomial of degree greater than or equal to 2, summing the terms in \cref{eq:AppTb1Final} over smaller window sizes can lead to much smaller buffer times. 

Note that in \cref{fig:SlideWind2}, we assumed that $T_l < r_j T_s$. In the large latency regime, where $T_l > r_j T_s > T^{(r_{j-1})}_{\text{DEC}}$ for all $j$, a quick calculation shows that $T^{b_1} = T_l + \sum_{i=2}^{n_w} r_i T_s$ and so the result is unchanged.

\section{Dependence of the surface code distance on $d_m$}
\label{appendix:SurfDvsDm}

In \cref{section:AlgoRun} we showed how the buffer times $T^{b_j}$ can increase with the number of consecutive non-Clifford gates in a quantum algorithm. One may be concerned that a large increase in buffer times could result in a much larger surface code distance in order to maintain a target logical failure rate $\delta$ set by a particular quantum algorithm. In this appendix we show that the code distance increase logarithmically with increasing buffer times.

\begin{figure}
    \centering
    \includegraphics[width=0.9\columnwidth]{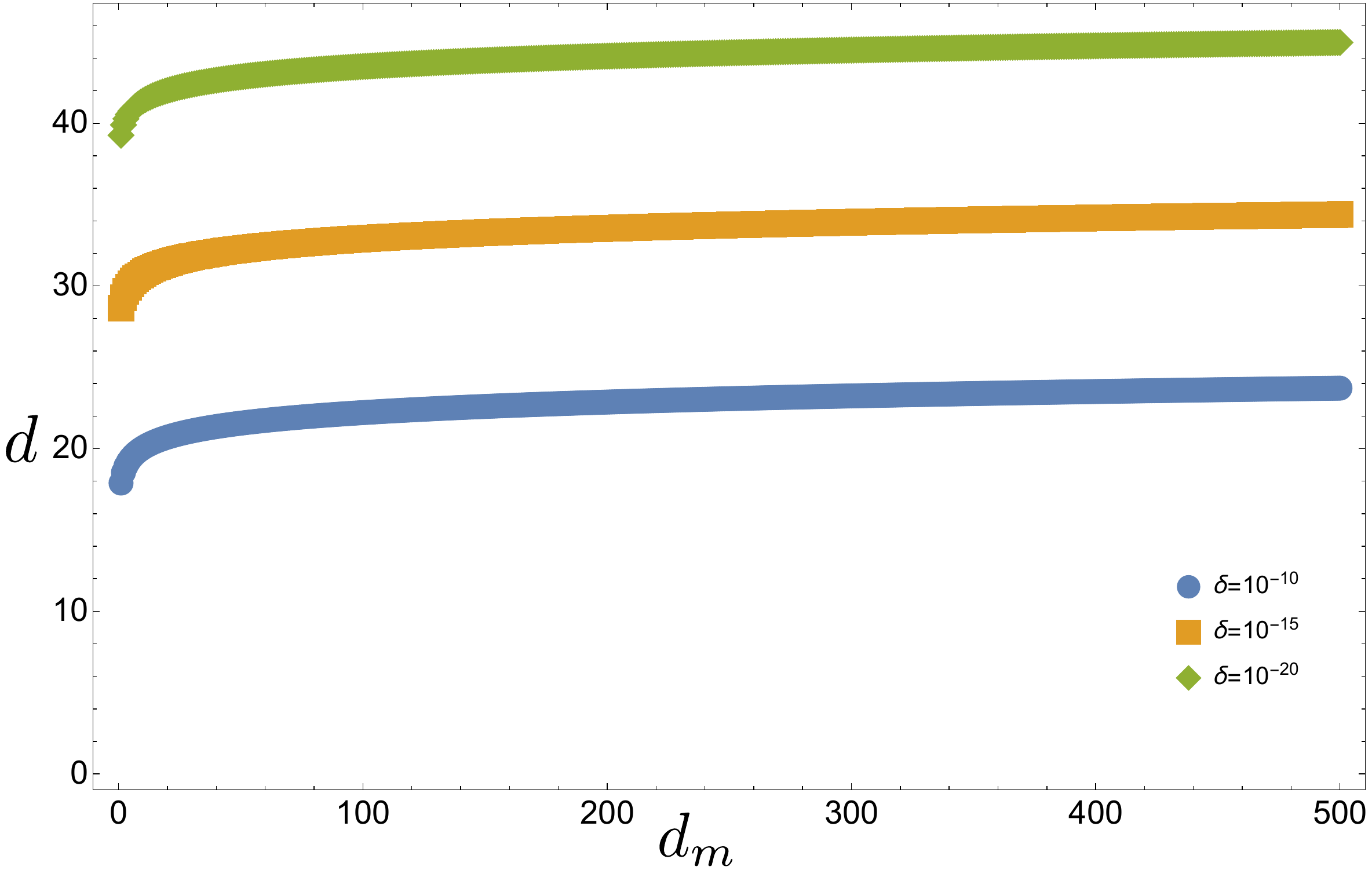}
    \caption{Plot of the surface code distance $d$ as a function of $d_m$ using the logical error rate polynomial $p^{(X;\text{vc})}_{L;\text{11l}}(p)$ given in \cref{eq:11layerPolyVC}. We set $p=10^{-3}$ and plot for different values of $\delta$ with the requirement that $p^{(X;\text{vc})}_{L;\text{11l}}(p) < \delta$.}
    \label{fig:dAsAfunctionOfDm}
\end{figure}

\begin{figure*}
	\centering
	\subfloat[\label{fig:HlatticeV1}]{%
		\includegraphics[width=0.45\textwidth]{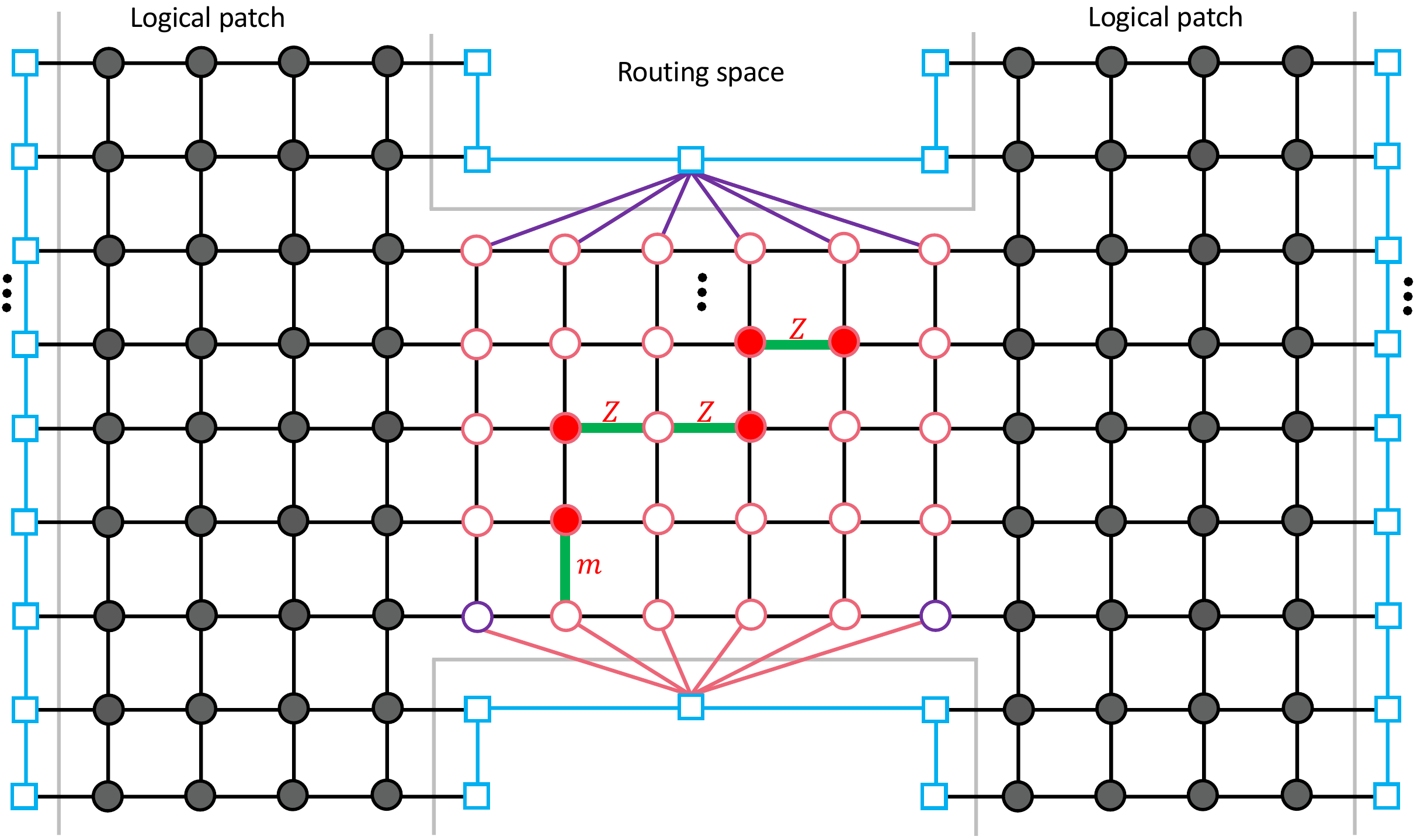}
	}
	\subfloat[\label{fig:HlatticeV2}]{%
		\includegraphics[width=0.45\textwidth]{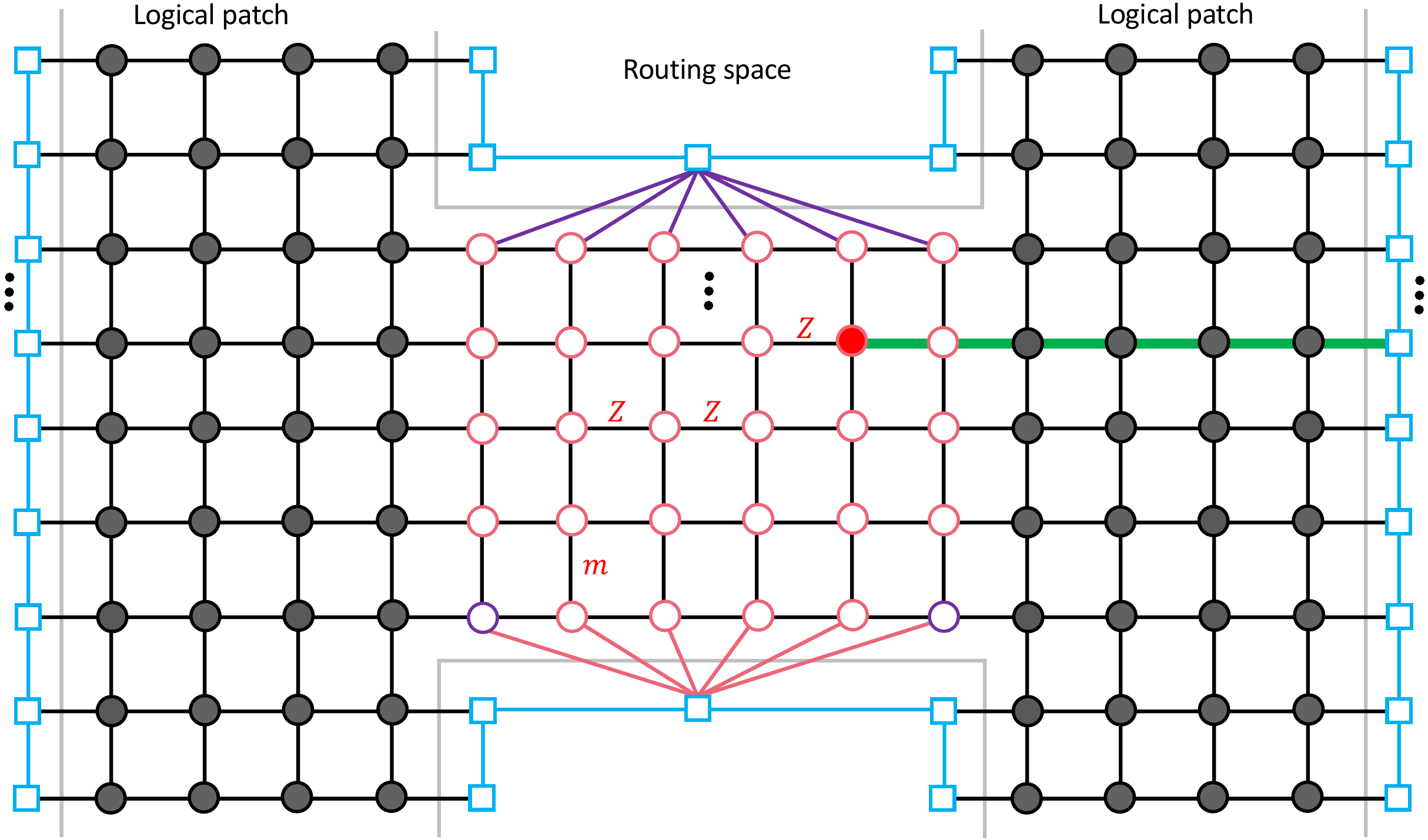}
	}

	\caption{\label{fig:HLattice} Illustration of a slice of a matching graph used to correct errors during an $X \otimes X$ multi-qubit Pauli measurement performed via lattice surgery. Highlighted vertices are shown in red, and we include temporal edges (shown in pink and purple) incident to vertices in the routing region. (a) Series of $Z$ errors and measurement errors occurring in the routing space region. The measurement error flips the parity of the multi-qubit Pauli measurement. The corrections are shown by the edges highlighted by thick green lines. The $Z$ errors are removed by performing MWPM, and the parity of $X \otimes X$ is flipped to the correct value. (b) Same as (a) but where we perform a vertical cleanup. In this case, MWPM can perform a string of $Z$ corrections to a $Z$ boundary, resulting in a logical $Z$ error on one of the surface code patches. Another option is to match to the top temporal boundary, which results in a timelike failure. In both (a) and (b), a local NN decoder is not applied, in order to illustrate the effects of performing a vertical cleanup during a lattice surgery protocol. }
\end{figure*}

Recall that the logical $X$ error rate polynomial for a surface of distance $(d_x,d_z)$ is given by
\begin{align}
    p^{(X)}_L = u d_m d_z (bp)^{(c d_x + k)},
\end{align}
for some constants $u$, $b$, $c$ and $k$ and where we assume that $d_m$ syndrome measurement rounds were performed. A quantum algorithm will have some target logical error rate $\delta$ with the requirement that $p^{(X)}_L < \delta$. Setting $d_x = d_z = d$ and solving for $d$ results in
\begin{align}
    d = \frac{\text{ProductLog}((c (b p)^{-k} \delta \log{b p})/(u d_m))}{c \log{(b p)}},
    \label{eq:ProductLog}
\end{align}
where $\text{ProductLog}(x)$ gives the principle solution for $w$ in the equation $x = w e^{w}$. 

In \cref{fig:dAsAfunctionOfDm} we show a plot of $d$ as a function of $d_m$ for various values of $\delta$ and fix $p$ to be $p=10^{-3}$. We used the logical $X$ error rate polynomial $p^{(X;\text{vc})}_{L;\text{11l}}(p)$ given in \cref{eq:11layerPolyVC} obtained by applying the 11-layer local NN decoder followed by a vertical cleanup and MWPM. As can be seen in \cref{fig:dAsAfunctionOfDm}, a large increase in $d_m$ results in a very modest increase in $d$, showing that increasing buffer times will not have a large effect on the surface code distance.

\section{Effects of performing a vertical cleanup during a parity measurement implemented via lattice surgery}
\label{appendix:VertCleanLatticeSurgery}

\begin{figure*}
	\centering
	\subfloat[\label{fig:VertTimeLikeCleanV1}]{%
		\includegraphics[width=0.6\textwidth]{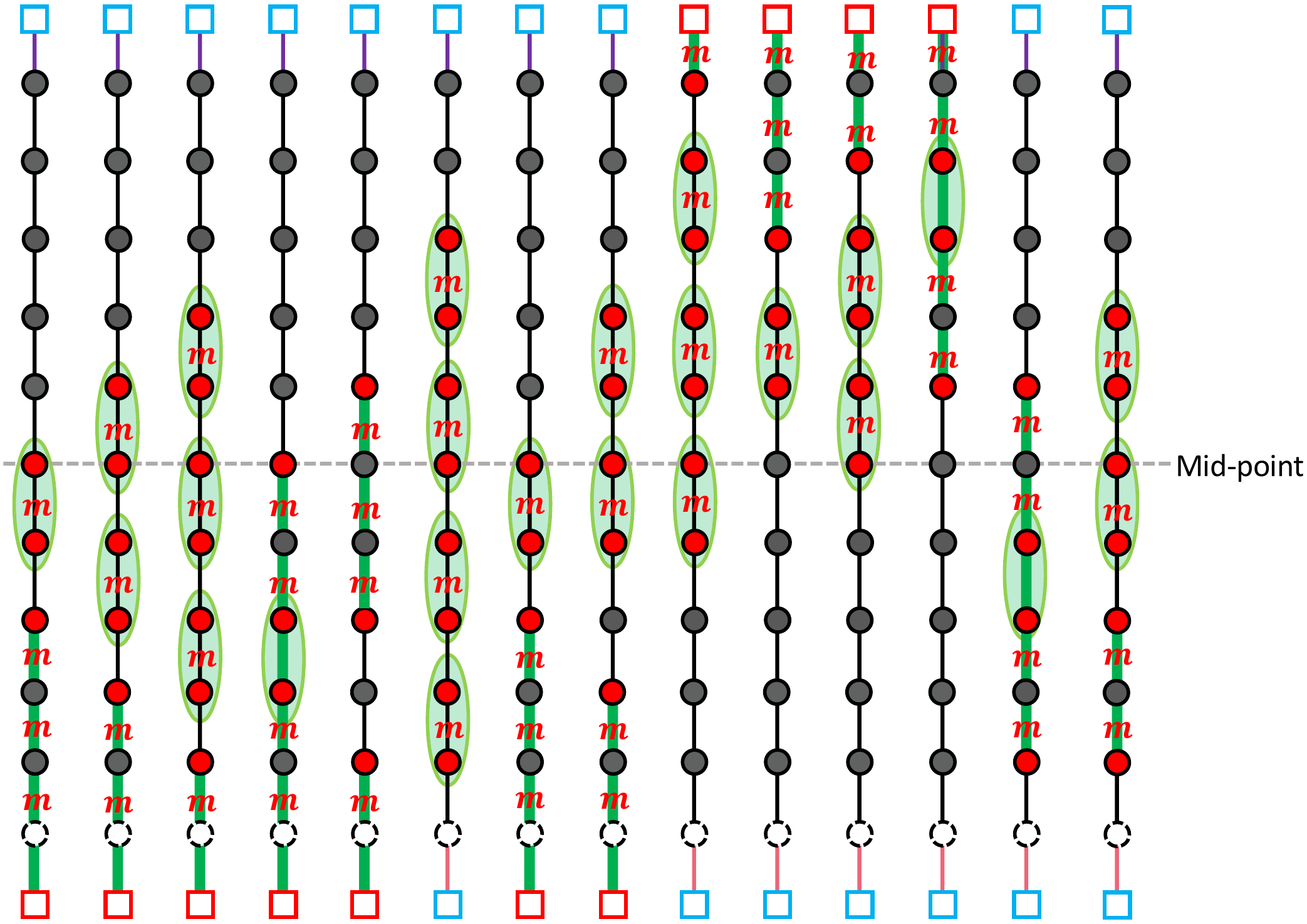}
	}
	\subfloat[\label{fig:VertTimeLikeCleanV2}]{%
		\includegraphics[width=0.023\textwidth]{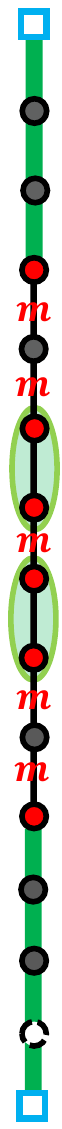}
	}

	\caption{\label{fig:VertTimeLikeClean} (a) Figure showing various configurations of measurement errors (labelled $m$) of 11 syndrome measurement rounds on the same stabilizer during a multi-qubit parity measurement implemented via lattice surgery. Each column of vertices corresponds to a different configuration of measurement errors for a given ancilla qubit. Time flows from the bottom of the figure to the top, and we illustrate temporal edges connecting to the boundary vertices following the same convention as in \cref{fig:HLattice}. Vertices in the first syndrome measurement round are shown in white since the measurement outcomes are random, thus such vertices are never highlighted. Pairs of highlighted vertices circled in green are removed when performing a vertical cleanup, and edges covered by thick green lines indicate the path chosen by a MWPM decoder after performing a vertical cleanup. For $d_m$ syndrome measurement rounds (with $d_m$ odd), the round labelled mid-point is the $(d_m+1)/2$ round. The goal of the figure is to illustrate that if the syndrome density above the mid-point is greater than the one below the mid-point, the vertical cleanup is done from bottom to top. On the other hand, if the syndrome density below the mid-point is greater than above, the vertical cleanup is performed from top to bottom. If they are the same, then a direction for the vertical cleanup is chosen at random. (b) Sequence of measurement errors for 13 syndrome measurement rounds where after performing a vertical cleanup, the minimum-weight correction  matches to the temporal boundary thus incorrectly flipping the parity. }
\end{figure*}

In this appendix we review the effects of performing a vertical cleanup when implementing a multi-qubit Pauli measurement via lattice surgery. For full details on the derivation of the matching graph, and the effects of timelike failures, the reader is referred to Ref.~\cite{CC21}.

We consider the simple case of performing an $X \otimes X$ multi-qubit Pauli measurement using two surface code patches. When performing the $X \otimes X$ measurement, the two surface code patches are merged into one patch by preparing qubits in the routing region in the $|0\rangle$ state, and performing a gauge fixing step where the $X$-type operators are measured \cite{Vuillot_2019}. A two-dimensional slice of the matching graph used to correct $Z$-type errors during the lattice surgery protocol is shown in \cref{fig:HLattice}. In particular, in the first round of the merge, the $X$-type measurements performed in routing space region are random, but the product of all such measurements encode the parity of the logical $X \otimes X$ operator being  measured. However, measurement errors can result in the wrong parity being measured. More generally, any fault mechanism resulting in an error which anticommutes with the $X \otimes X$ operator being measured will cause the wrong parity to be measured, and is referred to as a timelike failure. As such, repeated rounds of syndrome measurements are performed on the merged surface code patches, with the timelike distance given by the number of syndrome measurement rounds. 

When performing a vertical cleanup however, timelike failures which were correctable if no vertical cleanup were performed may no longer be correctable, with an example given in \cref{fig:HLattice}. In particular, strings of measurement errors starting from the first round of the merge patch would be unaffected by the implementation of a vertical cleanup, since a single vertex at the end of the measurement error string would be highlighted. The problematic cases arise when such error strings are combined with data qubit errors resulting in vertical pairs (as shown in \cref{fig:HlatticeV1}). 

We now show that there is preference in the ordering in which a vertical cleanup is performed which depends on the syndrome density either below or above some mid-point round. We also show the minimum temporal distance required to deal with a set of malignant failures, and discuss modifications to the vertical cleanup protocol to mitigate such effects. Note that in what follows, we do not remove vertical pairs between a highlighted vertex and a highlighted temporal boundary vertex. 

Suppose we perform $d_m$ syndrome measurement rounds when merging surface code patches to perform a multi-qubit Pauli measurement via lattice surgery. Consider the following sequence of measurement errors which occur when measuring some stabilizer $g_i$. In the first syndrome measurement round, a measurement error occurs resulting in the wrong parity of the multi-qubit Pauli measurement. Afterwords, a measurement error occurs every two syndrome measurement rounds, until there are a total of $(d_m-3)/2$ measurement errors. An example of such a sequence of faults is given in the third column of \cref{fig:VertTimeLikeCleanV1}. Performing a vertical cleanup starting from the first syndrome measurement round would result in a single highlighted vertex separated to the top temporal boundary by 4 vertical edges. Clearly for large $d_m$ and assuming all vertical edges have unit weight, MWPM would choose a path matching to the top temporal boundary resulting in timelike failure. However, if we performed a vertical cleanup starting from the last syndrome measurement round and moving downwards (i.e. towards the first syndrome measurement round), then the remaining highlighted vertex would be separated to the bottom temporal boundary by a single edge of unit weight. In this case, MWPM would correctly identify the parity measurement error. More generally, suppose $d_m$ syndrome measurement rounds are performed (with $d_m$ being odd) on a merged surface code patch part of parity measurement implemented via lattice surgery. We define the mid-point round to be the round $(d_m+1)/2$. As can be seen in \cref{fig:VertTimeLikeCleanV1}, for a given vertex of the syndrome measurement graph corresponding to particular stabilizer, if such a vertex is highlighted a larger number of times below the mid-point than above, a vertical cleanup on that vertex should be performed from top to bottom (i.e. starting from the round where the data qubits in the routing space are measured, and moving towards the round where they are initialized). On the other hand, if the density above the mid-point is greater than below, a vertical cleanup is performed in the opposite direction. Various configurations of measurement errors are illustrated in \cref{fig:VertTimeLikeCleanV1} showing that choosing the ordering for the vertical cleanup scheme as describe above avoids logical timelike failures.

Despite the above, there is still a sequence of measurement errors where regardless of the direction in which a vertical cleanup is performed, a timelike failure will occur. Consider a sequence of measurement errors occurring in two consecutive rounds (where the first round is after the surface code patch has been merged), followed by $m-4$ measurement errors every two rounds, and terminating with two consecutive measurement errors again, so that the total number of measurement errors is $m$. An example is shown in \cref{fig:VertTimeLikeCleanV2}. After performing a vertical cleanup, there will be two remaining highlighted vertices, associated with the first and last rounds of the sequence of measurement errors. The number of vertical edges connecting the two vertices which don't go through temporal boundary vertices is $n_v = 2m-3$, and the number of vertical edges connecting the two vertices which go through the temporal boundary is $n^{c}_v = d_m - 2m + 2$. As such, to ensure that MWPM does not map to a temporal boundary, thus incorrectly flipping the parity of the multi-qubit Pauli measurement, we must choose a large enough value of $d_m$ such that $n^{c}_v > n_v$ resulting in $d_m > 4m - 5$. Such an increase in $d_m$ has the effect of roughly doubling the runtime of a quantum algorithm. This increase in $d_m$ should be expected since performing a vertical cleanup is equivalent to adding additional measurement errors to the system  ``by hand", thus requiring a doubling in the code distance to have the same protection compared to a scheme which doesn't  perform a vertical cleanup. 

Two variations of the vertical cleanup protocol during a lattice surgery merge may maintain the full timelike distance and thus require fewer syndrome measurement rounds. The first variation would consist of identifying vertical pairs \textit{prior} to applying the local NN decoder, and only removing \textit{new} vertical pairs which are created after the local NN decoder is applied. In this case, vertical pairs due to measurement errors would not be removed, although this comes at the the cost of a higher syndrome density. Another approach would be to re-weight vertical edges incident to highlighted vertices which were removed from a vertical cleanup protocol, so that MWPM would give preferences to paths which go through such vertices. Lastly, using a TELS protocol described in Ref.~\cite{CC21} would allow larger timelike failure rates and thus could be used to avoid having to use a large value of $d_m$ when performing a vertical cleanup. We leave the numerical analysis of such protocols, along with using TELS alongside a vertical cleanup strategy, to future work.

\bibliography{NNbibliography}

\end{document}